\documentclass[twocolumn]{aastex631}

\usepackage[super]{nth}
\usepackage{hyperref}

\usepackage{color}

\usepackage{url}

\newcommand\kms{\ifmmode{\rm km\thinspace s^{-1}}\else km\thinspace s$^{-1}$\fi}
\newcommand\MAa{6.85~}
\newcommand\MAb{6.11~}
\newcommand\MB{7.90~}

\usepackage{xcolor}

\long\def\comment#1{}

\begin{document}

\title{TIC 290061484: A triply-eclipsing triple system with the shortest known outer period of 24.5 days}

\correspondingauthor{Veselin B. Kostov}
\email{veselin.b.kostov@nasa.gov}
\author[0000-0001-9786-1031]{V.~B.~Kostov}
\affiliation{NASA Goddard Space Flight Center, 8800 Greenbelt Road, Greenbelt, MD 20771, USA}
\affiliation{SETI Institute, 189 Bernardo Ave, Suite 200, Mountain View, CA 94043, USA}
\author[0000-0003-3182-5569]{S. A. Rappaport}
\affiliation{Department of Physics, Kavli Institute for Astrophysics and Space Research, M.I.T., Cambridge, MA 02139, USA}
\author[0000-0002-8806-496X]{T. Borkovits}
\affiliation{Baja Astronomical Observatory of University of Szeged, H-6500 Baja, Szegedi út, Kt. 766, Hungary}
\affiliation{HUN-REN -- SZTE Stellar Astrophysics Research Group,  H-6500 Baja, Szegedi út, Kt. 766, Hungary}
\affiliation{Konkoly Observatory, Research Centre for Astronomy and Earth Sciences, H-1121 Budapest, Konkoly Thege Miklós út 15-17, Hungary}
\author[0000-0003-0501-2636]{B. P. Powell}
\affiliation{NASA Goddard Space Flight Center, 8800 Greenbelt Road, Greenbelt, MD 20771, USA}
\author[0000-0002-5665-1879]{R. Gagliano}
\affiliation{Amateur Astronomer, Glendale, AZ 85308}
\author{M. Omohundro}
\affiliation{Citizen Scientist, c/o Zooniverse, Department of Physics, University of Oxford, Denys Wilkinson Building, Keble Road, Oxford, OX13RH, UK}
\author{I. B. B\'{\i}r\'o}
\affiliation{Baja Astronomical Observatory of University of Szeged, H-6500 Baja, Szegedi út, Kt. 766, Hungary}
\affiliation{HUN-REN -- SZTE Stellar Astrophysics Research Group,  H-6500 Baja, Szegedi út, Kt. 766, Hungary}
\author{M. Moe}
\affiliation{Department of Physics \& Astronomy, University of Wyoming, Laramie, WY 82072}
\author[0000-0002-2532-2853]{S.~B.~Howell}
\affiliation{NASA Ames Research Center, Moffett Field, CA 94035}
\author{T. Mitnyan}
\affiliation{HUN-REN -- SZTE Stellar Astrophysics Research Group,  H-6500 Baja, Szegedi út, Kt. 766, Hungary}
\author[0000-0002-2361-5812]{C.~A.~Clark}
\affil{Jet Propulsion Laboratory, California Institute of Technology, Pasadena, CA 91109}
\affil{NASA Exoplanet Science Institute, IPAC, California Institute of Technology, Pasadena, CA 91125}

\author[0000-0002-2607-138X]{M.~H.~Kristiansen}
\affil{Brorfelde Observatory, Observator Gyldenkernes Vej 7, DK-4340 T\o{}ll\o{}se, Denmark}
\author{I. A. Terentev}
\affiliation{Citizen Scientist, Planet Hunter, Petrozavodsk, Russia}
\author{H. M. Schwengeler}
\affiliation{Citizen Scientist, Planet Hunter, Bottmingen, Switzerland}
%
%
\author{A. P\'al}
\affiliation{Konkoly Observatory, Research Centre for Astronomy and Earth Sciences, MTA Centre of Excellence, Konkoly Thege Mikl\'os  \'ut 15-17, H-1121 Budapest, Hungary}
\author[0000-0001-7246-5438]{A. Vanderburg}
\affiliation{Department of Physics, Kavli Institute for Astrophysics and Space Research, M.I.T., Cambridge, MA 02139, USA}
%
%

\begin{abstract}

We have discovered a triply eclipsing triple star system, TIC 290061484, with the shortest known outer period, $P_{\rm out}$, of only 24.5 days. This `eclipses' the previous record set by $\lambda$ Tauri at 33.02 days which held for 68 years \citep{1956ApJ...124..507E}. The inner binary, with an orbital period of $P_{\rm in}=1.8$ days, produces primary and secondary eclipses, and exhibits prominent eclipse timing variations with the same periodicity as the outer orbit. The tertiary star eclipses, and is eclipsed by, the inner binary with pronounced asymmetric profiles. The inclinations of both orbits evolve on observable timescales such that the third-body eclipses exhibit dramatic depth variations in TESS data. A photodynamical model provides a complete solution for all orbital and physical parameters of the triple system, showing that the three stars have masses of \MAa M$_\odot$, \MAb M$_\odot$, and \MB M$_\odot$, radii near those corresponding to the main sequence, and $T_{\rm eff}$ in the range of 21,000 K - 23,700 K. Remarkably, the model shows that the triple is in fact a subsystem of a hierarchical 2+1+1 quadruple with a distant fourth star. The outermost star has a period of $\sim3,200$ days and a mass comparable to the stars in the inner triple. In $\sim20$ Myr, all three components of the triple subsystem will merge, undergo a type II supernova explosion, and leave a single remnant neutron star. At the time of writing, TIC 290061484 is the most compact triple system, and one of the tighter known compact triples (i.e., $P_{\rm out}/P_{\rm in} = 13.7$).

\end{abstract}


\section{Introduction} \label{sec:intro}

Binary stars are ubiquitous in the Galaxy, with more than half of the Sun-like stars having a stellar companion, in many cases more than one \citep[e.g.][and references therein]{Raghavan2010, Tokovinin2021}. In fact, the nearest star to the Sun -- Proxima Centauri \citep{Innes_1917} -- is part of the $\alpha$ Centauri triple system \citep{Kervella_2017}. The Proxima Centauri system is rather wide and all three components are visually resolved. The inner binary, composed of $\alpha$ Centauri A and B, has an orbital period of nearly 80 years, while Proxima takes about 550,000 years to complete one orbit around the common center of mass. With an outer orbital eccentricity of about 0.5, the physical separation between the inner binary and the outer tertiary is about 4,300 AU at periastron and up to 13,000 AU at apastron \citep{Kervella_2017}. Quite a wide orbit indeed -- Proxima Centauri is nearly 0.1 light years closer to Earth than $\alpha$ Centauri A and B. Thus it is perhaps unsurprising that it was not until recently that the triple nature of the system was demonstrated with high confidence \citep{Kervella_2017}. 

Triple star systems cover an enormous range of physical parameters, stellar types, and orbital configurations. The long-period ones, like $\alpha$ Centauri, represent one end of the spectrum where the interactions between the individual components occur on such colossal timescales that a human observer is unlikely to witness an exciting event. By contrast, compact, short-period systems that reside at the other end of the spectrum of stellar triples can exhibit a multitude of detectable dynamical interactions, in many cases quite dramatic. Naturally, the shorter the outer period the stronger the interactions between the individual components, such that the most interesting systems are usually those with outer orbital periods of less than 1,000 days -- typically referred to as compact hierarchical triples (CHT) \citep[e.g.][]{Tokovinin2021, Borkovits2020}. 

CHTs that are also triply-eclipsing, i.e., the tertiary star eclipses, and is eclipsed by, the inner binary, are exceptionally valuable because they often enable a complete diagnosis of the system parameters. The stellar masses, radii, $T_{\rm eff}$, and ages of these systems can be accurately determined, as well as the complete set of orbital parameters, including the outer period and eccentricity, and the mutual inclination angle. Remarkably, most of these can be estimated {\it without} radial velocity measurements. Instead, one can utilize various effects encoded in the information-rich eclipse timing variations (ETVs; i.e., light-travel time delays, dynamical delays, dynamically-driven apsidal motion, and forced precession of the orbital planes) as well as in the timing and profile of the tertiary eclipses \citep[e.g.][and references therein]{Borkovits2020,Borkovits_2022a,rappaport22a,Rappaport2023a,Rappaport2024}. 

The technique of using such information from a multistellar object, coupled with a comprehensive `photo-dynamical' modeling analysis, was pioneered by \citet{2011Sci...331..562C} for the case of the triple star KOI-126 found with {\it Kepler}. The model enabled sub-percent precision measurements of the masses and radii of the stars in KOI-126, which was particularly valuable for the M-dwarfs of the inner binary. Compared to the relative handful of triply-eclipsing triples that were known from the {\it Kepler} mission \citep{2016MNRAS.455.4136B}, TESS has already enabled the discovery of dozens of such systems \citep[e.g.,][]{rappaport22a, Rappaport2023a, Borkovits2019, Borkovits2020}. These systems inform theoretical formation models indicating that such compact, and typically flat, multiples likely form and evolve differently compared to their wider counterparts \citep[e.g.][]{Tokovinin2021, Docobo2021}.

\begin{figure}
    \centering
    \includegraphics[width=1.01\columnwidth]{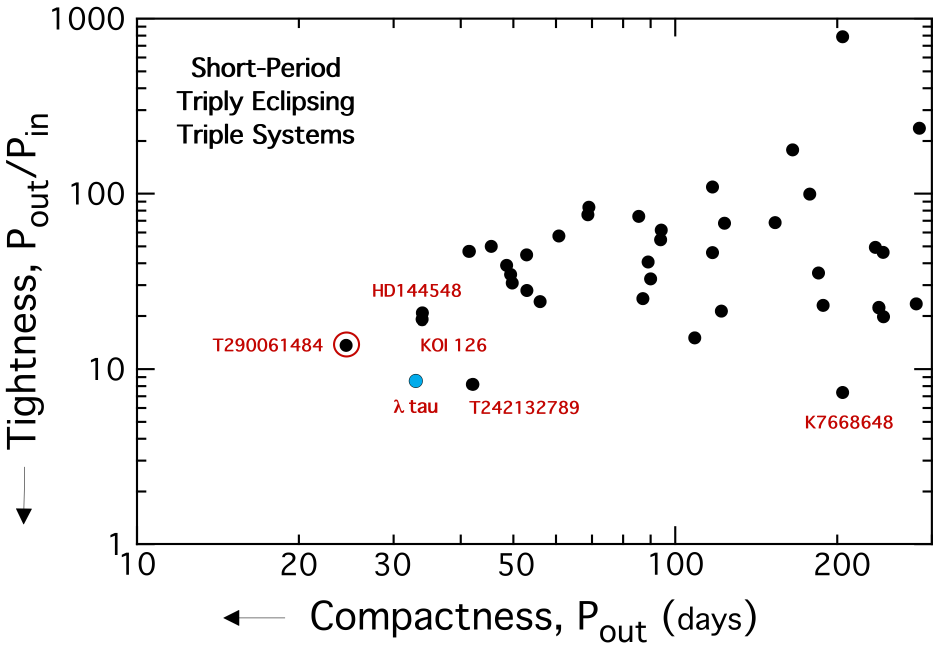}
    \caption{Tightness of triply eclipsing triple systems ($P_{\rm out}/P_{\rm in}$) as a function of the system compactness. $\lambda$ Tau (marked by a blue symbol) is not triply eclipsing, but is otherwise such a noteworthy benchmark system that we include it for reference. TIC 290061484, announced here and highlighted with a large open circle, is the most compact triple system known to date. }
    \label{fig:tightness_porb}
\end{figure}

Yet, despite the large leap in the state of our knowledge of CHTs enabled by these discoveries, one thing had remained unchanged since the 1950s. For more than 68 years, $\lambda$ Tauri has reigned supreme as the CHT with the shortest outer period -- 33.02 days \citep{1956ApJ...124..507E}. As highlighted in Fig.\,\ref{fig:tightness_porb}, two systems came close to dethroning it -- KOI-126 in 2011 (outer period of 33.92 days) and HD 144548 in 2015 (outer period 33.95 days, \citealt{Alonso2015}). Two other CHTs were dynamically tighter (smaller $P_{\rm out}/P_{\rm in}$), but nothing came even close to pushing the record to much shorter outer periods\footnote{Importantly, tight triples are not necessarily compact triples. As pointed out by \citep{Borkovits_2022a}, the triple system LHS 1070 with $P_{\rm out}/P_{\rm in} = 88/18$ years is close to the theoretical limit for dynamical stability -- yet hardly compact.}. Today, thanks to NASA's {\em TESS} mission we were able to do so with the discovery and confirmation of TIC 290061484 -- a triply-eclipsing CHT with an outer period of only 24.5 days, nearly 9 days shorter than $\lambda$ Tauri. 

This paper is organized as follows. In Section 2 we provide an overview of the system, outline the initial discovery and discuss the observations. Section 3 presents a comprehensive photometric-dynamical solution of the system properties. We highlight the stellar and orbital properties of this system in Section 4, and discuss our results in Section 5. 

\section{Discovery, Observational Material and Preliminary Analysis}

\subsection{Discovery}

We are continuously searching through the TESS full-frame image (FFI) photometry data for triple and higher-order multiple star systems \citep[see, e.g.,][]{Borkovits2020,Borkovits_2022a,rappaport22a,Rappaport2023a,Rappaport2024,Kostov2024}, generally following a three-stage process. 

In the first phase, we download the FFIs from the Mikulski Archive for Space Telescopes (MAST).  Then, with target lists assembled by using the {\tt tess-point} \citep{tess-point} code on the {\em TESS} Input Catalog (TIC) to determine which sources are present in a given sector, we extract the light curves for stars with TESS magnitude $T\le15$ mag from the raw FFIs on the NASA Center for Climate Simulation {\em Discover} supercomputer\footnote{\url{https://www.nccs.nasa.gov/systems/discover}} using the \textsc{eleanor} \citep{eleanor} code.

In the second phase, we use a neural network binary classifier trained to find eclipses in {\em TESS} light curves \citep[see][]{2021AJ....161..162P,Powell2022_eleanor_lite,eleanor_lite}.  The neural network pre-selects light curves that contain eclipses or features resembling eclipses for further visual examination.  Generally, this process results in $\sim$1\% of the total number of light curves being pre-selected, such that the fraction varies slightly depending on the density of stars in a particular sector (which causes blending of eclipses in the light curves of neighboring stars), and on systematic effects resembling eclipses. Thus out of hundreds of millions of light curves that we have produced from {\em TESS} sectors 1-76 to date, this process has yielded `merely' a few million for further visual survey.   

In the third phase of the search, our `Visual Survey Group' (VSG; \citealt{Kristiansen2022}) looks through these light curves by eye for extra eclipses which might indicate third body eclipses in a 2+1 triple system, or extra sets of regular eclipses which could suggest another eclipsing binary, potentially indicating a 2+2 quadruple system.  In addition, the VSG searches for other unusual behavior associated with the eclipsing binary (EB) such as pulsations that vary with the orbit (see, e.g., \citealt{Handler2020}). The visual surveying of the light curves is done with Allan Schmitt's {\tt LcTools} and {\tt LcViewer} software \citep{Schmitt2019}. This software makes it feasible to inspect a typical light curve in just a matter of a few seconds.  

From our searches through EB light curves obtained thus far from the TESS observations, we have found more than 70 triply eclipsing triples (see, e.g., \citealt{Borkovits2020b}; \citealt{Mitnyan2020}; \citealt{Borkovits2022b}; \citealt{rappaport22a}; and \citealt{Rappaport2023a}), over 200 eclipsing quadruples \citep{2022ApJS..259...66K,Kostov2024}, the first fully-eclipsing sextuple star system \citep{2021AJ....161..162P}, and other particularly unique systems \citep{2021AJ....162..299P,Powell2022}. In this work we report the discovery of a new triply eclipsing triple star system, TIC 20061484, with an outer orbital period of only 24.5 days -- the shortest ever found by a wide margin. The inner binary period itself is 1.792 days.  This is such an unusual system, including various kinds of interesting dynamical effects, that we report it separately here.

\begin{figure*}
    \centering
    \includegraphics[width=0.99\textwidth]{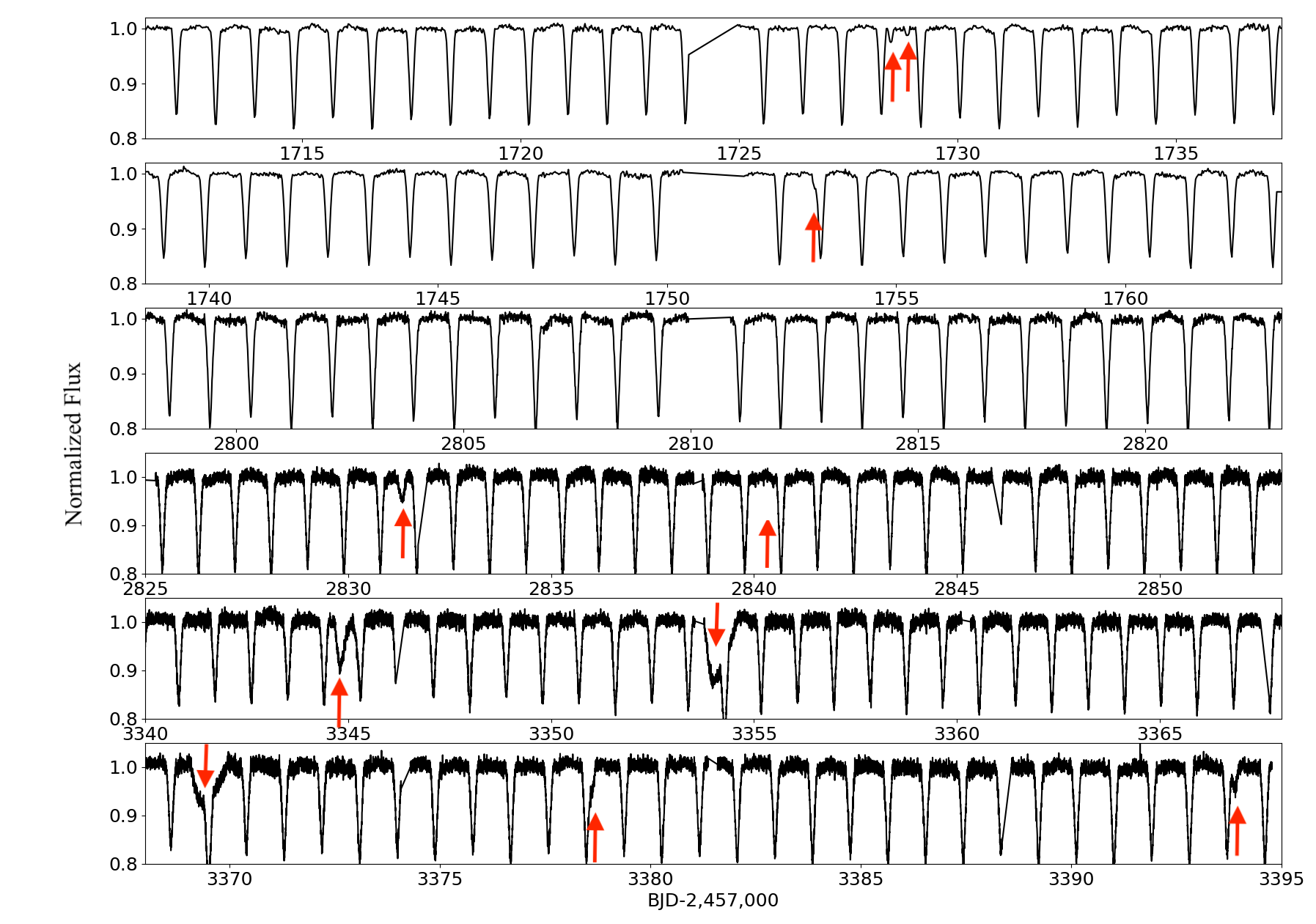}
    \caption{TESS FITSH light curve of TIC 290061484. Each panel represents a different sector. The deep regular eclipses are from the 1.792-d inner EB. The tertiary eclipses are highlighted with red arrows. Note that the tertiary eclipse near day 2840 is difficult to see on this scale. }
    \label{fig:FITSH_LC}
\end{figure*}

\subsection{The TESS light curve and `Extra' Eclipses}

TESS observed TIC 290061484 \citep{TIC} at a cadence of 1800 s in Sectors 15 and 16; 600 s in Sector 55; and 200 s in Sectors 56, 75, and 76. Once TIC 290061484 was noted by VSG as being special in the \textsc{eleanor} data, we extracted a somewhat improved light curve using the FITSH pipeline \citep{Pal2012}, shown in Fig.~\ref{fig:FITSH_LC}. The panels, arranged vertically, each represent a different TESS sector. The eclipses of the 1.792-day inner EB are readily apparent. In addition to these regular eclipses detected by the neural network, one of us (RG) noticed 9 `extra' features (a couple of which are quite shallow) that do not belong to the EB. These are marked in the figure with red arrows and their approximate times are listed in Table \ref{tbl:extra_times}. Closer inspection of these events revealed that six of them occur with a spacing equal to an integer of $\approx$24.5 days, while the remaining three events follow the same implied underlying period, but phased by 0.37 periods (rather than midway between the other group). This indicated a triply eclipsing triple system where the outer orbit has an eccentricity of $\sim$0.2.

When we fit a linear ephemeris to the set of 6 eclipses mentioned above, we find a period of 24.489 days with a statistical uncertainty of 0.004 d. Of course, this did not immediately confirm the triple nature of TIC 290061484 as there is always the possibility that a second periodicity in an EB light curve might be due to another EB that is either physically related to the target (yet unresolved) or a resolved, unrelated field EB that accidentally contaminates the TESS aperture of the target. However, it was immediately clear that this is not the case for TIC 290061484 because: (i) the shapes of the extra eclipses are quite asymmetric; (ii) one tertiary conjunction shows {\it two} extra eclipses separated by only $\approx0.4$ days; and (iii) the eclipse timing variations (ETV) in the 1.792-d EB, discussed below, have a non-linear periodic structure with a period of 24.5 days. 

\begin{figure}
    \centering
    \includegraphics[width=0.99\columnwidth]{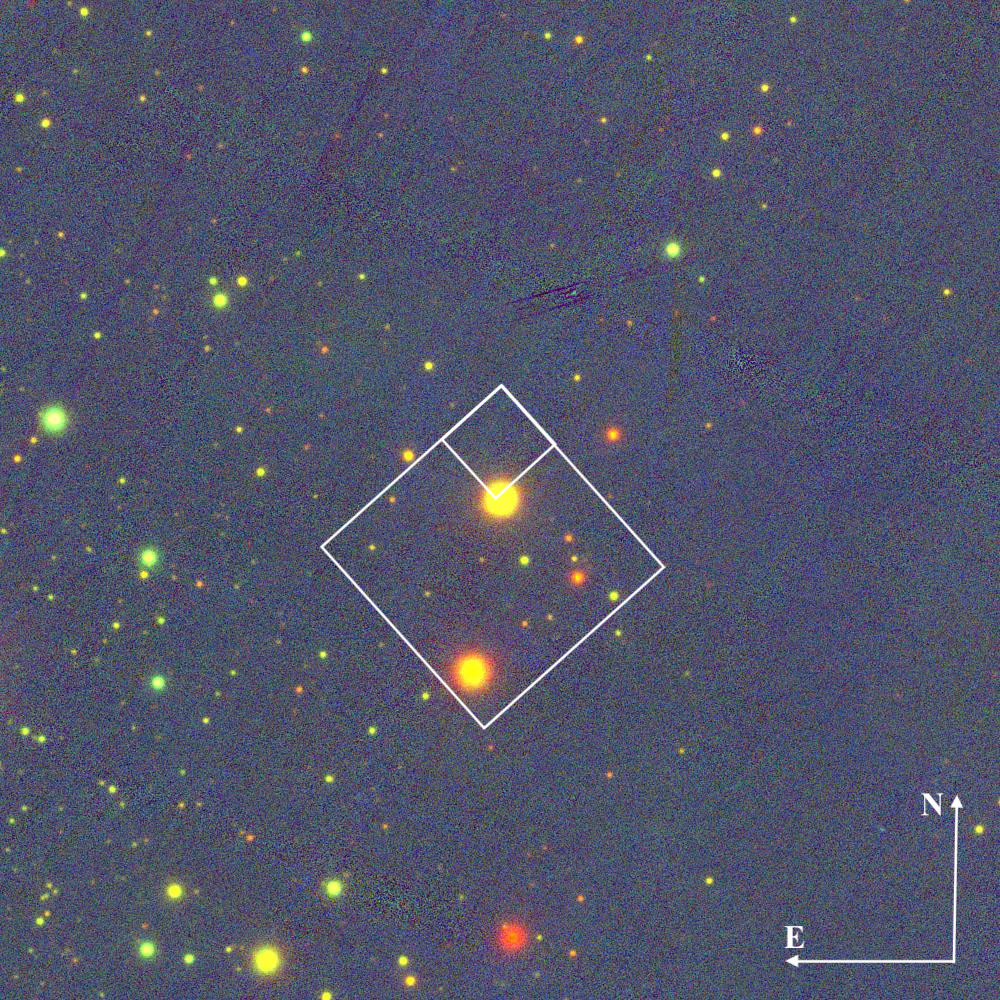}
    \caption{{\it Upper panel}: $4' \times 4'$ PanSTARRS image \citep{panstarrs2016} centered on TIC 290061492. The small white contour represents one TESS pixel ($21 \times 21$ arcsec) orientated according to Sector 16, and the large white contour represents a $1' \times 1'$ region. N is up and E is to the left. 
    }
    \label{fig:tpfplotter}
\end{figure}

It is important to note that TIC 290061484 is rather close to the Galactic mid-plane ($b^{II} = 2.9^\circ$). A $4' \times 4'$ PanSTARRS image of the region surrounding TIC 290061484 is shown in the upper panel of Fig.~\ref{fig:tpfplotter}. The nearest resolved star is TIC 290061492, at a separation of about 17 arcsec and ${\Delta m = 5.4\ \rm mag}$ (too faint to be the source of either the EB or the tertiary eclipses), while the nearest comparably-bright source is TIC 290061506 at a separation of about 43.8 arcsec and ${\Delta m = 0.5\ \rm mag}$. Thus contamination in the TESS aperture of TIC 290061484 is quite low and the observed eclipse depths represent the true depths. To pinpoint the origin of the eclipses, we performed two complementary tests: (i) a pixel-by-pixel analysis of the TESS light curve; and (ii) measurements of the center-of-light motion during the detected eclipses, using the methods described in \citep[e.g.][and references therein]{Kostov2024}. Both tests confirm that the target is the source of the eclipses (see Fig.~\ref{fig:tpfplotter}, lower panel).

\begin{table}[h]
\centering
\caption{Approximate Times of Third-Body Events$^a$}
\begin{tabular}{ccccc}
\hline
\hline
Event & Primary$^{b}$ & Secondary$^{c}$ & Cycle$^{d}$ & Deviation$^{e}$  \\
\hline
   1 & 8728.46 & ... & 1.00 & $-0.33$ \\
   2  & 8753.25  &... & 2.00 &$-0.03$ \\
   3  & 9831.33  & ... & 46.00& +0.54\\
   4  & ... & 9840.48 & 46.37 & +0.58\\
   5 & 10344.80 & ... & 67.00 & $-0.25$\\
   6  & ...  & 10353.98 & 67.37 & $-0.18$\\
   7  & 10369.38 & ... & 68.00 & $-0.16$\\
   8  & ... & 10378.60 & 68.37 & $-0.05$\\
   9 & 10393.89   & ... & 69.00 & $-0.13$\\
 \hline
\hline 
\label{tbl:extra_times} 
\end{tabular}

{Notes. (a) Eyeball estimates only due to the facts that the eclipses are asymmetric, sometimes incomplete, and not strictly periodic (i.e., they depend on the phasing of the inner EB near the times of the outer inferior and superior conjunctions). The uncertainty is on the order of 0.2 d.  (b) Times of the primary outer eclipses relative to BJD 2,450,000.  (c) Times of the secondary outer eclipses relative to BJD 2,450,000. (d) The fractional cycle numbers account for the eccentricity of the outer orbit, $e_{\rm out} \simeq 0.2$. (e) The deviations (in days) from a linear ephemeris with $P_{\rm out} = 24.489 \pm 0.004$ days.} 

\end{table} 

\subsection{Archival Data}

\begin{figure}
    \centering
    \includegraphics[width=0.95\linewidth]{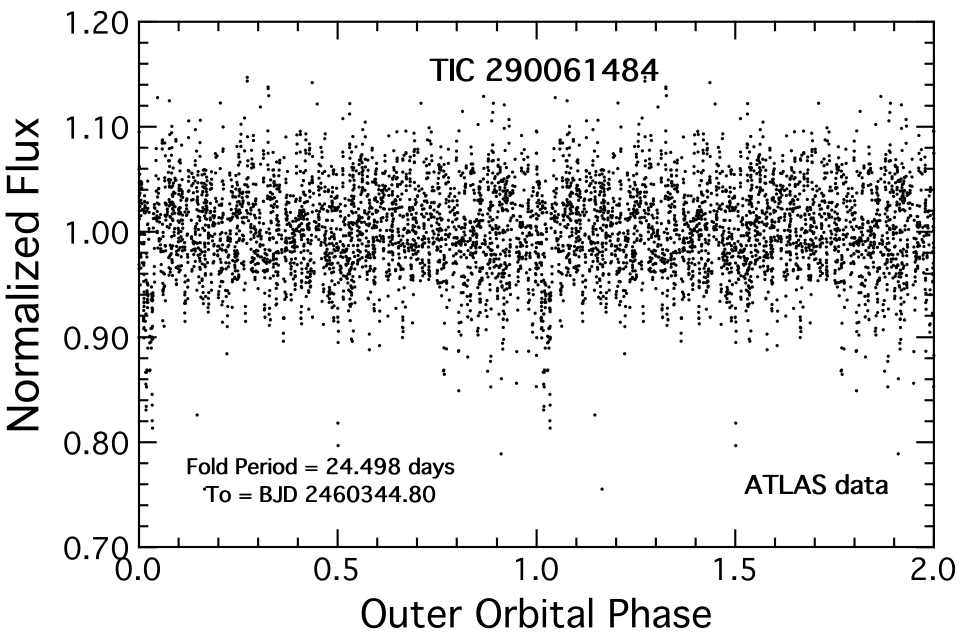}
    \caption{Point-wise fold of 2800 ATLAS archival data points for TIC 290061484. The fold period is 24.498 days which represents the highest peak in the BLS transform for periods between 10 and 100 days. The eclipses of the inner binary have been removed prior to the BLS transform. Only one of the two outer eclipses appears (near phase 1.0), the other likely being smeared due to the large jitter inherent in third-body eclipses which depend on both the inner and outer orbital phases. The uncertainty in the mean outer period from the ATLAS data alone is $\pm 0.005$ d.}
    \label{fig:folded_ATLAS}
\end{figure}

To complement the TESS observations of TIC 290061484, we investigated the archival photometry from ASAS-SN \citep{2017PASP..129j4502K}, ATLAS \citep{2018AJ....156..241H}, and Zwicky Transient Facility (ZTF,\citep{Graham2019}). There are insufficient ZTF observations to be useful for the purpose of tracking the outer eclipses. The ASAS-SN database contains 2350 observations, but the source is too faint (G = 14.4 mag) to detect even the 1.792-day EB. The ATLAS database contains 2800 observations. To analyze these, we first removed the eclipses of the inner binary by subtracting 50 harmonics of the 1.792-day eclipse structure, and then subjected the cleaned data set to a BLS transform \citep{kovacs02} covering the period range 1-1000 days. The BLS analysis shows that the highest peak in the period range of 10-100 d occurs at a period of $24.498 \pm 0.005$ d. A point-wise fold of the cleaned ATLAS data about this period is shown in Fig.~\ref{fig:folded_ATLAS}. 

We note that only one of the two sets of outer eclipses is detected in the ATLAS data. This is perhaps to be expected for the following reason. Even though there is a precise {\it underlying} outer period, the times of the outer eclipses are subject to substantial deviations from a linear ephemeris (of $\sim$ half a day) because those eclipses depend not only on the outer orbital phase, but on the relative phases of the inner EB at the times of the superior and inferior outer conjunctions. Thus, it is perhaps not surprising that we are able to detect (weakly, at that) only one of the two sets of outer eclipses in the archival data.   

\begin{deluxetable}{l r r}[!ht]
\tabletypesize{\scriptsize}
\tablecaption{Basic parameters for TIC 290061484\label{tab:EBparameters}}
\tablewidth{0pt}
\tablehead{
\colhead{Parameter} & \colhead{Value}  &\colhead{Source}
}
\startdata
\multicolumn{3}{l}{\bf Identifying Information} \\
\hline
TIC ID & 290061484 & 1\\
Gaia DR3 ID & 2169382208774963072 & 2 \\
$\alpha$ (J2000, degrees) & 316.109369 & 2\\
$\delta$ (J2000, degrees) & 51.225478 & 2\\
\\
\multicolumn{3}{l}{\bf Gaia Measurements} \\
\hline
$\mu_{\alpha}$ (mas~yr$^{-1}$) & -3.1094 & 2 \\
$\mu_{\delta}$ (mas~yr$^{-1}$) & -4.1678 & 2 \\
$\varpi$ (mas) & 0.6147 & 2\\
RUWE & 1.05 & 2\\
$astrometric\_excess\_noise$ (mas) & 0.12 & 2\\
$astrometric\_excess\_noise\_sig$ & 5.13 & 2\\
$non\_single\_star$ & 0 & 2\\
${\rm T_{eff}~(K)}$ & 15392 & 2\\
\\
\multicolumn{3}{l}{\bf Photometric Properties} \\
\hline
$T$ (mag) & 12.79 & 1\\
$G$ (mag) & 14.44 & 2\\
$B$ (mag) & 18.86 & 1\\
$V$ (mag) & 16.27 & 1\\
$J$ (mag) & 10.7 & 3\\
$H$ (mag) & 9.95 & 3\\
$K$ (mag) & 9.58 & 3\\
$W1$ (mag) & 9.4 & 4\\
$W2$ (mag) & 9.28 & 4\\
$W3$ (mag) & 9.22 & 4\\
$W4$ (mag) & 7.42 & 4\\
\hline
\enddata
Sources: (1) TIC-8 \citep{TIC}, (2) Gaia DR3 \citep{Gaia2021}, (3) 2MASS All-Sky Catalog of Point Sources \citep{2MASS}, (4) AllWISE catalog \citep{WISE}
\tablenotetext{}{}
\end{deluxetable}

\section{Followup Observations of TIC 290061484}
\label{sec:followup}

\subsection{Ground-based Follow-up Observations}
\label{sec:baja}

In order to extend the observational window for this remarkable triple system, we initiated photometric follow-up observations with the 80-cm RC telescope of Baja Astronomical Observatory, Hungary (BAO80), using a standard Sloan $r'$-band filter. The purpose of these observations is to (i) monitor the rapid, large amplitude eclipse timing variations of the inner EB; (ii) detect expected third-body eclipses and refine the photodynamical solution; and (iii) characterize the prominent eclipse depth variations of both the inner and the outer eclipses. By the time of submission of this manuscript we have just recorded one additional new third-body eclipse. The corresponding light curve was included in our detailed photodynamical analysis, as discussed below.

We note that the target is fairly faint (V = 16.27 mag) and obtaining radial velocity measurements would be highly challenging. However, as discussed below, we are able to determine the parameters of the system without such measurements, and also detail how this has been done numerous times in the past. Furthermore, it is important to point out that the key issue of the outer period of the triple, which is a major focus of this manuscript, does not depend in any way on radial velocity measurements.

\vspace{15pt}
\subsection{Speckle Interferometry}
\label{sec:speckle}

Additionally, to search for any wide companion in the TIC 290061484 system, we used the ‘Alopeke high resolution speckle imager mounted on the Gemini North 8-m telescope \citep{scott2021}. ‘Alopeke provides simultaneous speckle imaging in two bands of 562\,nm and 832\,nm. The output data products include reconstructed images in the two bands and robust contrast limits for companion detections. Twelve sets of $1000 \times 0.06$ sec exposures, with an EMCCD gain of 1000, were collected for TIC 290061484, as well as three sets of $1000 \times 0.06$ sec exposures with an EMCCD gain of 10 for the PSF standard star HR 8072. The PSF standard star observations were taken immediately after TIC 290061484, and the observations were subjected to Fourier analysis in our standard reduction pipeline \citep{howell2011}.
 
\begin{figure*}
    \centering
    \includegraphics[width=0.40\textwidth]{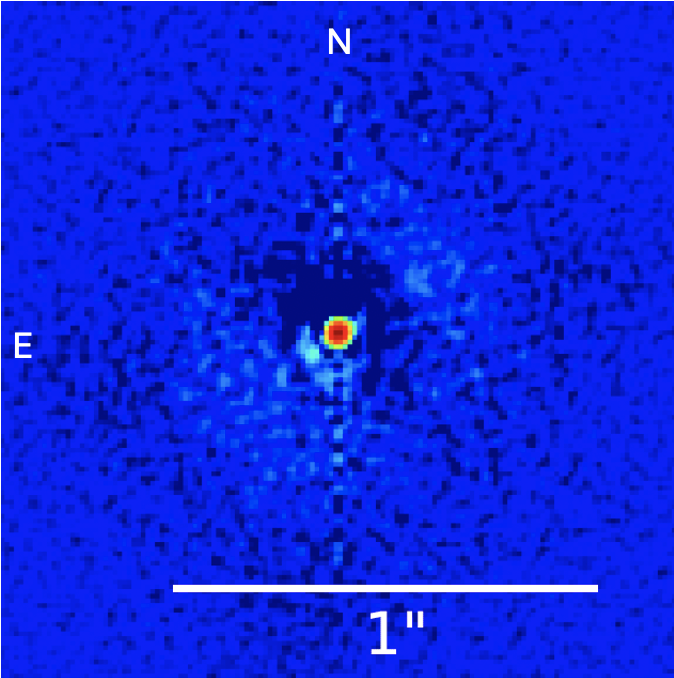}
    \includegraphics[width=0.50\textwidth]{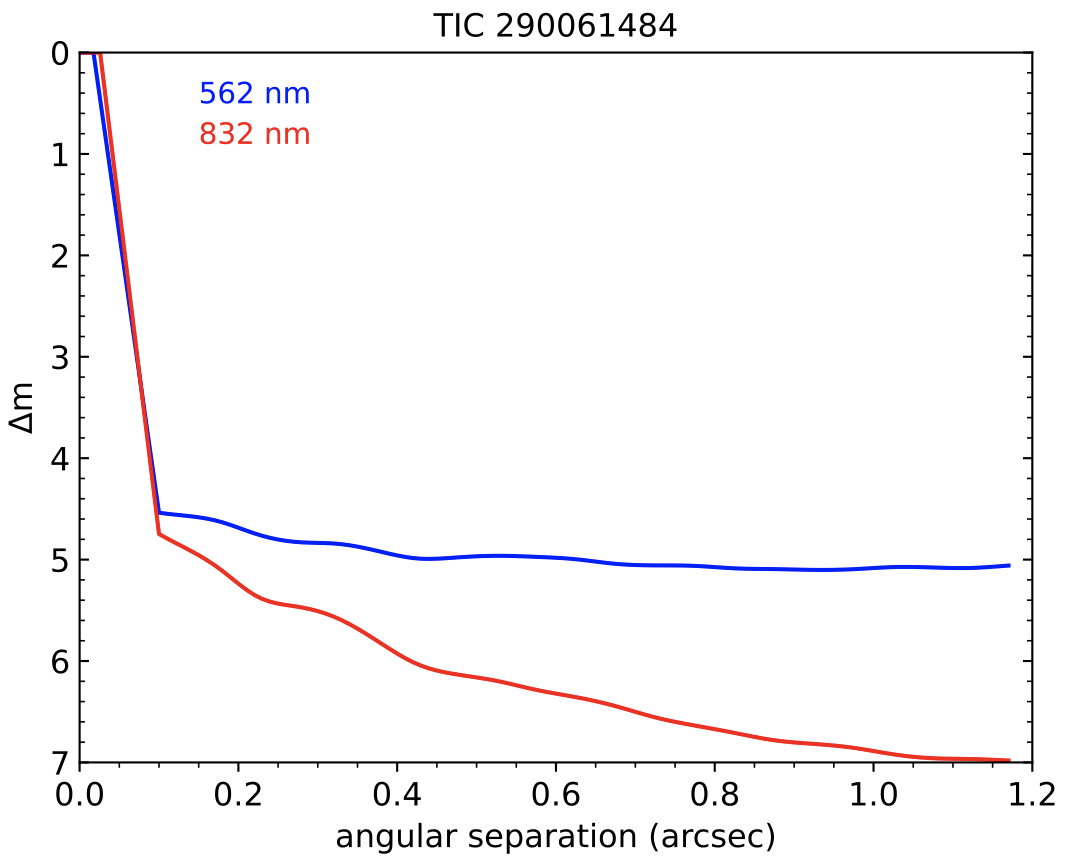}
    \caption{Contrast curves showing the 5$\sigma$ magnitude limits for the 562 nm and 832 nm high resolution speckle observations of TIC 290061484 {\it Left:}  The reconstructed 832 nm image. {\it Right:} close companions can be excluded within 20 mas to 1.2 arcsec of TIC 290061484 at the specified contrast limits.}
    \label{fig:speckle}
\end{figure*}

Figure \ref{fig:speckle} shows the reconstructed 832 nm image and resultant 5-$\sigma$ magnitude contrast limits obtained for the TIC 290061484 observations.  TIC 290061484 was found to harbor no close companion stars of any significant flux within the range of the angular diffraction limit of 0.02$''$ out to 1.2$''$. At the distance to TIC 290061484 ($d=1519$ pc) the angular limits correspond to spatial limits of 30 to 1822 AU. The most stringent contrast limits achieved by the observations were 5-7 magnitudes for the angular distance range 0.1$''$ to 1.2$''$. 

\section{Photodynamical Modeling of the System}
\label{sec:photodynamics}

\subsection{General description of the code}
\label{sec:general_info}

In order to extract all of the physical and orbital parameters of TIC 290061484, we utilized the software package {\sc lightcurvefactory} \citep[see, e.g.][and references therein]{Borkovits2019,Borkovits2020}. The code was developed to analyze multiple star systems, including binaries, triples, and quadruple stars of both the 2+2 and 2+1+1 architecture. A detailed description of the inner workings of {\sc lightcurvefactory}, the steps involved in the analysis procedure, and its application to a wide range of eclipsing multi-stellar systems can be found in \citep[e.g.][and references therein]{Borkovits2018a, Borkovits2019, Borkovits2020, Borkovits2020b, Mitnyan2020}. Briefly, the code contains four main features: (i) emulators for multi-passband light curve(s), the corresponding ETVs, and radial velocities (if available); (ii) an ability to calculate stellar masses (as proxies, in the absence of RV measurements), radii, temperatures and several passband magnitudes (for fitting the net SED of the system) in an iterative manner, with the use of built-in \texttt{PARSEC} isochrone tables \citep{PARSEC}; \texttt(iii) a seventh-order Runge-Kutta-Nystr\"om numerical integrator designed to calculate the perturbed 3-D coordinates and velocities of all bodies in the system; and (iv) a Markov Chain Monte Carlo (MCMC)-based search routine for determining the best-fit system parameters and corresponding statistical uncertainties. The routine uses our custom implementation of the generic Metropolis-Hastings algorithm \citep[see, e.g., ][]{ford05}.
 
Essentially all the details of how {\sc lightcurvefactory} is used to analyze compact triply eclipsing triple systems discovered with TESS can be found in \citet{rappaport22a}. Here we provide only a high-level overview of the inputs to the code and of the parameters that are either fitted or constrained by the MCMC routine. Altogether, in the case of a hierarchical triple configuration, there are 25 -- 27 system parameters: (i) 9 stellar parameters (3 $\times$ mass, radius, $T_{\rm eff}$); (ii) 12 orbital parameters\footnote{In the case of available radial velocity data, the systemic radial velocity can be added as another parameter.}; and (iii) 4 general parameters -- distance, interstellar extinction, metallicity and age. Finally, the code allows for passband-dependent contaminated (extra) light $\ell_4$.
 
In turn, the input information can be broadly split into two basic categories. First, there are the `data' that include the (i) EB eclipse profiles; (ii) tertiary eclipse profile(s); (iii) eclipse times of the EB; (iv) archival SED values; and (v) radial velocity measurements (not yet available for TIC 290061484). Second, we utilize \texttt{PARSEC} model stellar evolution tracks, isochrones, and atmospheres \citep{PARSEC}.  The evolution tracks enable us to find the stellar radius and $T_{\rm eff}$ for a given stellar mass, age and metallicity, while the isochrones allow us to compute stellar magnitudes in different passbands in order to fit the SED curve. 

\subsection{``Reader's Digest'' description of the photodynamics code}

The entire photometric light curve, and especially the eclipse fitting using the lightcurve emulator for the inner binary, yield dimensionless quantities such as $R_{\rm 
Aa}/a$, $R_{\rm Ab}/a$ and $T_{\rm Aa}/T_{\rm Ab}$, as well as $e_\mathrm{in}$, $
\omega_\mathrm{in}$ and $i_\mathrm{in}$. Likewise, fitting of the outer eclipses yields another independent set of such ratios -- and even more, as the profiles and timings of these outer eclipses are especially sensitive to more of the geometrical and dynamical quantities, including e.g., the mass ratios of both the inner and outer pairs, their mutual inclination, etc. In the case of tight and compact systems, where the gravitational third-body interactions perturb the motions of all three stars on very short timescales (weeks and months), several further parameters can be inferred with great accuracy, including even the individual masses, which effect the timescales of such variations. This is especially true when both orbits are eccentric, and the two orbital planes are inclined to each other, as is the case for TIC 290061484. 

The SED fitting, given a Gaia distance, relates the absolute stellar radii and $T_{\rm eff}$ values to the composite system fluxes, which can be matched to the SED. In turn, for a given chemical composition, the stellar masses and system age are directly given by the radii and $T_{\rm off}$ values, and vice versa, via the stellar evolution tracks. 

Finally, the ETV fitting from the binary eclipse timing, among other things, yields the same information as RV studies that would follow the gamma velocity of the binary; this measures the orbit of the binary center of mass around the tertiary star. But, in addition to that, there are other dynamical effects encoded in the ETV curve, e.g., which again determine such things as the eccentricity of both orbits, the orientation of the two orbits relative to each other, as well as to the 
observer, and also both mass-ratios.  These latter effects are basically due to the way in which the tertiary star periodically alters the period of the inner binary.  In most cases, the latter effect is unmeasureable with RV studies. 

All these effects are extensively documented and verified in the various references given in Section \ref{sec:general_info}.

\subsection{Detailed description of the photodynamical analysis of TIC 290061484}
To model the light curve of TIC 290061484 we used the Full-Frame Image FITSH photometry \citep{Pal2012}. To save computational time, we binned the 200-sec and 600-sec cadence data to 30-min cadence\footnote{While {\sc lightcurvefactory} is able to handle directly finite exposure times, we found that finite exposure (more, strictly speaking cadence-) time corrections were unnecessary to apply even for the 30-min cadence light curves, due to the relatively long durations of both the inner and outer eclipses compared to the cadence times.}, and only used the $\pm0\fp15$ phase-domain regions around the EB eclipses. However, whenever a data segment contains a third-body (i.e., `outer') eclipse, we keep the data for an entire binary period before and after the first and last contacts of that particular third-body eclipse. 

We note that the lack of RV data does not prevent derivations of absolute stellar masses, temperatures, and radii. The light curve of a triply-eclipsing triple contains a wealth of information that encodes combinations of the stellar masses. Additionally, the ETV curve and (to a lesser extent) the light curve contain signatures of the dynamical delays which probe the mutual inclination and most of the inner and outer orbital elements\footnote{In general, the measured modulations in the ETVs also depend on the light travel time effect (LTTE), which is equivalent to an SB1 RV measurement of the outer orbit. For the case of TIC 290061484, the LTTE is almost negligible relative to the dynamical delays.}. In turn, these depend primarily on $q_{\rm out}$ and, through higher order perturbations, on $q_{\rm in}$ \citep[see, e.g.,][]{borkovitsetal15}. Additionally, the geometry and timing of the outer eclipses provide significant information about the ratio of $q_{\rm out}/q_{\rm in}$ \citep[see Appendix A of][]{borkovitsetal13}. The inner and outer eclipses constrain $R_{\rm Aa,Ab}/a_{\rm in}$ and $R_{\rm Aa,Ab,B}/a_{\rm out}$, respectively, and their combination leads to the above mentioned ratio of mass ratios.

To compensate for the lack of spectroscopically-determined temperatures and metallicities, we combine SED-derived estimates of absolute temperatures (as employed by e.g., \citealt{Miller2020} and \citealt{stassun16}) with the results of the simultaneous light curve fits. The latter provide combinations of the ratios of the surface brightnesses and, hence, indirectly the absolute temperatures of the three stars. We note that the ratio of the surface brightnesses of the inner EB components can be obtained not only from the mutual eclipses of the inner components, but also from the tertiary eclipses. Hence from a light curve containing both inner and outer eclipses, the information which can be extracted is not simply the sum of the parameters that can be determined from two independent eclipsing light curves, but much more. 

Finally, we combine the SED information with theoretical, coeval stellar isochrones which provide information on the radii and $T_{\rm eff}$ of the stars (assuming the three stars have the same age), and also on the masses for a given age\footnote{Naturally, such masses are no longer independent of astrophysical assumptions and, hence, may be somewhat inferior to those dynamical masses which can be directly inferred from high quality RV data \citep[see e.g.,][]{Borkovits2020,Borkovits_2022a}.}. And, of course, knowledge of the masses sets the size scales of the system, which then provides for absolute determinations of semi-major axes and stellar radii.

\section{System parameters}
\label{sec:results}

\begin{table*}
 \centering
\caption{Orbital and astrophysical parameters from the joint 2+1+1 quadruple photodynamical light curve, ETV, and SED and \texttt{PARSEC} isochrone solutions.}
 \label{tab:syntheticfit}
\scalebox{0.91}{\begin{tabular}{@{}lllll}
\hline
\multicolumn{5}{c}{orbital elements} \\
\hline
   & \multicolumn{4}{c}{subsystem}  \\
   & \multicolumn{2}{c}{Aa--Ab} & A--B &  AB--C  \\
  \hline
  $t_0$ [BJD - 2400000]& \multicolumn{4}{c}{$58711.0$} \\
  $P$$^a$ [days] & \multicolumn{2}{c}{$1.79364_{-0.00013}^{+0.00013}$} & $24.6347_{-0.0019}^{+0.0021}$ & $3205_{-152}^{+140}$ \\
  $a$ [R$_\odot$] & \multicolumn{2}{c}{$14.59_{-0.15}^{+0.11}$} & $98.18_{-1.17}^{+0.68}$ & $2734_{-103}^{+102}$ \\
  $e$ & \multicolumn{2}{c}{$0.00250_{-0.00034}^{+0.00033}$} & $0.2011_{-0.0024}^{+0.0022}$ & $0.44_{-0.18}^{+0.17}$ \\
  $\omega$ [deg]& \multicolumn{2}{c}{$151_{-12}^{+14}$} & $92.2_{-2.0}^{+2.1}$ & $155_{-23}^{+23}$ \\ 
  $i$ [deg] & \multicolumn{2}{c}{$85.50_{-0.36}^{+0.27}$} & $94.43_{-0.13}^{+0.15}$ & $82_{-22}^{+34}$ \\
  $\mathcal{T}_0^\mathrm{inf/sup}$ [BJD - 2400000]& \multicolumn{2}{c}{$58713.0340_{-0.0020}^{+0.0021}$} & ${58728.733_{-0.041}^{+0.043}}^*$ & $-$ \\
  $\tau$ [BJD - 2400000]& \multicolumn{2}{c}{$58712.442_{-0.059}^{+0.070}$} & $58704.228_{-0.093}^{+0.165}$ & $58931_{-66}^{+78}$ \\
  $\Omega$ [deg] & \multicolumn{2}{c}{$0.0$} & $1.70_{-1.02}^{+0.80}$ & $-1_{-17}^{+29}$ \\
  $\left(i_\mathrm{mut}\right)_\mathrm{A-...}$ [deg] & \multicolumn{2}{c}{$0$} & $9.17_{-0.42}^{+0.39}$ & $29_{-13}^{+17}$ \\
  $\left(i_\mathrm{mut}\right)_\mathrm{B-...}$ [deg] & \multicolumn{2}{c}{$9.17_{-0.42}^{+0.39}$} & $0$ & $33_{-18}^{+12}$ \\
  $\varpi^\mathrm{dyn}$ [deg]& \multicolumn{2}{c}{$332_{-12}^{+14}$} & $272.1_{-2.9}^{+3.8}$ & $334_{-23}^{+22}$ \\
  $i^\mathrm{dyn}$ [deg] & \multicolumn{2}{c}{$23_{-10}^{+15}$} & $27_{-14}^{+11}$ & $6.1_{-3.1}^{+1.9}$ \\
  $\Omega^\mathrm{dyn}$ [deg] & \multicolumn{2}{c}{$48_{-109}^{+135}$} & $129_{-160}^{+76}$ & $301_{-157}^{+82}$ \\
  $i_\mathrm{inv}$ [deg] & \multicolumn{4}{c}{$84_{-19}^{+28}$} \\
  $\Omega_\mathrm{inv}$ [deg] & \multicolumn{4}{c}{$-1_{-14}^{+23}$} \\
  \hline
  mass ratio $[q=M_\mathrm{sec}/M_\mathrm{pri}]$ & \multicolumn{2}{c}{$0.890_{-0.012}^{+0.015}$} & $0.610_{-0.014}^{+0.012}$ & $0.288_{-0.024}^{+0.019}$ \\
  $K_\mathrm{pri}$ [km\,s$^{-1}$] & \multicolumn{2}{c}{$193.4_{-2.6}^{+2.3}$} & $77.8_{-1.8}^{+1.2}$ & $9.9_{-1.7}^{+1.7}$ \\ 
  $K_\mathrm{sec}$ [km\,s$^{-1}$] & \multicolumn{2}{c}{$217.1_{-2.5}^{+2.1}$} & $127.4_{-1.4}^{+1.1}$ & $35.3_{-6.3}^{+4.4}$ \\ 
  \hline
  \multicolumn{5}{c}{Apsidal and nodal motion related parameters} \\
  \hline
$P_\mathrm{apse}$ [year] & \multicolumn{2}{c}{$2.75_{-0.12}^{+0.15}$} & $15.60_{-0.13}^{+0.14}$ & $37124_{-14606}^{+53557}$ \\ 
$P_\mathrm{apse}^\mathrm{dyn}$ [year] & \multicolumn{2}{c}{$1.34_{-0.04}^{+0.04}$} & $2.26_{-0.18}^{+0.13}$ & $3893_{-1124}^{+1651}$ \\ 
$P_\mathrm{node}^\mathrm{dyn}$ [year] & \multicolumn{3}{c}{$2.40_{-0.14}^{+0.34}$} & $4306_{-1502}^{+1224}$ \\
$\Delta\omega_\mathrm{3b}$ [arcsec/cycle] & \multicolumn{2}{c}{$4448_{-134}^{+131}$} & $38660_{-2111}^{+3246}$ & $2911_{-855}^{+1195}$ \\ 
$\Delta\omega_\mathrm{GR}$ [arcsec/cycle] & \multicolumn{2}{c}{$7.33_{-0.15}^{+0.11}$} & $1.834_{-0.043}^{+0.026}$ & $0.099_{-0.011}^{+0.030}$ \\ 
$\Delta\omega_\mathrm{tide}$ [arcsec/cycle] & \multicolumn{2}{c}{$296_{-16}^{+18}$} & $0.939_{-0.053}^{+0.056}$ & $0.0016_{-0.0004}^{+0.0010}$  \\ 
  \hline  
\multicolumn{5}{c}{stellar parameters} \\
\hline
   & Aa & Ab &  B & C \\
  \hline
 \multicolumn{5}{c}{Relative quantities} \\
  \hline
 fractional radius [$R/a$]  & $0.2111_{-0.0028}^{+0.0030}$ & $0.1922_{-0.0023}^{+0.0027}$ & $0.0356_{-0.0010}^{+0.0009}$ & $0.00102_{-0.00008}^{+0.00005}$  \\
 temperature relative to $(T_\mathrm{eff})_\mathrm{Aa}$ & $1$ & $0.9426_{-0.0069}^{+0.0083}$ & $1.0707_{-0.0116}^{+0.0100}$ & $0.9351_{-0.0394}^{+0.0286}$ \\
 fractional flux [in \textit{TESS}-band] & $0.2478_{-0.0059}^{+0.0054}$ & $0.1866_{-0.0037}^{+0.0039}$ & $0.3601_{-0.0222}^{+0.0253}$ & $0.1804_{-0.0315}^{+0.0265}$ \\
fractional flux [in SLOAN $r'$-band] & $0.2469_{-0.0092}^{+0.0065}$ & $0.1843_{-0.0068}^{+0.0077}$ & $0.3596_{-0.0236}^{+0.0296}$ & $0.1779_{-0.0311}^{+0.0254}$ \\
 \hline
 \multicolumn{5}{c}{Physical Quantities} \\
  \hline 
 $M$ [M$_\odot$] & $6.853_{-0.235}^{+0.151}$ & $6.106_{-0.206}^{+0.142}$ & $7.903_{-0.333}^{+0.234}$ & $6.009_{-0.587}^{+0.399}$ \\
 $R$ [R$_\odot$] & $3.077_{-0.045}^{+0.048}$ & $2.803_{-0.050}^{+0.053}$ & $3.499_{-0.113}^{+0.086}$ & $2.779_{-0.215}^{+0.154}$ \\
 $T_\mathrm{eff}$ [K]& $22154_{-407}^{+292}$ & $20887_{-462}^{+362}$ & $23703_{-472}^{+391}$ & $20690_{-981}^{+615}$ \\
 $L_\mathrm{bol}$ [L$_\odot$] & $2030_{-128}^{+155}$ & $1339_{-122}^{+122}$ & $3482_{-417}^{+320}$ & $1271_{-373}^{+305}$ \\
 $M_\mathrm{bol}$ & $-3.50_{-0.08}^{+0.07}$ & $-3.05_{-0.09}^{+0.10}$ & $-4.08_{-0.10}^{+0.14}$ & $-2.99_{-0.23}^{+0.38}$ \\
 $M_V           $ & $-1.29_{-0.05}^{+0.05}$ & $-0.98_{-0.05}^{+0.07}$ & $-1.70_{-0.07}^{+0.10}$ & $-0.94_{-0.17}^{+0.27}$ \\
 $\log g$ [dex] & $4.295_{-0.013}^{+0.011}$ & $4.327_{-0.011}^{+0.009}$ & $4.248_{-0.020}^{+0.018}$ & $4.331_{-0.024}^{+0.026}$ \\
 \hline
\multicolumn{5}{c}{Global system parameters} \\
  \hline
$\log$(age) [dex] & \multicolumn{4}{c}{$7.116_{-0.115}^{+0.092}$} \\
$[M/H]$  [dex]    &\multicolumn{4}{c}{$-0.394_{-0.066}^{+0.089}$} \\
$E(B-V)$ [mag]    &\multicolumn{4}{c}{$2.901_{-0.013}^{+0.014}$} \\
extra light $\ell_\mathrm{x}$ [in \textit{TESS}-band] & \multicolumn{4}{c}{$0.018_{-0.012}^{+0.031}$} \\
extra light $\ell_\mathrm{x}$ [in SLOAN $r'$-band] & \multicolumn{4}{c}{$0.028_{-0.019}^{+0.027}$} \\
$(M_V)_\mathrm{tot}$  &\multicolumn{4}{c}{$-2.77_{-0.05}^{+0.06}$} \\
distance [pc]           & \multicolumn{4}{c}{$1519_{-39}^{+39}$} \\  
\hline
\end{tabular}}

\textit{Notes}. (a) These are the instantaneous osculating periods, referred to time $t_0$. The long-term mean `observational' period for the triple is $24.498 \pm 0.005$ d.

\end{table*}

\begin{table}
\centering
\caption{Definitions of the parameters listed in Table~\ref{tab:syntheticfit}}
\label{tbl:definitions}
\small
\begin{tabular}{lc}
\hline
\hline
Parameter$^a$ & Definition   \\
\hline
$t_0$ & Epoch time for osculating elements    \\
$P$ & Orbital period  \\ 
$a$ & Orbital semimajor axis  \\
$e$ & Orbital eccentricity \\
$\omega$ & Argument of periastron (of secondary) \\
$i$ & Orbital inclination angle \\
$\mathcal{T}_0^\mathrm{inf/sup}$ & Time of conjunction of the secondary$^b$ \\
$\tau$ & Time of periastron passage  \\
$\Omega$ & Longitude of the node relative to \\
& the node of the inner orbit \\
$i_{\rm mut}$ & Mutual inclination angle$^c$   \\
$q$ & Mass ratio (secondary/primary)  \\ 
$K_\mathrm{pri}$ & ``K'' velocity amplitude of primary \\
$K_\mathrm{sec}$ & ``K'' velocity amplitude of secondary \\
$R/a$ & Stellar radius divided by semimajor axis \\
$T_{\rm eff}/T_{\rm eff,Aa}$ & Temperature relative to EB primary \\
fractional flux  & Stellar contribution in the given band \\
$M$ & Stellar mass  \\
$R$ & Stellar radius   \\
$T_\mathrm{eff}$ & Stellar effective temperature  \\ 
$L_\mathrm{bol}$ & Stellar bolometric luminosity  \\
$M_\mathrm{bol}$ & Stellar absolute bolometric magnitude \\
$M_V$ & Stellar absolute visual magnitude \\
$\log g$ & log surface gravity (cgs units) \\
$[M/H]$ & log metallicity abundance to H, by mass \\
$E(B-V)$ & Color excess in B-V bands  \\
extra light, $\ell_4$  & Contaminating flux in the given band   \\
$(M_V)_\mathrm{tot}$ & System absolute visual magnitude   \\ 
distance & Distance to the source  \\
\hline   
\end{tabular}
\textit{Notes}. (a) The units for the parameters are given in Table~\ref{tab:syntheticfit}. (b) The superscript of ``inf/sup'' indicates inferior vs.~superior conjunctions. (By default we give inferior conjunctions. Superior conjunctions are indicated by asterisk.) (c) More explicitly, this is the angle between the orbital planes of the inner binary and the outer triple orbit.
\end{table} 

In the first stage of our analysis we searched for a photodynamical solution without the use of any additional astrophysical constraints such as the system SED or the use of \texttt{PARSEC} tables to constrain stellar masses, radii and temperatures. In such a manner we have found some seemingly good solutions, but without very accurate information about the individual stellar masses and temperatures. Nonetheless, we were able to find good constraints on such dimensionless quantities as the mass and temperature ratios, as well as the fractional stellar radii. 

We quickly realized, however, that our astrophysically unconstrained model was clearly unphysical, which led to the following problem. The two (robustly obtained) mass ratios of $q_\mathrm{in}=0.89$ and $q_\mathrm{out}=0.58$ revealed that the tertiary star should be more massive by $\approx10\%$ than the primary of the inner EB, and hence it is formally the most massive star in the triple, though not by much. Despite this fact, our solution resulted in a largely oversized tertiary star and, at the same time, substantially smaller, almost spherical stars for the inner binary\footnote{This latter finding comes from the fact that the inner EB exhibits only minor ellipsoidal light variations (ELV) during the out-of-eclipse sections.}. Moreover, the tertiary star's effective temperature was found to be very close to that of the primary of the EB.  At first glance, this apparent contradiction can be resolved by assuming that the tertiary star has just evolved off the MS, while the EB's primary is still close to the zero age main sequence (ZAMS). However, the problem with this interpretation is that we were unable to find a suitable combination of masses, ages, and metallicities which would have resolved all these contradictions. In other words, either the primary of the inner EB became too large, producing a more significant ELV on the light curve (in contradiction with the TESS observations), or the tertiary star was not large enough (and, moreover, it was too hot) for a satisfactory solution. 

To resolve this issue, we employed two different strategies: we allowed for (i) the tertiary to be older than the two inner binary members, and adjusted the age of the tertiary star independent of the age of the binary; and (ii) a significant amount of extra light contamination of the system, which enabled a coeval solution.

\begin{figure*}
    \centering
    \includegraphics[width=0.49\textwidth]{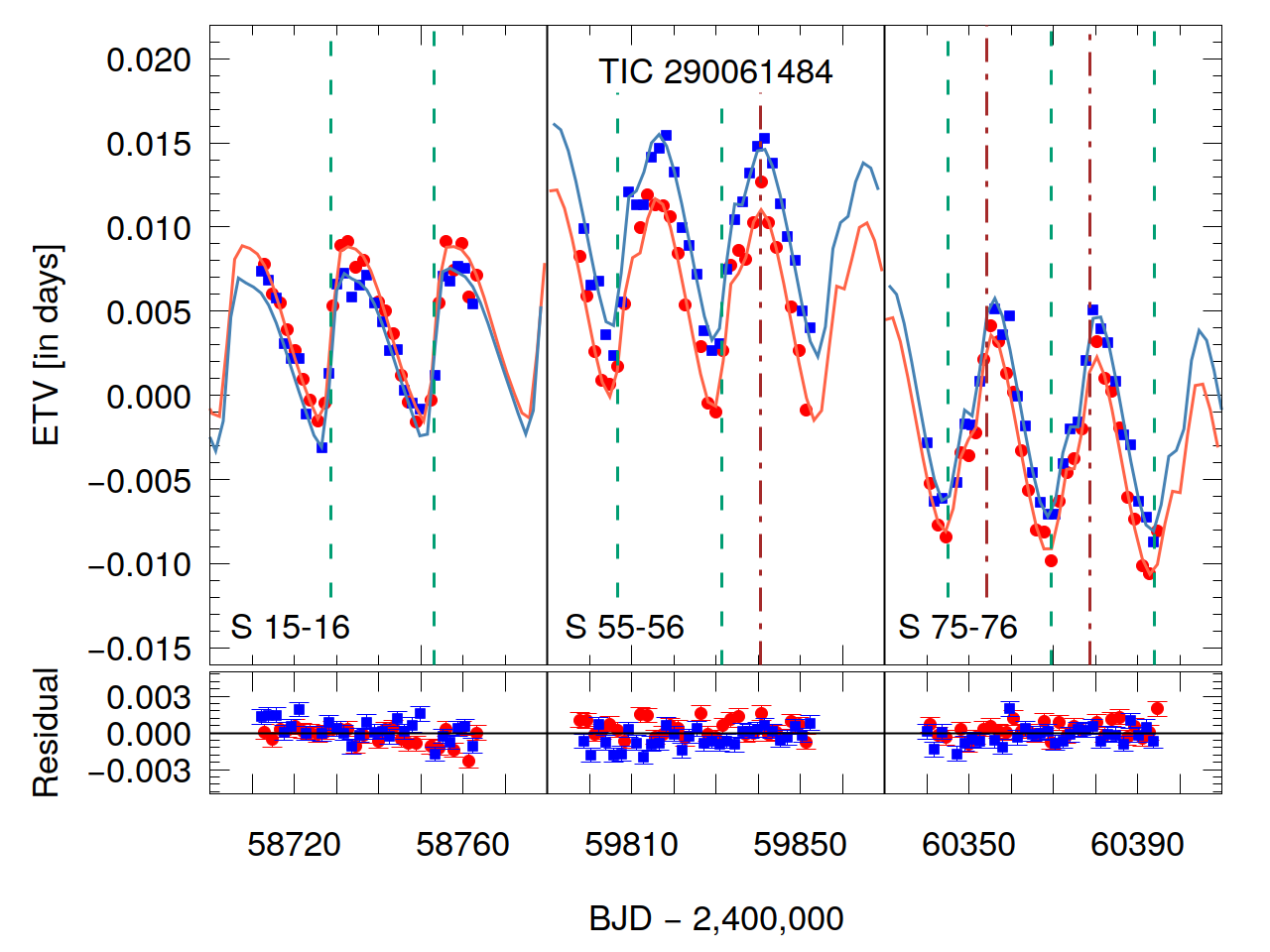}
    \includegraphics[width=0.49\textwidth]{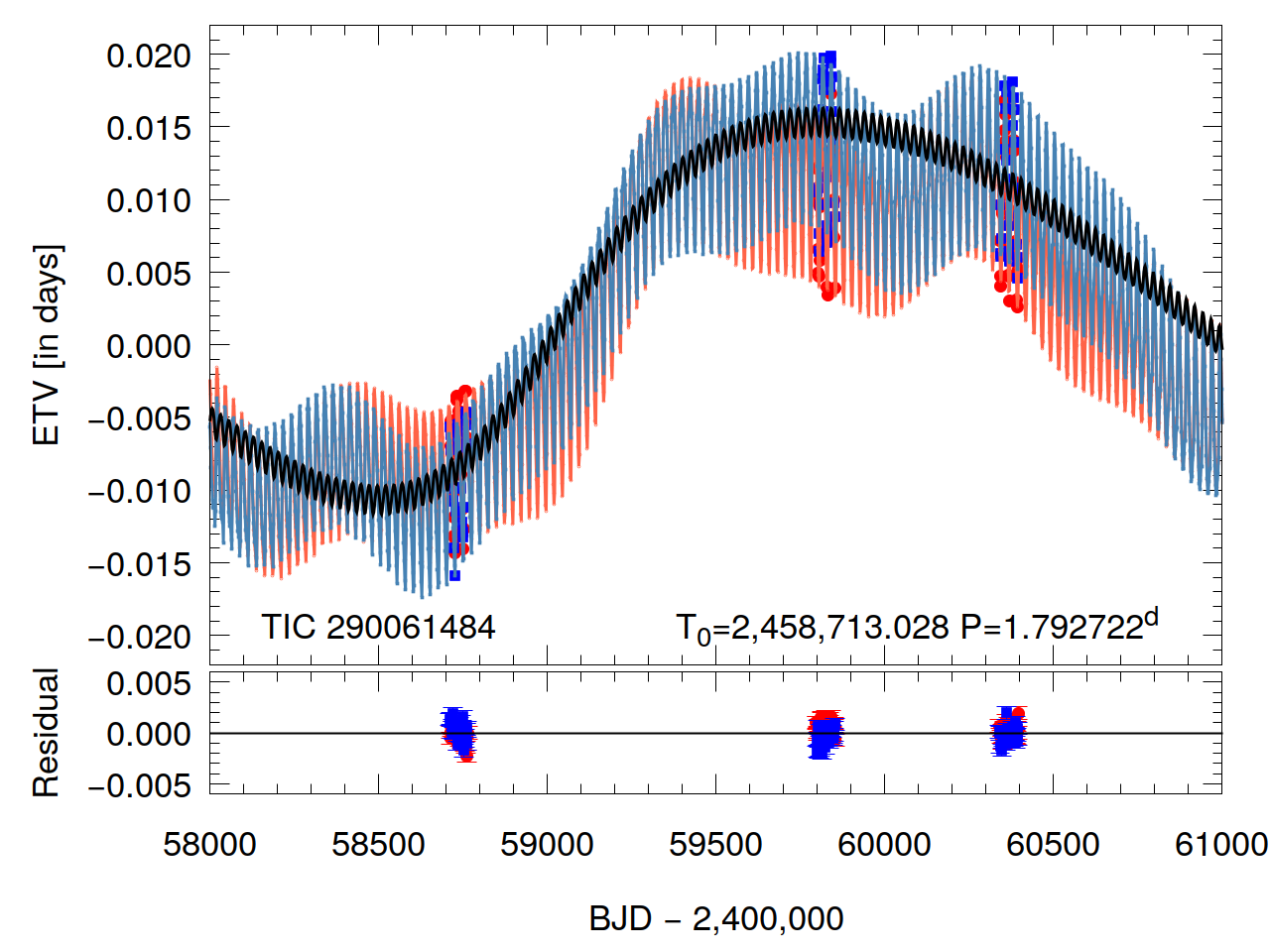}
    \caption{{\it Left}: Primary and secondary ETVs of the inner binary constructed from the TESS \textsc{FITSH} data. The three panels each cover two sectors of TESS data (as labeled in the lower left corners), i.e., about two cycles of the outer orbit. Red (blue) points represent primary (secondary) eclipses. The smooth curves are model fits from the photodynamical analysis (see Sect.~\ref{sec:photodynamics}). The vertical dashed and dot-dashed lines represent the locations of TESS observed eclipses where the binary eclipses the distant tertiary and vice versa, respectively. The lower panels represent the corresponding residuals. {\it Right}: Same as left, but zoomed out to cover an interval of about eight years. The black, sinusoidal curve represent the pure LTTE contribution. This nicely illustrates that on a yearly timescale the ETV is chiefly dominated by the third-body perturbation effects (including the dynamically driven apsidal motion, as well, but on a bit longer timescale, the LTTE, caused primarily by the fourth component, becomes dominant.}
    \label{fig:ETV_segments}  
\end{figure*}

Yet even after allowing for these two possibilities, a significant additional issue remained unresolved. Specifically, as one can notice in Fig.~\ref{fig:ETV_segments}, the medians of the TESS Year 2, 4, and 6 ETVs are offset relative to each other. The origin of such a period variation might be intrinsic (i.e., astrophysical), such as mass transfer between the binary components, mass loss, or some kind of magnetic effects \citep[e.g.][]{Applegate1992}. The physics of these phenomena, and their mathematical form, as they are manifested in an ETV curve, are summarized in \citep[e.g.][]{2011A&A...535A.126N,2015A&A...575A..64N}. Such effects are less likely in the current system due to the detached nature of the inner binary, as well as to the fact that these are hot stars and, thus, we do not expect strong magnetic fields. Therefore, in the current situation, it is more likely that the additional timing variations are caused by some extra dynamical effects. These might have arisen either (i) from some higher-order dynamical perturbations and, perhaps, some interactions between dynamical and tidally forced apsidal motion or, the simpler and more likely assumption (ii) from the effect of an additional, more distant, fourth companion star.

First, we checked the former case. Satisfactory modelling of these effects, however, required very strong fine-tuning of the system parameters\footnote{Formerly, we have detected and successfully modelled the effects of such higher order perturbations in the ETV curves of some tight and compact hierarchical triple and quadruple systems \citep[][]{2023MNRAS.524.4220P,BorkovitsMitnyan2023}.}. Hence, to make it easier for {\sc lightcurvefactory} to find a satisfactory ETV solution, we allowed the code to adjust the first apsidal motion constant ($k_2$) for the two members of the inner binary. This solution resulted in $k_2^{Aa}=0.022\pm0.002$ and $k_2^{Ab}=0.006\pm0.003$. While the latter value is theoretically acceptable, the large value obtained for the primary star appears to contradict the models for such hot main-sequence stars \citep[see, e.g.,][]{2023A&A...674A..67C}. 

Finally, however, we realized that we can obtain a much better ETV solution {\it and} resolve all the other problems discussed above, i.e., the large extra light contamination, the unphysical apsidal motion constant for the primary of the EB, and even the non-coeval evolution issue, by assuming that this extremely compact triple system is actually a subsystem of a quadruple system of (2+1)+1 hierarchy.  In other words, we hypothesize that the system contains a fourth, somewhat more distant star, which (i) produces the missing flux ($\approx20-25\%$ of the total flux) and, moreover, (ii) generates a light-travel time effect (LTTE) (and, to a lesser extent four-body perturbations) that may well explain the longer term structure of the ETV curve. Hence, we switched {\sc Lightcurvefactory} into its (2+1)+1 quadruple-type mode, fixed the apsidal motion constants of the inner pair at a physically realistic value ($k_2^\mathrm{(Aa,Ab)}=0.01$), and reiterated the entire photodynamical analysis with these settings. The 4-star model produced an excellent fit to all available data and indeed settled all the above-mentioned issues associated with the initial 3-star model.

In this regard, we note that it is not exceptional that a tight triple system has a more distant fourth stellar component as well. One example is the HIP 41431 system \citep{2019MNRAS.487.4631B}, observed by the \textit{K2} mission. The tight triple nature of this system was identified independently via spectroscopic, RV analysis of the three inner stars (with orbital periods of $P_\mathrm{in}=2.9$\,d and $P_\mathrm{mid}=59$\,d, respectively), as well as the ETV analysis of the innermost, eclipsing pair. Then, historical RV measurements led to the certain discovery of a fourth component star on a wider orbit with a period of $P_\mathrm{out}=3.9$\,yr, the presence of which was also confirmed with follow-up eclipse timing measurements \citep{2019MNRAS.487.4631B}.

In Table \ref{tab:syntheticfit} we present the results for the system parameters (stellar and orbital) obtained from the photodynamical fit for this (2+1)+1 quadruple model. The following subsection discusses the results. The definitions of the parameters in Table \ref{tab:syntheticfit} are listed in Table \ref{tbl:definitions}.

\subsection{Discussion of the (2+1)+1 quadruple model solution}
\label{sec:2+1+1}

Our results show that TIC 290061484 contains four massive stars: $M_\mathrm{Aa}=6.85\pm0.15\,\mathrm{M}_\sun$ and $M_\mathrm{Ab}=6.11\pm0.15\,\mathrm{M}_\sun$, representing the inner binary Aa+Ab, $M_\mathrm{B}=7.90\pm0.27\,\mathrm{M}_\sun$, representing the tertiary star B, and, bringing up the outer 2+1+1 configuration, the much more weakly constrained quaternary star C with $M_\mathrm{C}=6.01\pm0.5\,\mathrm{M}_\sun$. The corresponding stellar radii are not too far from the zero age main sequence, at $R_{\rm Aa} = 3.08 \pm 0.06$\,R$_\odot$, $R_{\rm Ab} = 2.80 \pm 0.06$\,R$_\odot$, $R_{\rm B} = 3.5 \pm 0.12$\,R$_\odot$, and $R_{\rm C} = 2.78 \pm 0.22$\,R$_\odot$.  Not surprisingly, the stars are quite hot, with $T_\mathrm{Aa}=22154\pm300\,\mathrm{K}$, $T_\mathrm{Ab}=20877\pm400\,\mathrm{K}$, $T_\mathrm{B}=23703\pm400\,\mathrm{K}$, and $T_\mathrm{C}=20690\pm1000\,\mathrm{K}$. The system is young, at $t\approx1.4\times10^7$\,yr-old.

We find a photometric distance to the system of $d=1519\pm39$\,pc with a very large color excess of E(B-V)=$2.90\pm0.01$\,mag, which leads to extreme extinctions, as $A_\mathrm{V}\approx7.9$ or, $A_\mathrm{G}\approx6.7$. Our distance is in essentially perfect agreement with the Gaia DR3-derived purely geometric distance of $1502 \pm 41$\,pc \citep{bailer-jonesetal21}. In our interpretation, this result provides strong support for our quadruple-system hypothesis. 

The extreme extinction to this source, which is located near the Galactic midplane at $\ell^{II} = 91^\circ$ and $b^{II} = 2.9^\circ$, makes the SED fitting difficult. As noted above, the photodynamical analysis indicates that $A_V \simeq 7.9$. This is at least qualitatively consistent with the hydrogen column density (to infinity) of $N_H = 1.03 \times 10^{22}$ cm$^{-2}$ provided by HEASARC\footnote{\url{https://heasarc.gsfc.nasa.gov/cgi-bin/Tools/w3nh/w3nh.pl?}}. Using a conversion factor of $N_H({\rm cm}^{-2}) \simeq 2.21 \times 10^{21} A_V$ \citep{Guver2009}, we would estimate an $A_V$ value that could be in the vicinity of 5. 

TIC 290061484 has a significant mutual inclination of $i_\mathrm{mut}=9.2\degr\pm0.5\degr$. As a natural consequence, both the inner and outer orbital planes precess with a (theoretical) period of $P_\mathrm{node}=2.40\pm0.2$\,yr. This means that the normal vectors of the inner and outer orbital planes (i.e., their orbital angular momentum vectors) move along the surfaces of cones with half-angles of $i^\mathrm{dyn}_\mathrm{in}=8.5\degr\pm0.4\degr$ and $i^\mathrm{dyn}_\mathrm{out}=1.77\degr\pm0.05\degr$ around the normal of the invariable plane of the whole triple subsystem (i.e., the net orbital angular momentum vector of the inner triple) with that period.\footnote{Here we consider the spin angular momenta of the three stars to be negligible.} 

The photodynamical model fits to the three segments of ETV results are shown in Fig.~\ref{fig:ETV_segments}. Each panel contains the ETVs from two adjacent sectors of TESS data, i.e., about 55 days in duration. Thus, the dominant non-linear behavior will occur on the timescale of the orbital period of the inner triple, and therefore, there should be approximately two cycles in each panel. And, indeed this is what we see. The semimajor axis of the EB orbit about the triple's center of mass is only about 33 R$_\odot$. Therefore, the classical light-travel time (LTTE) effect will have an amplitude of only $\sim$0.0009 days. However, we can see from Fig.~\ref{fig:ETV_segments} that the modulation amplitude of the ETV curves is nearly an order of magnitude larger. The explanation for this is that the ETV behavior is largely the result of dynamical delays caused by the varying distance of the EB to the tertiary star \citep[see, e.g.,][]{rappaport2013}. 

From the overall photodynamical fit, which includes the ETV curves as a major constraint on the system parameters, we find that the inner binary is nearly circular, with very small -- but significant -- non-zero eccentricity of $e_\mathrm{in}=0.0025\pm0.0003$. This can be readily seen in the small, but varying, offsets of the secondary eclipses relative to a photometric phase of $0.5$ (i.e., relative to the primary eclipses) in the three segments of the measured ETVs (Fig.~\ref{fig:ETV_segments}). This offset is a manifestation of the mainly third-body-forced, dynamical apsidal motion. As highlighted in the `Apsidal and nodal motion related parameters' section of Table~\ref{tab:syntheticfit}, the tidally forced apsidal motion contribution is about 6\% and, hence, it is non-negligible. In contrast, the relativistic contribution to the apsidal motion is nearly two orders of magnitude smaller and, thus, negligible. The theoretical net period of the apsidal motion is extremely short, $P_\mathrm{apse}=2.8\pm0.1$\,yr (see Table~\ref{tab:syntheticfit}), and its effects are readily visible in the longer timescale ETVs (Fig.~\ref{fig:ETV_segments}). 

The most dominant property of the ETV curve on the longer timescales, however is, the above mentioned extra offset between the different years of \textit{TESS} observations for which our fourth-body assumption provides a natural explanation. In Fig.~\ref{fig:ETV_segments}, besides the full photodynamical ETV solution for both the primary and secondary eclipses (red and blue, respectively), we also show the LTTE contribution caused by the orbit of the triple subsystem around the centre of mass of the whole quadruple system (black curve). As one can see, on a timescale of a few years, this LTTE contribution will be the most characteristic effect. The corresponding parameters of the outermost orbit and the fourth, more distant component, are also tabulated in Table~\ref{tab:syntheticfit}. 

With that said, we caution the reader that the {\it quantitative} results for the outermost orbit and fourth star are not very well constrained. There are several reasons for this caveat. First, an LTTE curve (which is what we are working with for the outermost orbit) does not contain any information about the spatial orientation of the orbital plane of the outermost orbit and, hence, one can say nothing about its inclination and node. Likewise, one can say very little about the mutual inclination between any of the two orbits of the inner triple subsystem and the outermost orbit. In the current situation one can see that the median value for the mutual inclination between the middle and the outermost orbits is $(i_\mathrm{mut})_\mathrm{AB--C}=29\degr\pm16\degr$, which would result in a high amplitude precession of the orbital plane of the inner triple subsystem.  In turn, that would lead to the disappearance of the tertiary eclipses on a timescale of $\sim10,000$ tertiary orbits, and even the binary eclipses after another $\sim20,000$ tertiary orbits. Such an outcome, however, is far from certain and long-term monitoring would be needed to confirm or reject it.

It is important to note that some of the outermost orbital elements ($e_\mathrm{out}$, $\omega_\mathrm{out}$ and $\tau_\mathrm{out}$), as well as the outermost period $P_\mathrm{out}$ can be determined, at least in principle, with considerable accuracy from the ETV curve if at least one full cycle is observed. Naturally, the more cycles the better the accuracy. Unfortunately, this is not the presence case as the available observations do not cover even a single outermost cycle. 

Strictly speaking, in the current situation we cannot be certain even about the outermost period, not to mention all the other orbital elements. Despite this, as we have discussed in the previous section, we are quite certain about the presence of the fourth, more distant component. Hence, the (2+1)+1 configuration {\it qualitatively} seems fairly robust. Further observations will be critical to characterize the properties of the outermost subsystem with higher confidence.

With these cautionary remarks in mind, in what follows we briefly discuss the immediate observational consequences of our photodynamical solution concerning the inner triple subsystem, regardless of the presence and properties of an inclined fourth component.

\begin{figure*}
    \centering
    \includegraphics[width=0.3\textwidth]{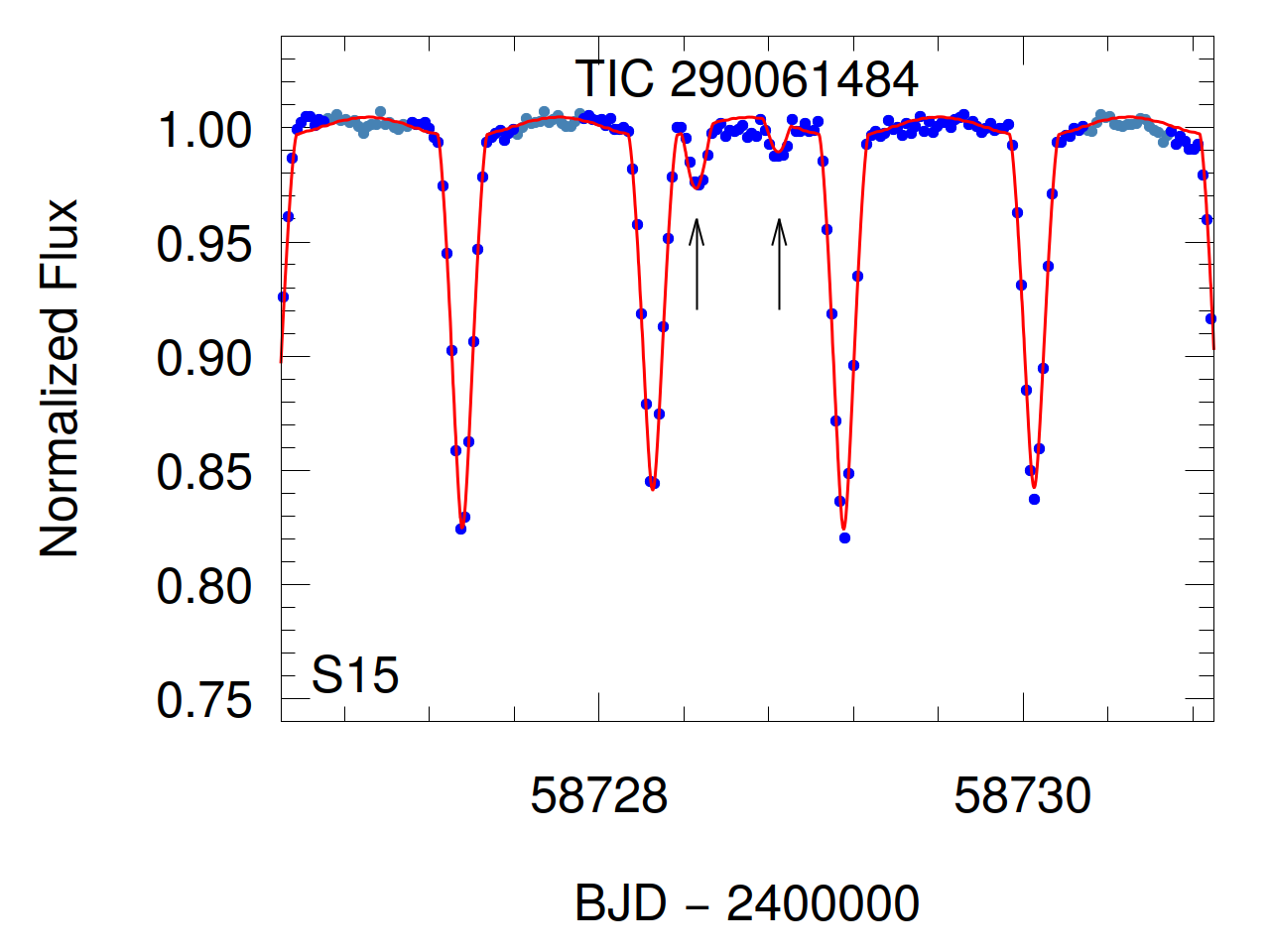}
    \includegraphics[width=0.3\textwidth]{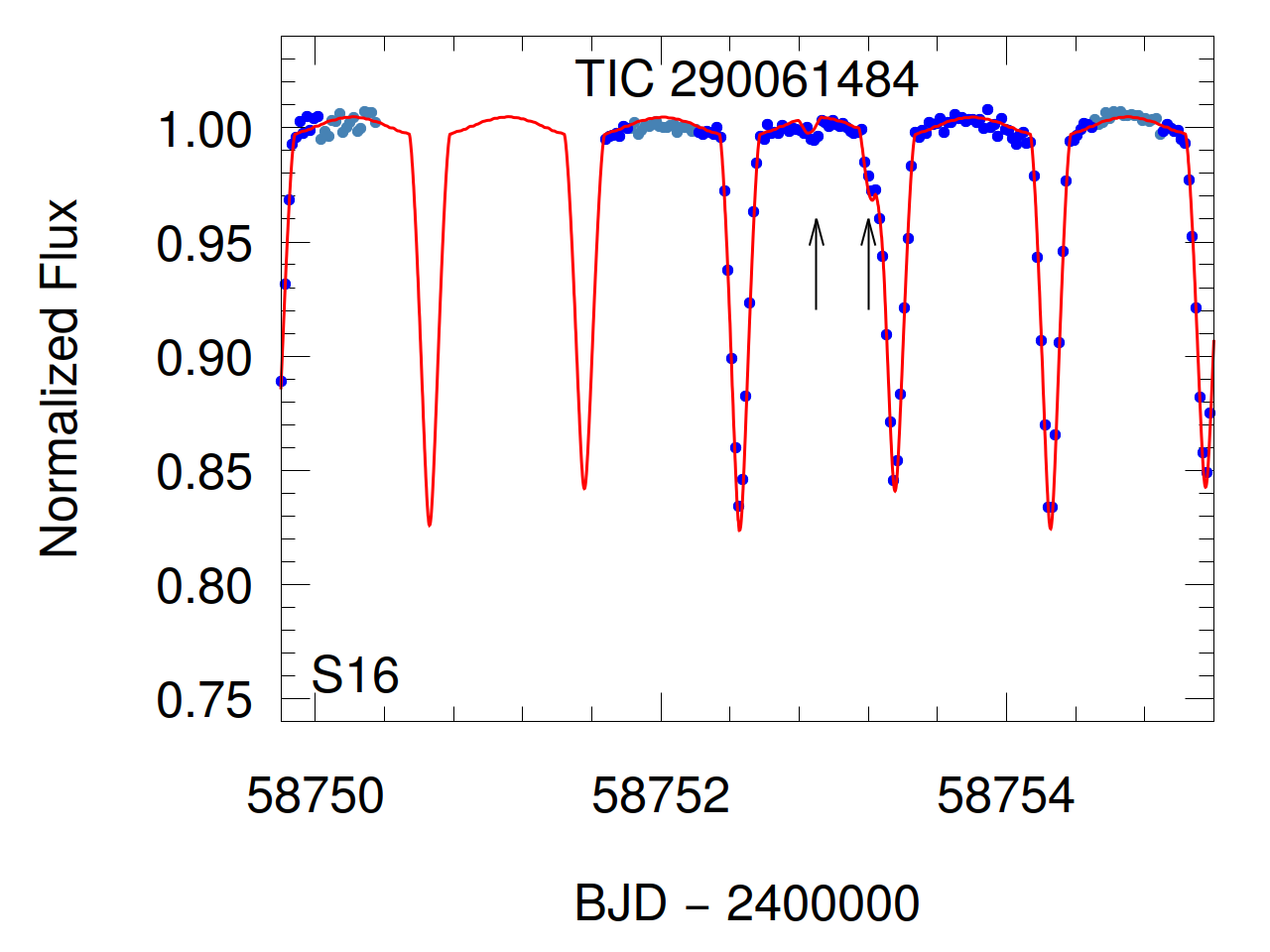}
    \includegraphics[width=0.3\textwidth]{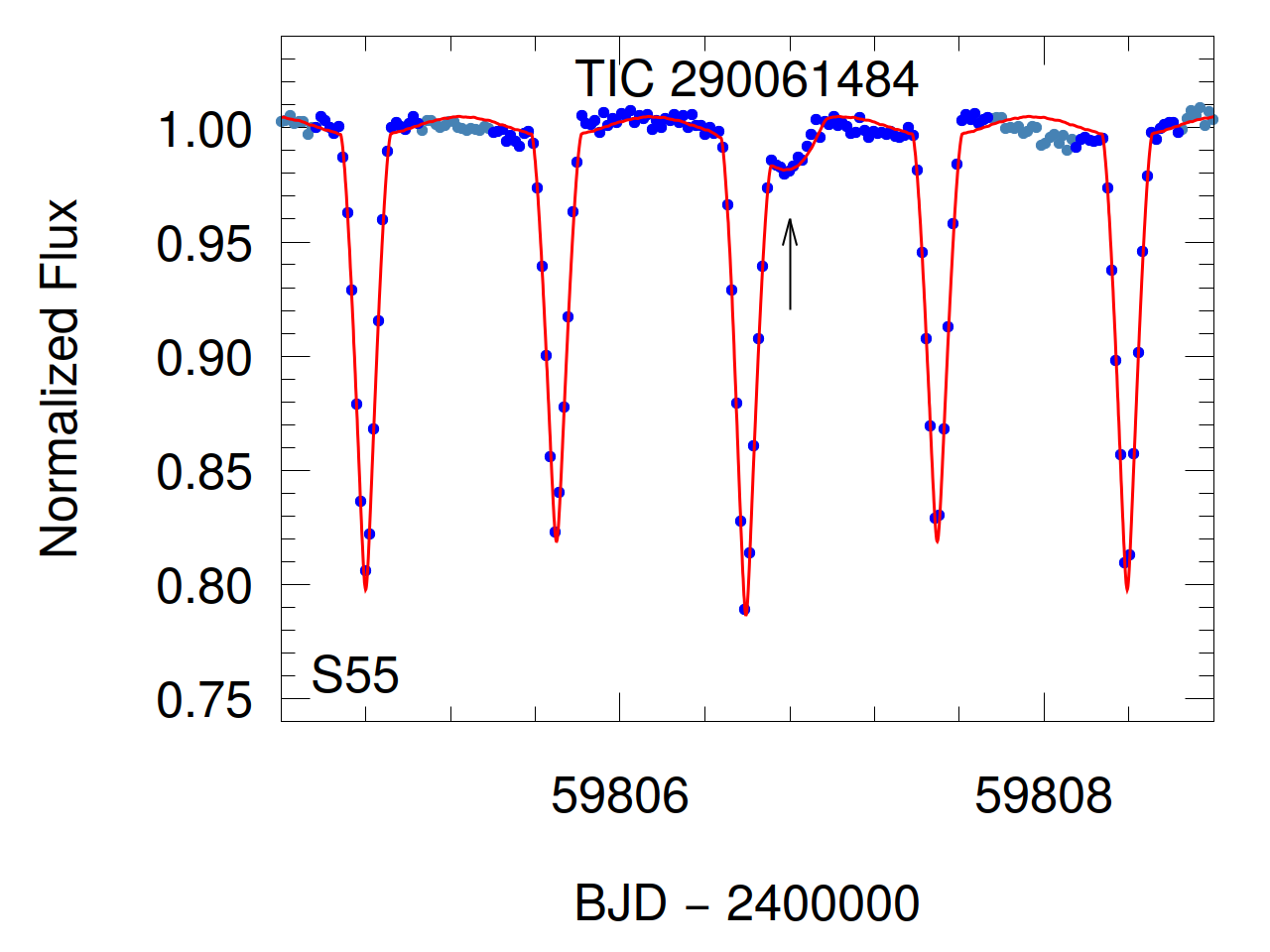}
    \includegraphics[width=0.3\textwidth]{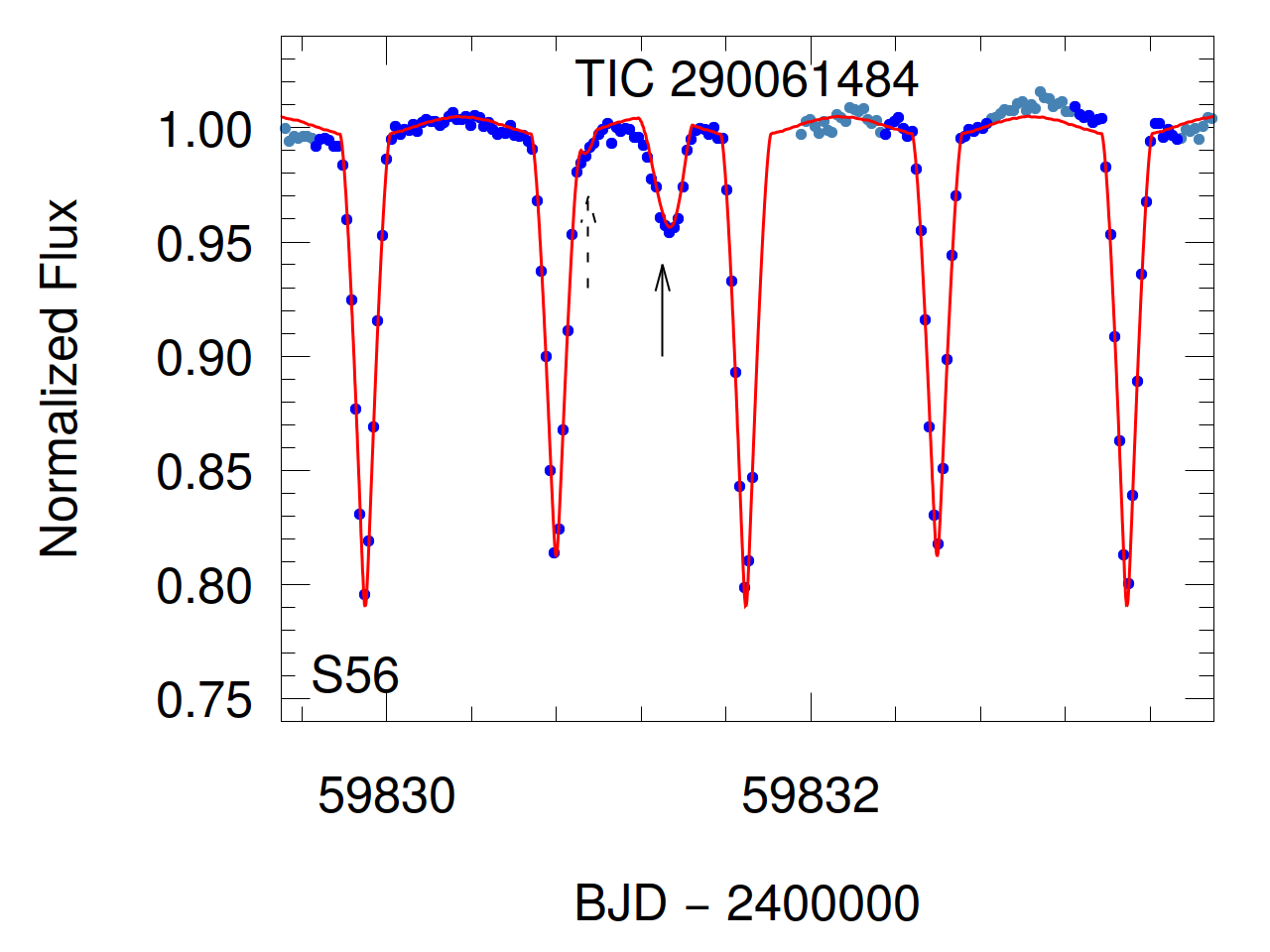}
    \includegraphics[width=0.3\textwidth]{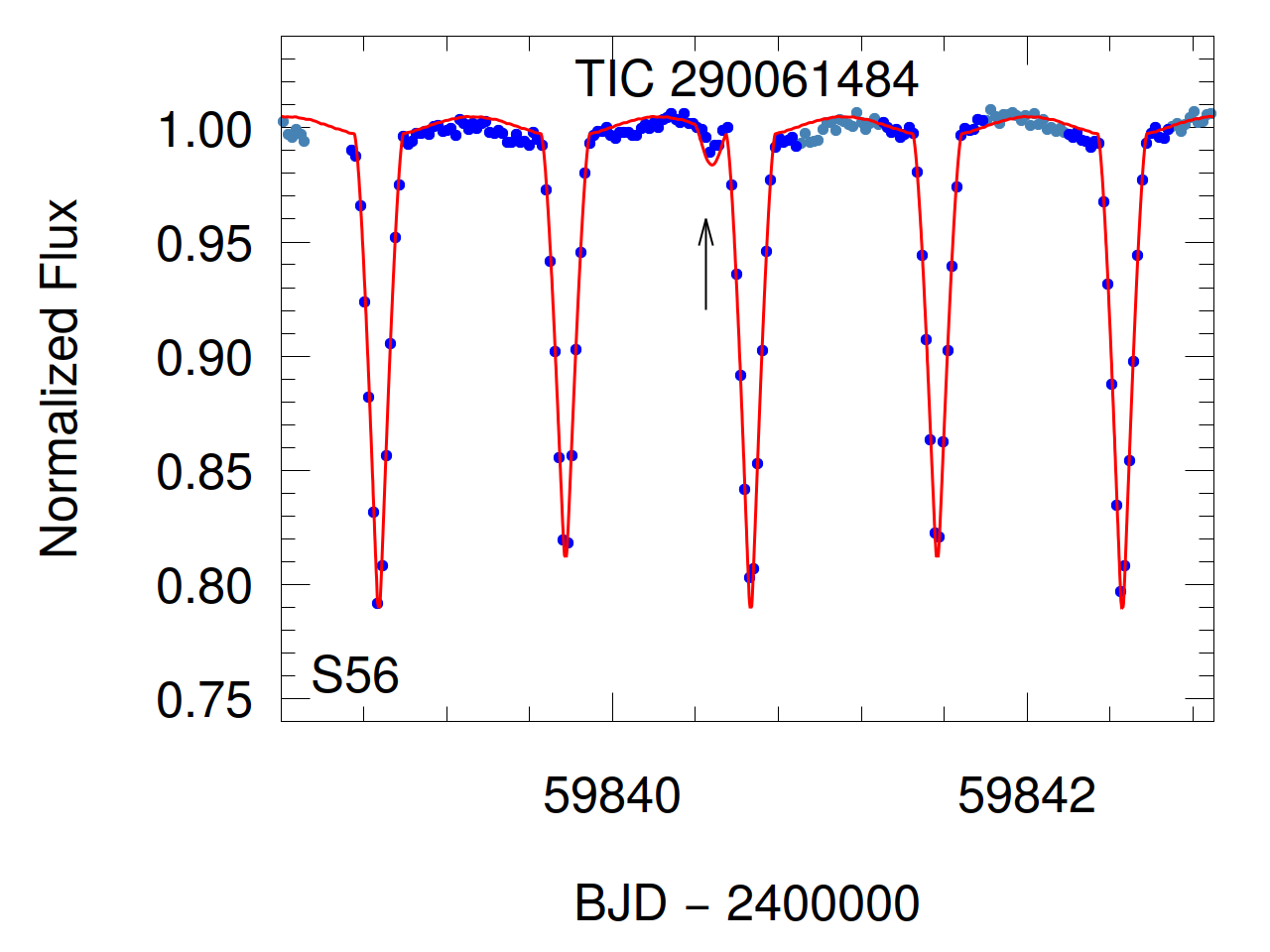}
    \includegraphics[width=0.3\textwidth]{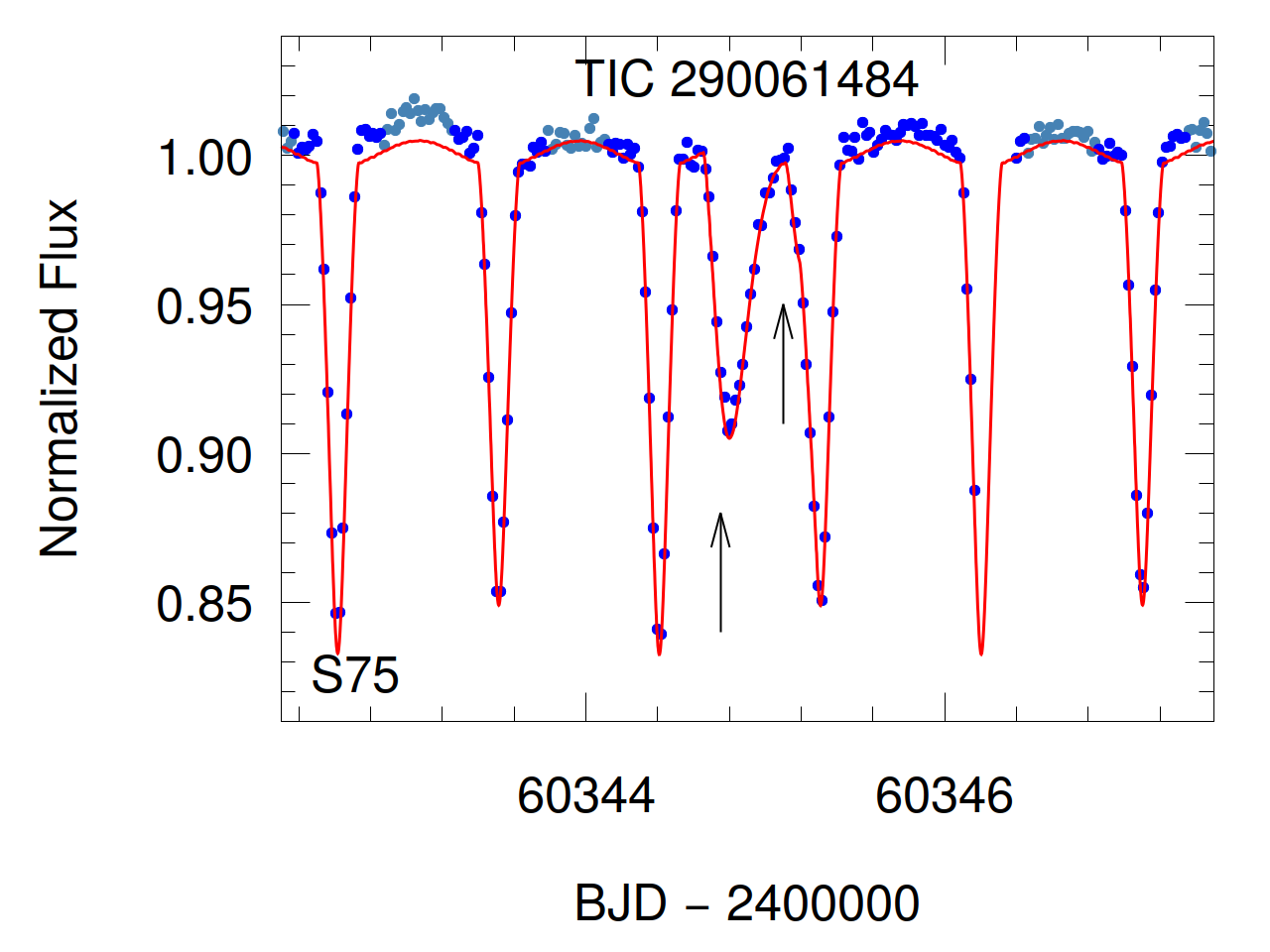}
    \includegraphics[width=0.3\textwidth]{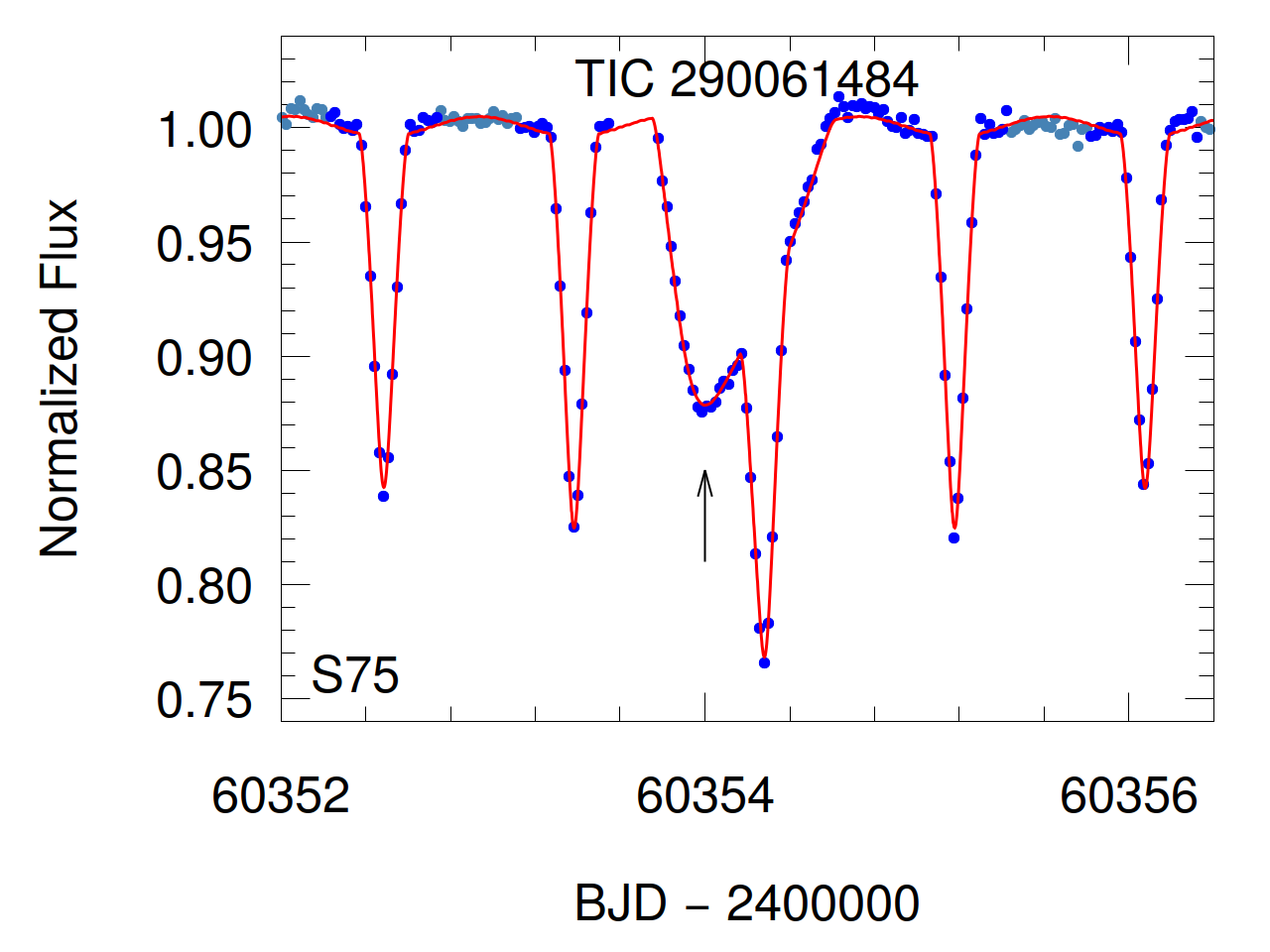}
    \includegraphics[width=0.3\textwidth]{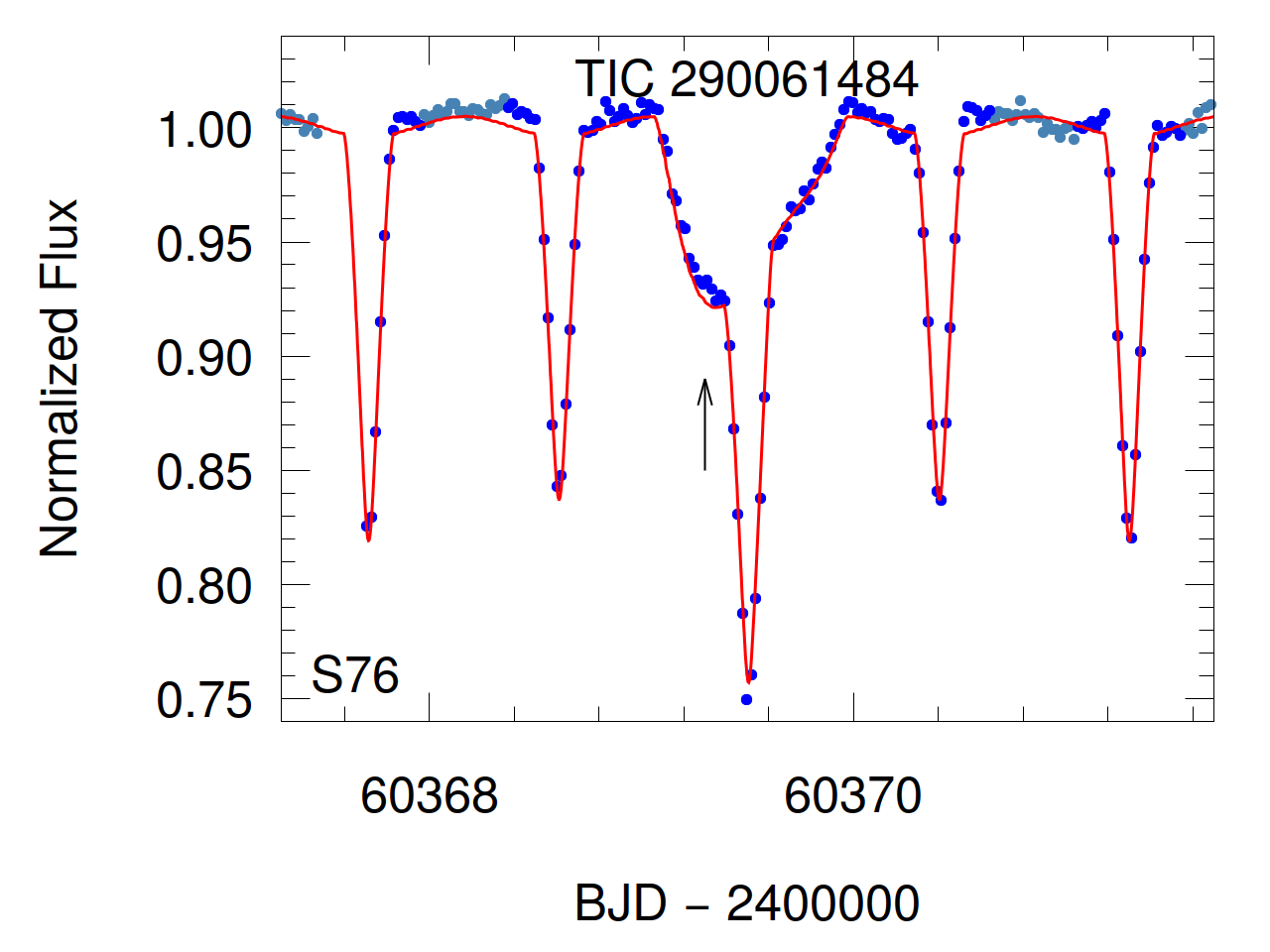}
    \includegraphics[width=0.3\textwidth]{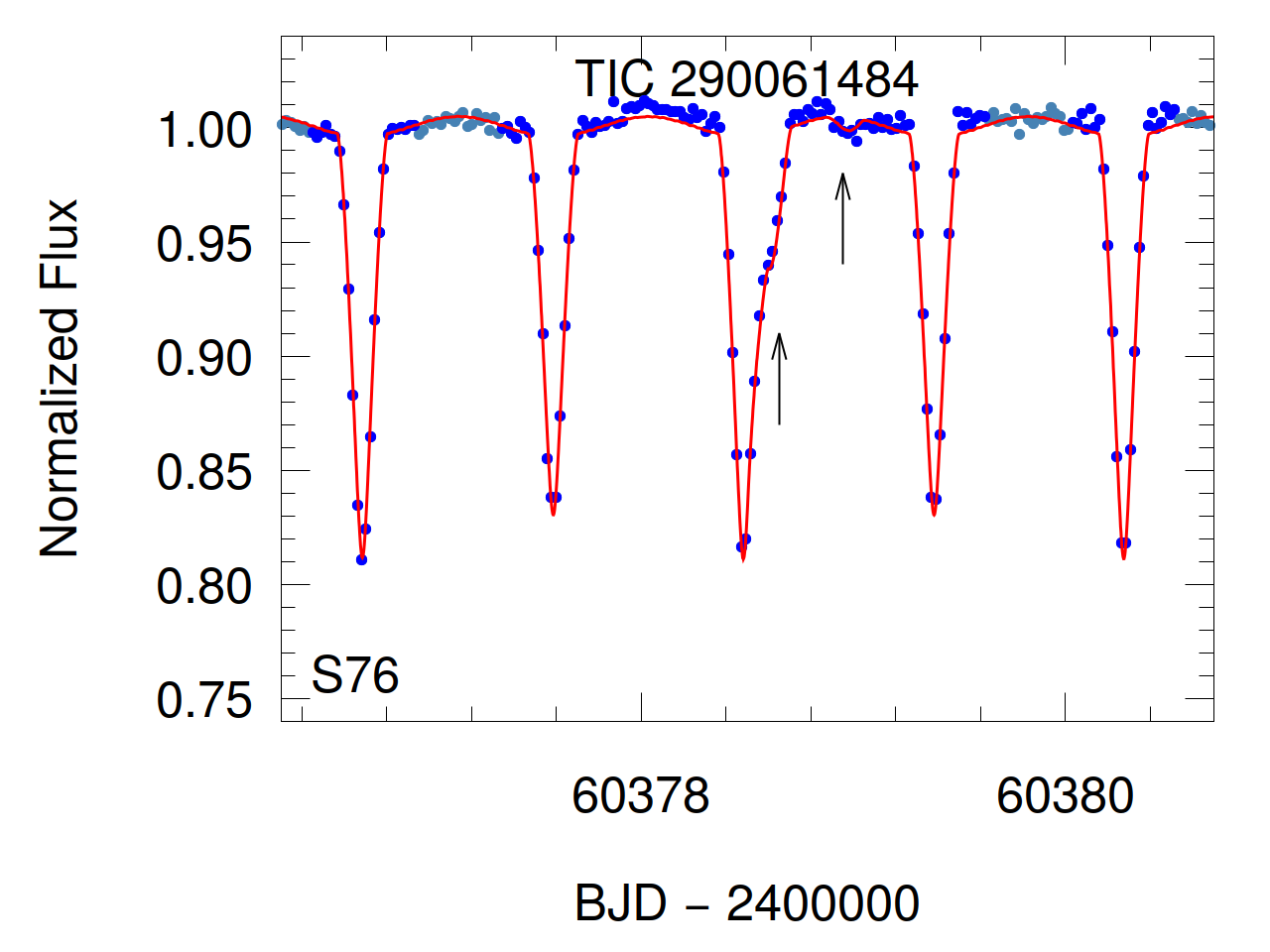}
    \includegraphics[width=0.3\textwidth]{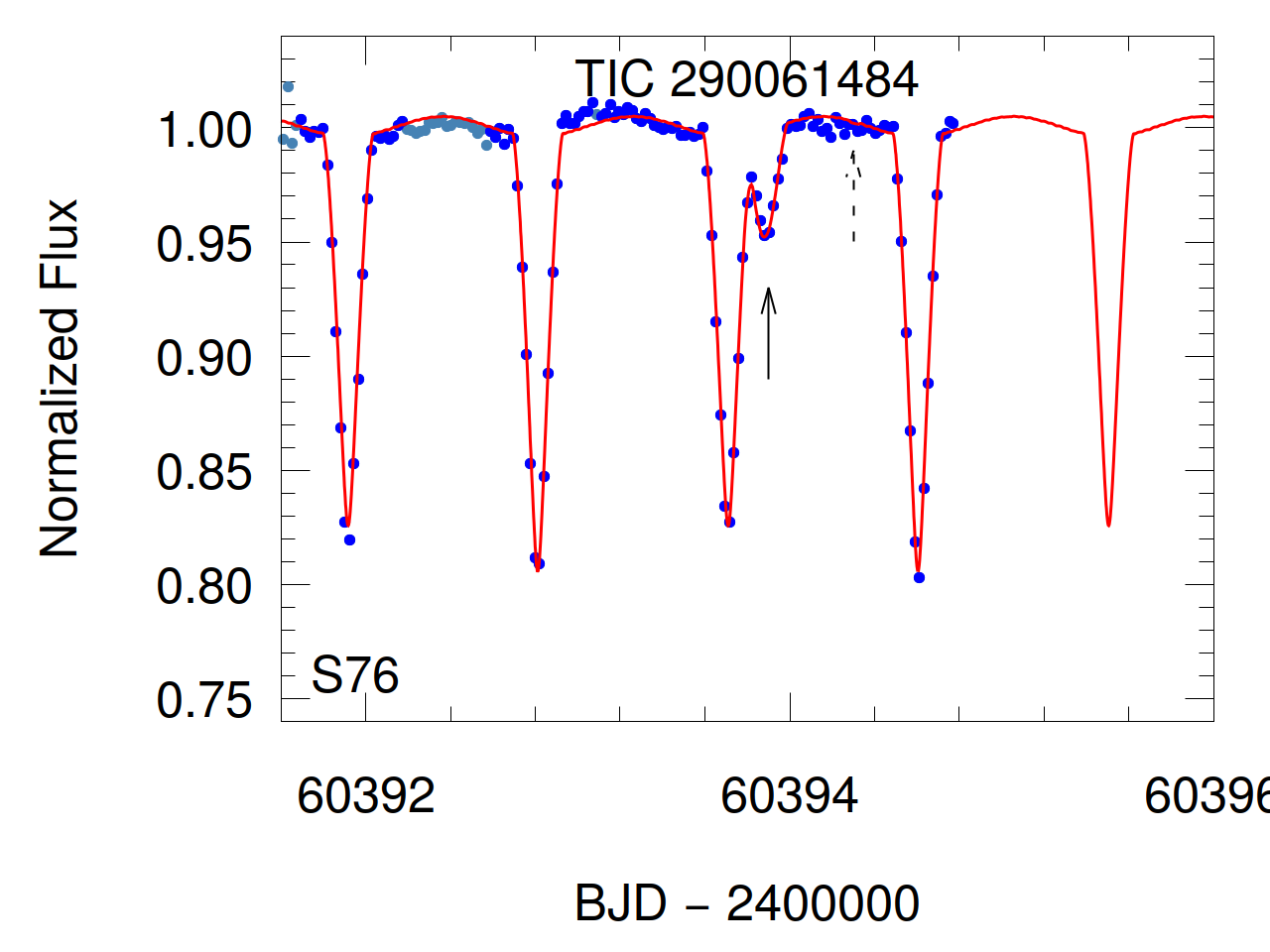}
    \includegraphics[width=0.3\textwidth]{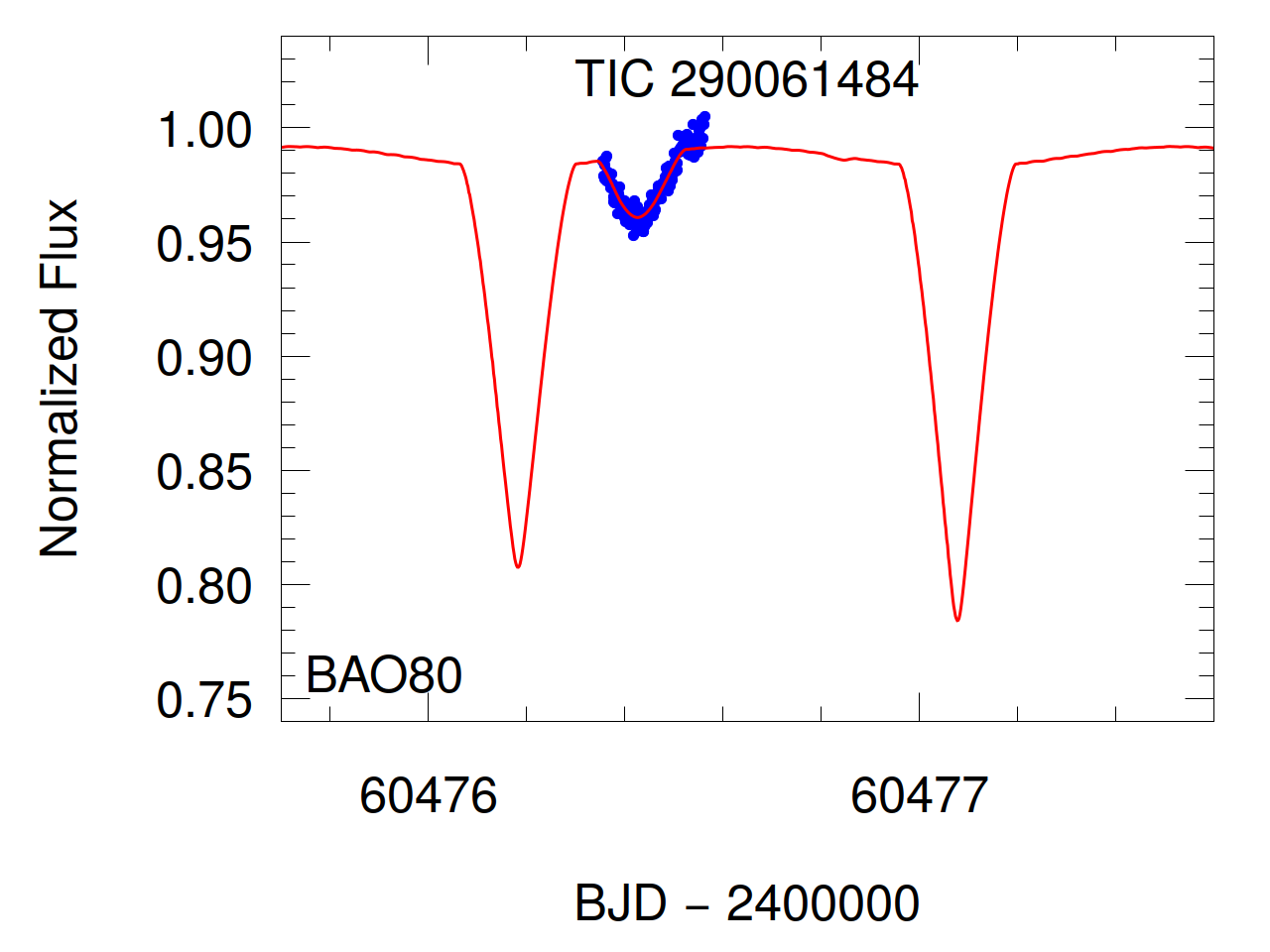}
    \includegraphics[width=0.3\textwidth]{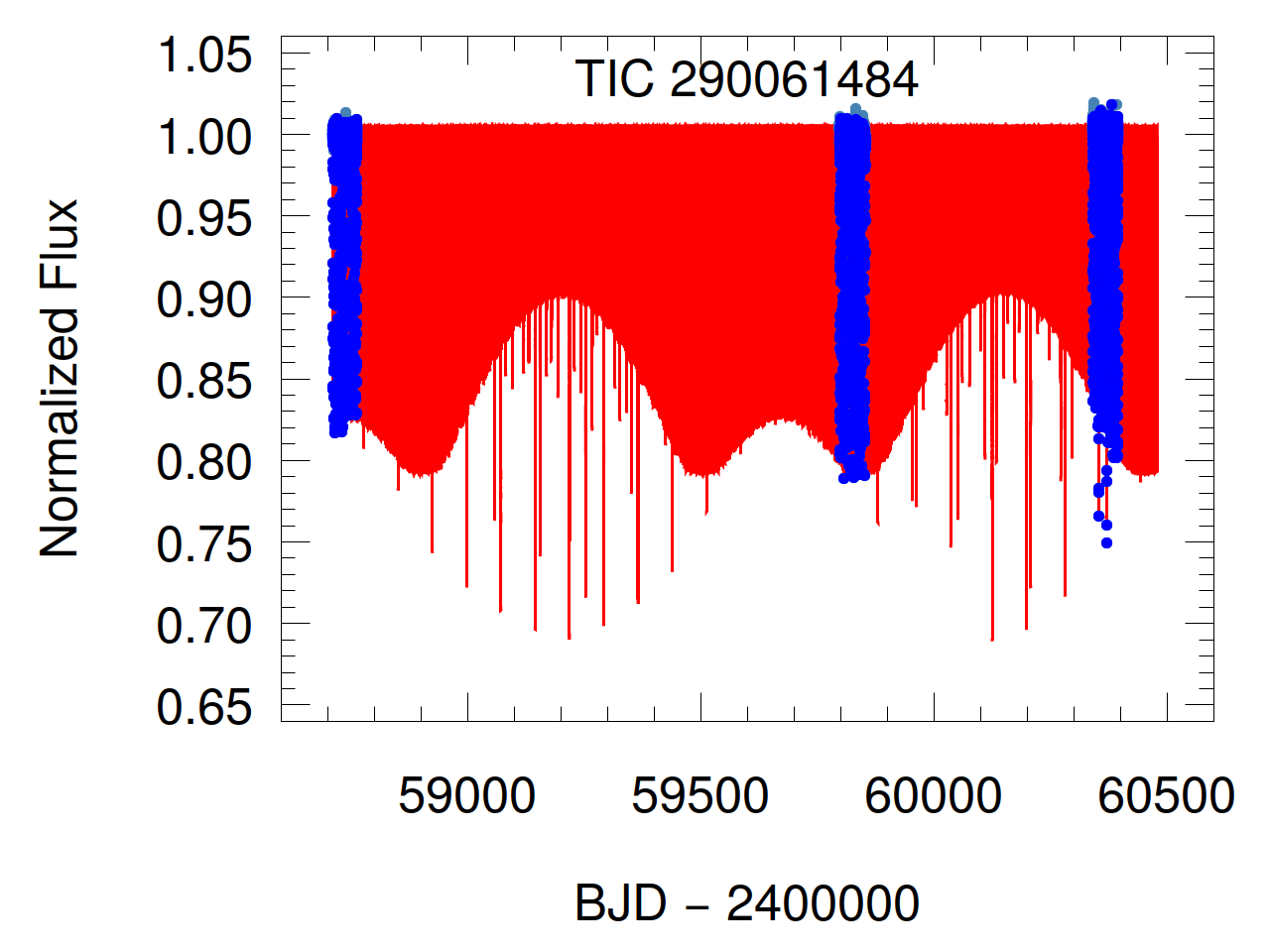}
    \caption{{\it Panels 1 through 11, counting from upper left}: Photodynamical model fits highlighting the tertiary eclipses of TIC 290061484 in Sectors 15, 16, 55, 56, 75 and 76. The blue points are the TESS photometric measurements, while the red curve is the model fit. The TESS sector is marked in the lower left corner of each panel. For an easier identification of the shallowest third-body eclipses, all tertiary events are marked with vertical arrows. Note, dashed arrows show two extremely shallow third-body eclipses which are present on the model light curve, but cannot be identified in the observations. The middle panel in the last row presents the first tertiary event which has been detected in our ground-based photometric follow up campaign, at Baja Astronomical Observatory (Hungary). {\it Lower right panel}: Simulated long-term light curve of TIC 290061484 (red), emphasizing the expected eclipse depth changes. The dense pattern represents the EB eclipses, the thin lines extending below the pattern represent the tertiary eclipses.}
    \label{fig: LCF_lc_fit}
\end{figure*}

The photodynamical model fits to the TESS light curves are shown in Fig.~\ref{fig: LCF_lc_fit}. The asymmetric profile of some of the tertiary events are readily seen as well as their dramatic depth changes between the three TESS epochs (Sector 15-16, Sectors 55-56, Sectors 75-76). The expected long-term light curve of the system is shown in Fig.~\ref{fig: LCF_lc_fit}, showcasing the changing eclipse depths of the inner binary due to the precession of its orbit.

\begin{figure}[ht]
    \centering
    \includegraphics[width=0.99\columnwidth]{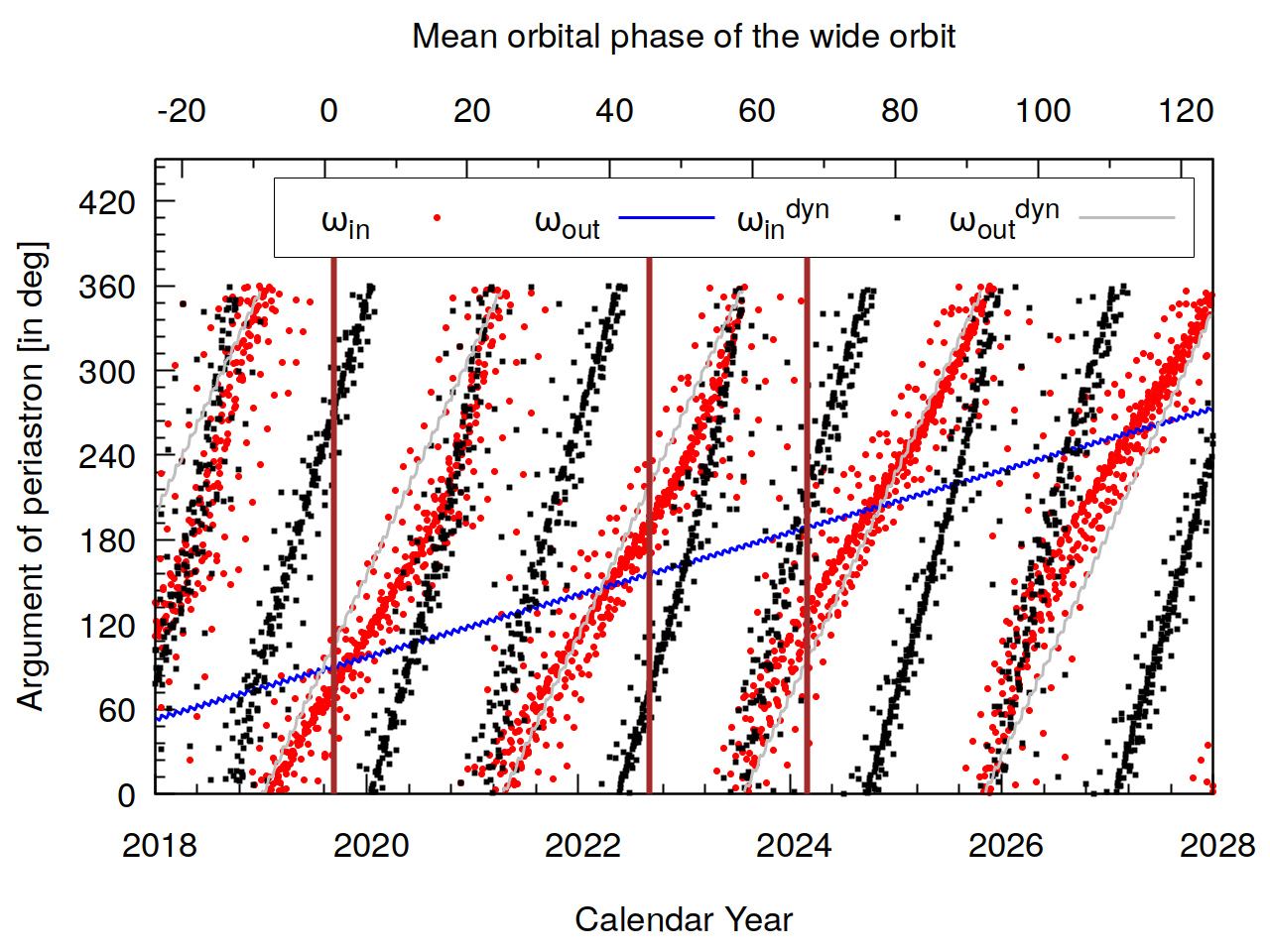}
    \includegraphics[width=0.99\columnwidth]{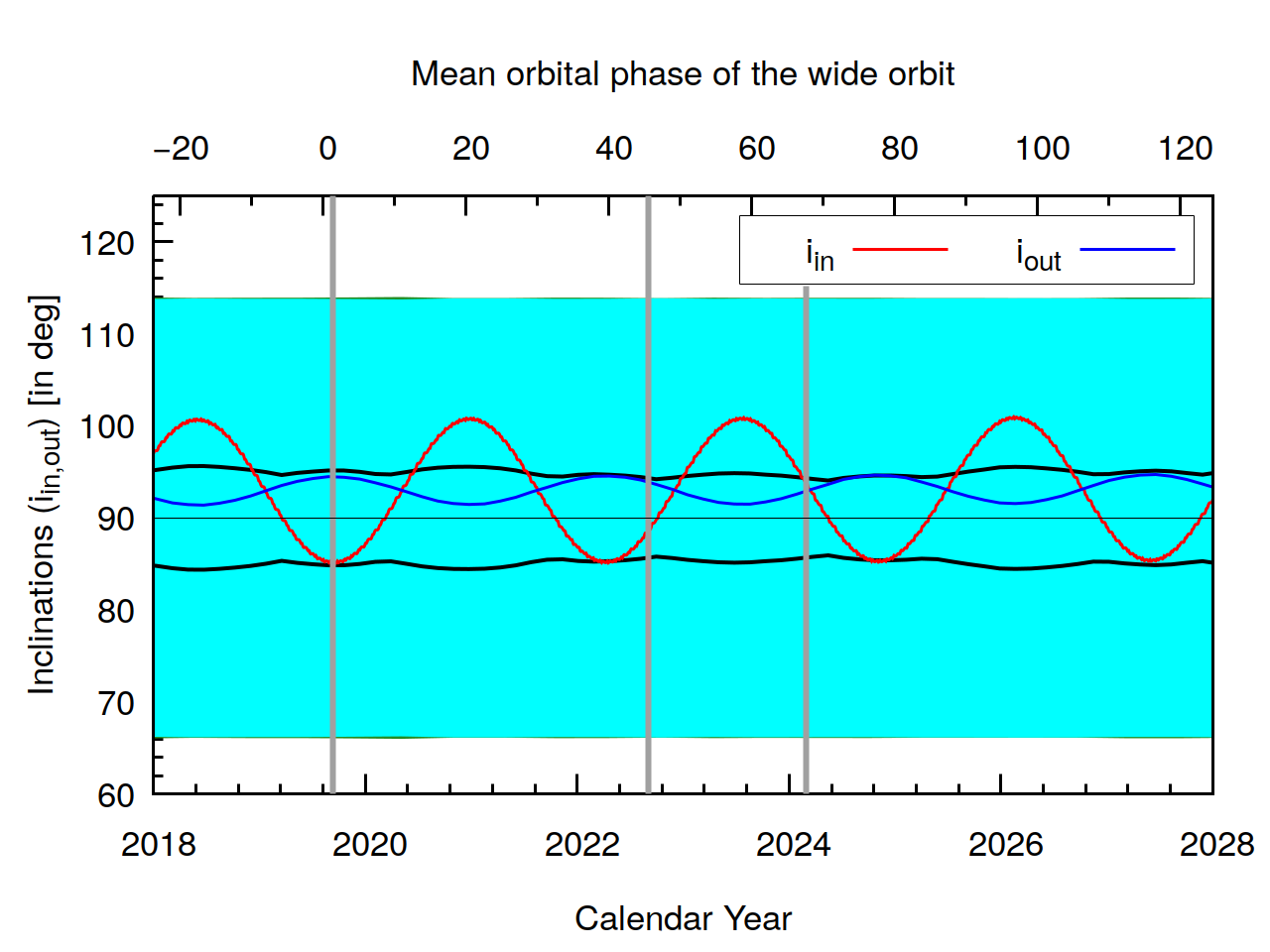}
    \caption{{\it Upper panel:} The variations of the observable and dynamical arguments of periastron of the inner orbit (red and black dots, respectively), as well as the same orbital elements of the outer orbit (blue and gray curves). The vertical, thick grey lines represent the mid times of the three segments of TESS observations. {\it Lower panel:} The variations of the observable inclinations of both the inner and outer orbits (red and blue curves, respectively). The cyan-colored area represent the domain of the (inner) inclination angle where regular, two-body eclipses may occur, while the two, mostly horizontal, thick, black curves represent the borders of the third-body eclipses. As one can see, the inner inclination remains continuously well within its eclipsing domain, hence, the inner EB exhibits permanently deep eclipses. On the other hand, the outer inclination cyclically reaches, and even intersects the border of the third-body eclipsing domain, which leads to very shallow, grazing and even disappearing third-body eclipses from time to time.}
    \label{fig:orbelems}
\end{figure}

The observable consequences of the orbital plane precession mentioned above and the corresponding periodic inclination variations are the {\it slightly} varying eclipse depths of the inner binary and, the {\it substantially} varying depths and durations of the tertiary eclipses between the three TESS epochs (see Fig.~\ref{fig: LCF_lc_fit}). Note, however, that besides the variations of the outer inclination, there is another effect which plays an important role in the duration and depth variations of the outer eclipses. This is the $P_\mathrm{apse,out}=15.7\pm0.2$\,yr-period apsidal motion of the outer orbit. This effect results in a $\sim120\degr$ variation of $\omega_\mathrm{out}$ during the $\sim$5\,yrs of the TESS observations (see Fig.~\ref{fig:orbelems}), which strongly influences the phase-offsets, durations and depths of the third-body eclipses.\footnote{A nice illustration of the effects of the orientation of the apsidal line to the different properties of the eclipses can be seen in Fig.~7 of \citet{BorkovitsMitnyan2023}.} Yet another aspect worth noting is the different shapes of the ETV curves during the three epochs of the TESS observations (Fig.~\ref{fig:ETV_segments}). This is also the direct consequence of the apsidal motion of both orbits (and hence, the variations in the parameters $\omega_{in,out}$), which can be seen using the analytic formulae of \citet{borkovitsetal15}.

\begin{figure}
    \centering
    \includegraphics[width=1.03\columnwidth]{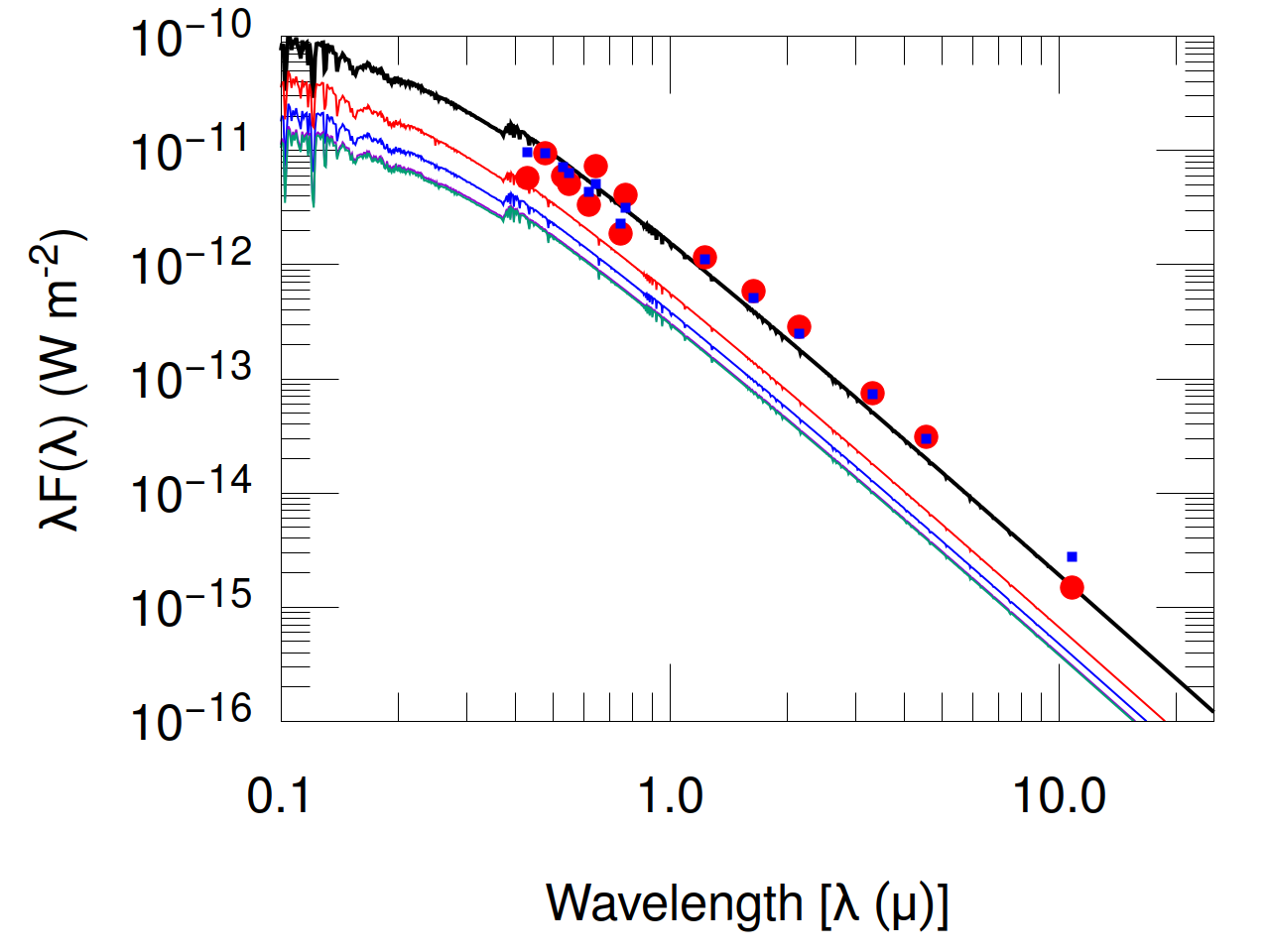}
    \caption{Cumulative SED of the TIC 290061484 system. Red points represent the (dereddened) catalog passband magnitudes that were used in the SED fitting part of the joint photodynamical analysis. Blue dots stand for the theoretical passband magnitudes interpolated from the \texttt{PARSEC} isochrone grids. For comparison we plotted the theoretical ATLAS model atmosphere SEDs \citep{Castelli2003}for the entire system (thick black line representing the cumulative SED), and for the individual stars (thin red line $-$ star B; thin blue $-$ Aa; thin green $-$ Ab and C).}
    \label{fig:SED}
\end{figure}

We are able to come full circle with the photodynamical modeling of the system by examining the system SED. In Fig.~\ref{fig:SED} we show how the measured SED points from the blue out to the 10 $\mu$m WISE 3 band, after correcting for the extreme interstellar extinction, are fit by model spectra for the four massive stars. In particular we compare the dereddened SED points to theoretical passband magnitudes taken from the \texttt{PARSEC} isochrone grids. These are in substantial agreement. We also overlay the theoretical ATLAS model atmosphere SEDs of \citet{Castelli2003}, showing the contributions of the four individual stars, and of the total system light.

Finally, we note that the sky-projected separation of the outermost orbit is $\sim$12 AU. This is considerably smaller than the $\sim$30 AU lower limit constraint achieved by the speckle imaging observations discussed above. Thus, while we cannot resolve the fourth star, at least we know that there is no fifth star present on an even wider orbit.  

\section{Summary and Discussion}

\subsection{System properties}
\label{Sect:system_properties}

In this work we reported the discovery of a new triply eclipsing triple star system, TIC 290061484, first detected in TESS data. The system is remarkable for several reasons. First, it has the shortest outer orbital period ($24.498\pm0.005$ days, long-term average period) of any known triple-star system by quite a wide margin. Second, it has a low ratio of $P_{\rm out}/P_{\rm in} = 13.7$, and thus is highly dynamically interactive. Last but not least, the system exhibits many of the standard dynamical interactions as well as several aspects that require follow-up observations and analysis for a more comprehensive understanding.

All of the system parameters, including those of all four stars and all three orbits, that we were able to derive from the photodynamical modeling are presented in Table \ref{tab:syntheticfit}.

\subsection{Interesting Dynamical Interactions}

Here we summarize the interactions that distinguish the TIC 290064184 triple system from others where the center of mass of the binary is in a simple, unperturbed, eccentric Keplerian orbit about the tertiary star.  The triple system consists of a binary of total mass 13 M$_\odot$ in an eccentric orbit with a nearly 8 M$_\odot$ tertiary star.  The triple's orbit has a mutual inclination angle (i.e., between the plane of the binary and that of the triple) of $\sim$$9^\circ$, and an eccentricity of 0.2.  
The eclipse timing variations on the inner binary exhibit small LTTE delays, and substantial dynamical delays with the period of the outer triple orbit of 24.5 days (see Fig.~\ref{fig:ETV_segments}).  As discussed in Sect.~\ref{sec:2+1+1} these dynamical delays are primarily due to physical changes in the binary period induced by the presence of the tertiary on its eccentric orbit.  The dynamical delays have an amplitude of $\sim$9 min and vary in a quasi-sinusoidal manner.  There is dynamically forced apsidal motion of the inner binary and outer triple orbits on timescales of $\sim$2.8 and $\sim$15.7 years, respectively (see Sect.~\ref{sec:2+1+1}).  A small, but non-negligible contribution to the binary's apsidal motion is from the classical tidal effect.  There is also dynamically forced precession of the orbital planes on cones of half width 1.8$^\circ$ and 8.5$^\circ$ for the binary and outer triple orbit, respectively.  The timescale for these precessions is $\sim$2.4 years (see Sect.~\ref{sec:2+1+1}).  These result in dramatic eclipse depth variations for the tertiary eclipses. 

Dynamical simulations indicate that the EB eclipses will continue for millennia, while the tertiary eclipses will range from prominent to barely grazing on the same timescales (see Figs.~\ref{fig: LCF_lc_fit} and \ref{fig:orbelems}). The behavior of the eclipses on longer timescales depends on the still-uncertain configuration of the outermost orbit. For the best-fit photodynamical solution presented here, the EB eclipses do not stop while the tertiary eclipses last for $\sim30,000$ outer periods, cease for the next $\sim35,000$ outer periods, and then start again.

The long-term ETV curve (see Fig.~\ref{fig:ETV_segments}) shows an LTTE variation of amplitude $\sim$18 min, that is unexplained in the context of the triple system.  For this reason, among a couple of others related to the lightcurve and SED fitting, we invoked the presence of a fourth star in the system (in a 2+1+1 configuration) with an outermost period of $\sim$3300 d. With that, the model for the long-term ETV curve is fit very well.  

Continuous monitoring of the ETVs might reveal some higher-order, smaller amplitude perturbations that we have not yet explored. Especially interesting would be perturbations on the intermediate timescale between the outer orbit of less than a month, and the longer timescale apse-node perturbations.

\subsection{Predictions of Future Eclipses}

As one can see from Fig.~\ref{fig:ETV_segments}, the long-term ETV curve, dominated by the LTTE contribution of the orbit of the triple around the fourth star, is uncertain because of the relatively sparse sampling and a total observational baseline that is only $\sim$60\% of the period of the outermost orbit (2730 days). Therefore, it would be extremely helpful to have continued ground-based follow up observations of this object. Measured times of primary and secondary eclipses are always helpful, but the times of just a few third-body eclipses over the next couple of years will allow the photodynamical model to be updated in a meaningful way.

In Table \ref{tbl:TIC_290061484_futureE3} we list the mid-times of all 33 outer eclipses predicted by the photodynamics model for the next 500 days, and deeper than $\Delta\,m\geq0.1$ mag. Any observations of these eclipses would be very helpful, and we will update our photodynamics model accordingly as new eclipse times are provided. 

The eclipse depths are also listed in Table \ref{tbl:TIC_290061484_futureE3}. Two thirds of the third-body eclipses are expected to be more than 5\% deep, and these should be readily observable with amateur telescopes. 

\begin{table}
 \centering
\caption{Predicted third-body eclipse events for the next 500 days.}
 \label{tbl:TIC_290061484_futureE3}  
\begin{tabular}{@{}lrllrl}
\hline
mid-time & width  & depth & mid-time & width  & depth \\ 
$BJD-t0^\dag$ & d &  $\Delta$ mag &$BJD-t0^\dag$ & d &  $\Delta$ mag  \\ 
\hline
500.9 &    0.3 & 0.028 & 893.0$^*$ & 0.1 & 0.136 \\ 
549.9 &    0.2 & 0.011 & 893.1 &     0.3 & 0.135 \\ 
697.3 &    0.2 & 0.017 & 907.4 &     0.1 & 0.014 \\ 
746.0$^*$ & 0.1 & 0.030 & 907.9 &     0.4 & 0.103 \\ 
746.3 &    0.3 & 0.022 & 917.3 &     0.2 & 0.044 \\ 
770.7 &    0.4 & 0.055 & 917.7 &     0.3 & 0.161 \\ 
795.1 &    0.3 & 0.045 & 931.8 &     0.2 & 0.026 \\ 
810.1 &    0.3 & 0.026 & 932.2 &     0.2 & 0.071 \\ 
819.5$^*$ & 0.1 & 0.053 & 941.7 &     0.3 & 0.078 \\ 
819.7 &    0.4 & 0.075 & 942.1 &     0.2 & 0.101 \\ 
834.5 &    0.2 & 0.036 & 956.1 &     0.3 & 0.034 \\ 
843.8 &    0.2 & 0.022 & 956.8$^*$ & 0.5 & 0.095 \\ 
844.2 &    0.3 & 0.110 & 966.5$^*$ & 0.9 & 0.209 \\ 
868.3 &    0.2 & 0.032 & 981.2$^*$ & 0.9 & 0.187 \\ 
868.6 &    0.2 & 0.075 & 990.8 &     0.2 & 0.075 \\ 
883.1$ˇ*$ & 0.1 & 0.068 & 991.2 &     0.4 & 0.191 \\ 
883.4 &     0.5 & 0.097 & \\ 
\hline
\end{tabular}

Note: $^\dag$ The reference time is $-2\,460\,000$; asterisks denote third-body event which are superposed with regular two-body eclipses.
\end{table}

\subsection{Comparison with previous observations of triples}

Compared to the collection of 33 compact triply eclipsing triples discussed in \citet[][see their Fig.~22]{Rappaport2024}, TIC 290061484 stands out mostly for the substantially higher masses of its components, as well as the very short outer period. Most of the stellar masses in the systems discussed in \citet{Rappaport2024} are in the range of 1-3 M$_\odot$, with a typical value of $\sim$2 M$_\odot$. In contrast, the three stars in TIC 290061484 each have a mass above 6 M$_\odot$. Aside from that, the inner and outer mass ratios ($q_{\rm in}$ and $q_{\rm out}$) for TIC 2900661484 are consistent with those of the \citet{Rappaport2024} collection. Likewise, the outer orbital eccentricity of TIC 290061484 (0.2) is rather typical. The 9$^\circ$ mutual orbital inclination angle of TIC 290061484 is exceeded by only three of 33 systems discussed in the \citet{Rappaport2024}. As highlighted here, the outer orbital period is shorter, by far, than for any other triple system known at the time of writing.

We note that the other noteworthy close triple involving massive stars is TIC 470710327 \citep{Eisner2022}, where the inner and outer periods are 1.1 d and 52 d, respectively. The system contains a tertiary of $\sim$15 M$_\odot$ and an inner binary of total mass $\sim$12 M$_\odot$. In this case, as a slight exception to the general rule that we have found for several dozen triples with less massive stars, TIC 470710327 has $q_{\rm out} \simeq 1.25 \pm 0.18$, i.e., marginally $\gtrsim 1$. As discussed in \citep{Eisner2022}, such massive stars are undoubtedly headed toward a type II supernova during some future phase of the evolution of this system. 

\subsection{Searches for triples with even shorter periods}
\label{sec:stability}

The discovery of TIC 290061484 with an outer orbital period of only 24.5 d, after the previously shortest period of 33 days (for $\lambda$ Tau) had stood for 68 years, raises the intriguing question of whether even more compact triple systems are possible. In order to have shorter outer periods such systems must (i) be able to form in the first place, and (ii) be long-term dynamically stable.

To check the long-term dynamical stability of triple systems, which is a minimum requirement, we make use of the stability criteria for nearly coplanar triple systems summarized by \citep{Mikkola2008}. In particular, we use the formalism of \cite{Mardling2001}, expressed in terms of the orbital periods:

\begin{eqnarray}
P_{\rm trip}  \gtrsim 4.7 \left(\frac{M_{\rm trip}}{M_{\rm bin}}\right)^{1/10} \frac{(1+e_{\rm out})^{3/5}}{(1-e_{\rm out})^{9/5}} ~ P_{\rm bin} ~.
\label{eqn:stableP1}
\end{eqnarray}  

If we ignore the very weak dependence on mass, and take $e_{\rm out}$ to be 0.2, as representative (and equal to the outer eccentricity in TIC 290061484), this expression comes to
\begin{eqnarray}
P_{\rm trip}  \gtrsim 7.8 ~ P_{\rm bin} ~.
\label{eqn:stableP2}
\end{eqnarray}  
For TIC 290061484, $P_{\rm out} = 13.66 \, P_{\rm in}$, and thus the system should be long-term dynamically stable with room to spare. If we reduce both periods of TIC 290061484 in half, we would have a perfectly conventional 0.9-d inner binary and 12-day outer period, which is similarly stable. Even outer periods of 8-10 days seem not implausible to contemplate.

\subsection {Formation Scenarios}

Close binary companions cannot form in situ \citep{Boss1986,Bate1998}. Stellar companions instead fragment on protostellar disk or molecular core scales beyond $a$~$>$~10~AU and subsequently migrate inward, probably through circumbinary disk/envelope accretion whereby the orbital energy is dissipated into the surrounding gas \citep{Bate2002,Moe2018,Tokovinin2020,Offner2023}.  Extremely compact triples likely form through a specific sequence of two fragmentation episodes and substantial circumbinary/triple accretion. Circumtriple accretion tends to dampen the eccentricities and reduce the mutual inclination of the orbits \citep{Bate2010,Bate2012}, which are necessary ingredients in maintaining dynamical stability while hardening toward an extremely compact configuration (see above). 

Moreover, as a binary migrates inward, most of the mass from the circumbinary disk is accreted by the companion, thereby driving the mass ratio toward unity \citep{Farris2014,Young2015}. Close solar-type binaries exhibit an excess of twins with $q$ = 0.96\,-\,1.00 (\citealt{Tokovinin2000}; \citealt{Moe2017}; see Fig.~14 in \citealt{ElBadry2019}) and a deficit of $q$ $<$ 0.1 companions known as the brown dwarf desert \citep{Grether2006}. Toy models of disk fragmentation, inward disk migration, and circumbinary disk accretion can successfully reproduce the observed brown dwarf desert and measured 25\% twin fraction within $P$ $<$ 100~days \citep{Tokovinin2020}.

An ultra-compact triple like TIC~290061484 with $P_{\rm out}$ = 24.6~days requires a delicate, fine-tuned, multi-staged formation process. If the companions fragment too late, the remaining gas in the surrounding disk/envelope would be insufficient to harden the components. Conversely, if the inner binary undergoes inward disk migration that is too efficient, the pair will merge during the pre-main-sequence (pre-MS). Similarly, if the tertiary migrates inward too quickly relative to the inner binary, the triple can become gravitationally unstable (see Eqn.~\ref{eqn:stableP1}), which typically results in the least massive component being thrown to large separations or ejected entirely \citep{Valtonen2006,Moe2018}.

We compile a list of 44 compact triples with $P_{\rm out}$ $<$ 300 days and measured component masses, including the triply eclipsing triples in Fig.~\ref{fig:tightness_porb} and slightly misaligned triples that have non-eclipsing tertiaries \citep[][Borkovits et al., in prep.]{Gaulme2022,Borkovits2022,Powell2022,Eisner2022,Orosz2023,BorkovitsMitnyan2023,Moharana2024}. We exclude $\lambda$~Tau because the inner binary is an evolved, semi-detached Algol that has widened its orbit as the subgiant donor has transferred most of its mass to the MS accretor.  We count nine extremely compact triples with $P_{\rm out}$ $<$ 50~days, of which eight have $P_{\rm in}$ = 0.8\,-\,1.8~days, including TIC~290061484 (the single exception is TIC~2421327789 with $P_{\rm in}$ = 5.1 days). Most extremely compact triples therefore have inner binaries that migrated to extremely close separations while narrowly avoiding a merger. In particular, the inner binary of TIC~290061484 contains early-type components, currently $R$ = 3\,R$_{\odot}$ but likely $R$ $\approx$ 5\,R$_{\odot}$ during the embedded pre-MS phase, and thus just barely escaped over-filling its Roche lobe. 

It should be noted that 96\% of very close binaries with $P$ $<$ 3~days are in triples \citep{Tokovinin2006}. It was originally speculated that the tertiary played an active role in hardening the inner binary via Kozai-Lidov cycles and tidal friction \citep{Kiseleva1998,Fabrycky2007}. However, \citet{Moe2018} demonstrated that most triples do not have the necessary orbital configurations to dynamically harden the inner binary to very short periods. Moreover, the timescales of dynamical hardening and tidal friction are too long to reproduce the observed population of young pre-MS binaries with $P$ $<$ 3 days, and thus they concluded that most very close binaries must derive from inward migration in a gaseous disk/envelope during the pre-MS phase. \citet{Tokovinin2020} subsequently explained the common origin responsible for the observed correlation between very close binaries and their large triple star fraction. The formation and migration of a binary to very short periods requires a massive disk/envelope, and such a massive disk/envelope is also more likely to fragment twice, thereby forming a triple. Similarly, the formation and migration of an ultra-compact triple like TIC~290061484 required an even more massive disk/envelope, which was prone to fragment into the compact 2+1+1 quadruple that we see today.

The average mutual inclination of triples/quadruples increases with increasing separation \citep{Tokovinin2017b}.  The outer component in TIC~290061484 is probably more misaligned than the ultra-compact triple. In fact, given the measurement uncertainties in our photodynamical fit (see Table~\ref{tab:syntheticfit}), the quaternary could conceivably have a sufficiently large mutual inclination $i_{\rm mut}$ $>$ 39$^{\circ}$ to excite Kozai-Lidov cycles in the tertiary, potentially responsible for its 9$^{\circ}$ misalignment with respect to the inner binary. However, the outer component cannot be so inclined to cause substantial pumping of the tertiary's eccentricity. The tertiary must remain below $e$ $<$ 0.38 to remain dynamically stable according to Eqn.~\ref{eqn:stableP1}, and thus the outer component must have $i_{\rm mut}$ $\lesssim$ 45$^{\circ}$ (quadrupole-order approximation; Eqn.~20 in \citealt{Naoz2016}).

\begin{figure}
    \centering
    \includegraphics[width=0.5\textwidth]{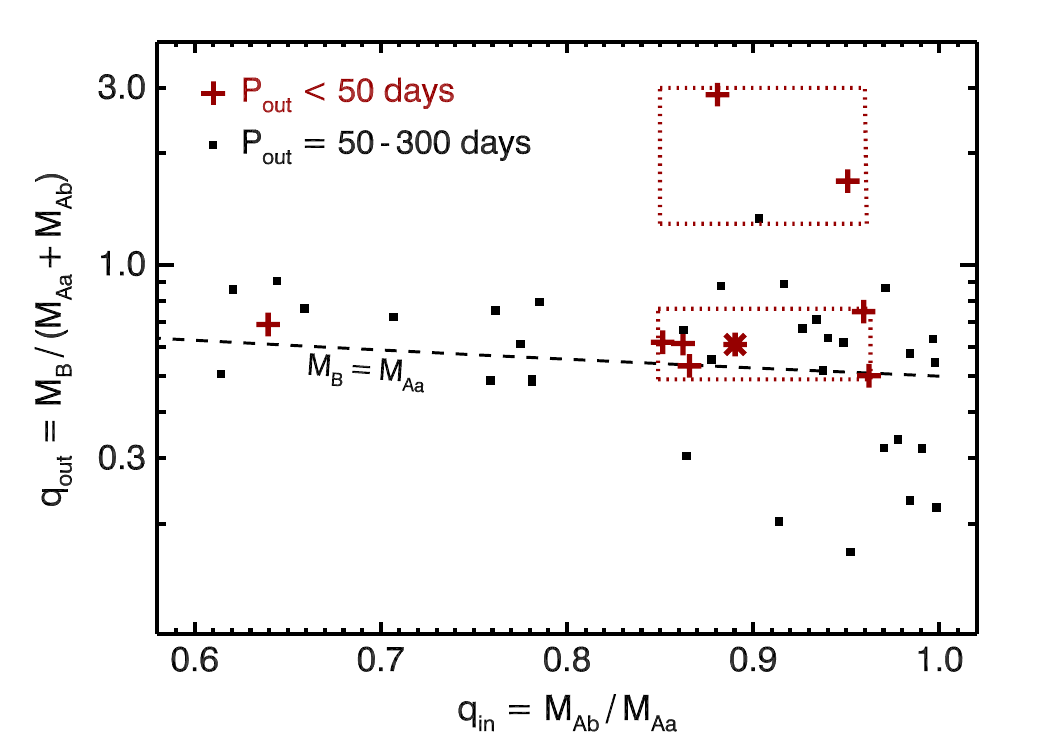}
    \caption{Outer versus inner mass ratios for compact triples with $P_{\rm out}$ $<$ 300 days. We highlight the nine extremely compact triples with $P_{\rm out}$ $<$ 50 days (red pluses), including TIC~290061484 (red asterisk), and we bracket their two main islands (dotted red lines). A triple where the tertiary is the most massive component lies above the dashed line.}
    \label{fig:massratios}
\end{figure}

The unique channel of circumtriple accretion that hardens a triple to below $P_{\rm out}$ $<$ 50 days imprints a unique signature on its mass ratios $q_{\rm in}$ = $M_{\rm Ab}/M_{\rm Aa}$ and $q_{\rm out}$ = $M_{\rm B}$/($M_{\rm Aa}+M_{\rm Ab}$). In Fig.~\ref{fig:massratios}, we display the inner and outer mass ratios for our 44 triples with $P_{\rm out}$ $<$ 300 days and measured component masses (2 with $q_{\rm in}$ $<$ 0.45 fall off the plot because they are to the left of the displayed domain). Of our nine extremely compact triples with $P_{\rm out}$ $<$ 50~days, all but one (TIC~332521671) have inner mass ratios that span a narrow interval $q_{\rm in}$ = 0.85\,-\,0.96. The inner binaries are relatively near equal mass but not quite true twins. As discussed in \citet{Tokovinin2020}, twins must have initially fragmented early in their formation process and experienced significant circumbinary disk accretion. Of our 44 triples with $P_{\rm out} < 300$ d, 9 (20\%) have $q_{\rm in}$ $>$ 0.96, consistent with the observed 25\% twin fraction among all very close solar-type binaries. The components of the inner binaries in extremely compact triples must have also fragmented relatively early and experienced substantial circumbinary disk accretion to exceed $q_{\rm in}$ $>$ 0.85. However, the presence and hardening of the compact outer tertiary interrupted the final accretion toward twin status, and thus extremely compact triples have inner binaries that mostly span $q_{\rm in}$ = 0.85\,-\,0.96. 

The majority (7/9) of the extremely compact triples have $q_{\rm out}$ = 0.50\,-\,0.75. As a low-mass tertiary migrates inward and accretes most of the mass from the circumtriple disk, the outer mass ratio increases. Just as there is a desert of close brown dwarf companions below $q$~$<$~0.1, there is a desert of tertiaries below $q_{\rm out}$ $<$ 0.5 within $P_{\rm out}$ $<$ 50 days. Conversely, it is more difficult to harden a companion that is already a twin. In the \citet{Tokovinin2020} model, most of the inward disk migration occurs when $q$ $<$ 0.2 and the migration halts once the companion accretes to $q$ = 1. It is thus not surprising that all known triples with $q_{\rm out}$ = 0.75\,-\,1.0 have tertiaries that remain beyond $P_{\rm out}$ $>$ 50 days. 

Disk fragmentation, inward migration, and circumtriple disk accretion can only form triples with $q_{\rm out}$ $<$ 1 \citep{Tokovinin2020}, and indeed the majority of compact triples have $q_{\rm out}$ $<$ 1. The three outliers with massive tertiaries spanning $q_{\rm out}$ = 1.3\,-\,3.0 in Fig.~\ref{fig:massratios} are KOI-126, HD\,181068 and TIC~470710327. The latter has an outer tertiary with $P_{\rm out}$ = 52~days, marginally wider than our extremely compact triples arbitrarily defined as those with $P_{\rm out}$ $<$ 50 days. These three systems mirror the main population of extremely compact triples but with inverted outer mass ratios, i.e., the inner binary components are similar in mass with $q_{\rm in}$ = 0.85\,-\,0.96 suggesting significant circumbinary disk accretion. The outer pairs of these three systems with $q_{\rm out}$ $>$ 1 likely formed via core fragmentation. Both KOI-126 and TIC~470710327 have modest outer eccentricities $e_{\rm out}$ = 0.3, further indicating a dynamical origin whereby the tertiary fragmented on larger core scales. 

A larger sample of compact triples with $q_{\rm out}$ $>$ 1 is needed to better understand their formation. Fortunately, our main island of extremely compact triples, including TIC~290061484, is better constrained. Circumbinary accretion hardened the inner binary to $P_{\rm in}$ = 0.8\,-\,1.8 days and $q_{\rm in}$ = 0.85\,-\,0.96, and circumtriple accretion hardened the tertiary to $P_{\rm out}$ $<$ 50 days and $q_{\rm out}$ = 0.50\,-\,0.75, all while maintaining dynamical stability and avoiding a merger of the inner binary.

In Appendix \ref{sec:popsyn} we simulate the formation of a very large population of multistellar systems, and we show that there is only one ultra-compact triple like TIC 290061484 for every $\sim$1.3 million star systems. We also estimate that for every triple like TIC~290061484 that could successfully thread the needle of migrating to $P_{\rm out}$ = 25~days while maintaining dynamical stability, there were likely 10$^4$ triples that became disrupted during their formation process.

\subsection{Long-term Dynamical Stability}

Based on the analytic fitting formulae for long-term dynamical stability in triple systems (see Eqn.~\ref{eqn:stableP1}), the triple subsystem of TIC 290061484 has a ratio of $P_{\rm out}/P_{\rm in}$ that is nearly twice that required for stability ($\sim13.7$ vs 7.8). The outermost orbit is comfortably stable as well, with a ratio of $P_{\rm out}/P_{\rm in}$ that is more than seven times larger than the minimum allowed ratio of ${\sim18.6}$. 

\begin{figure*}
    \centering
    \includegraphics[width=0.95\textwidth]{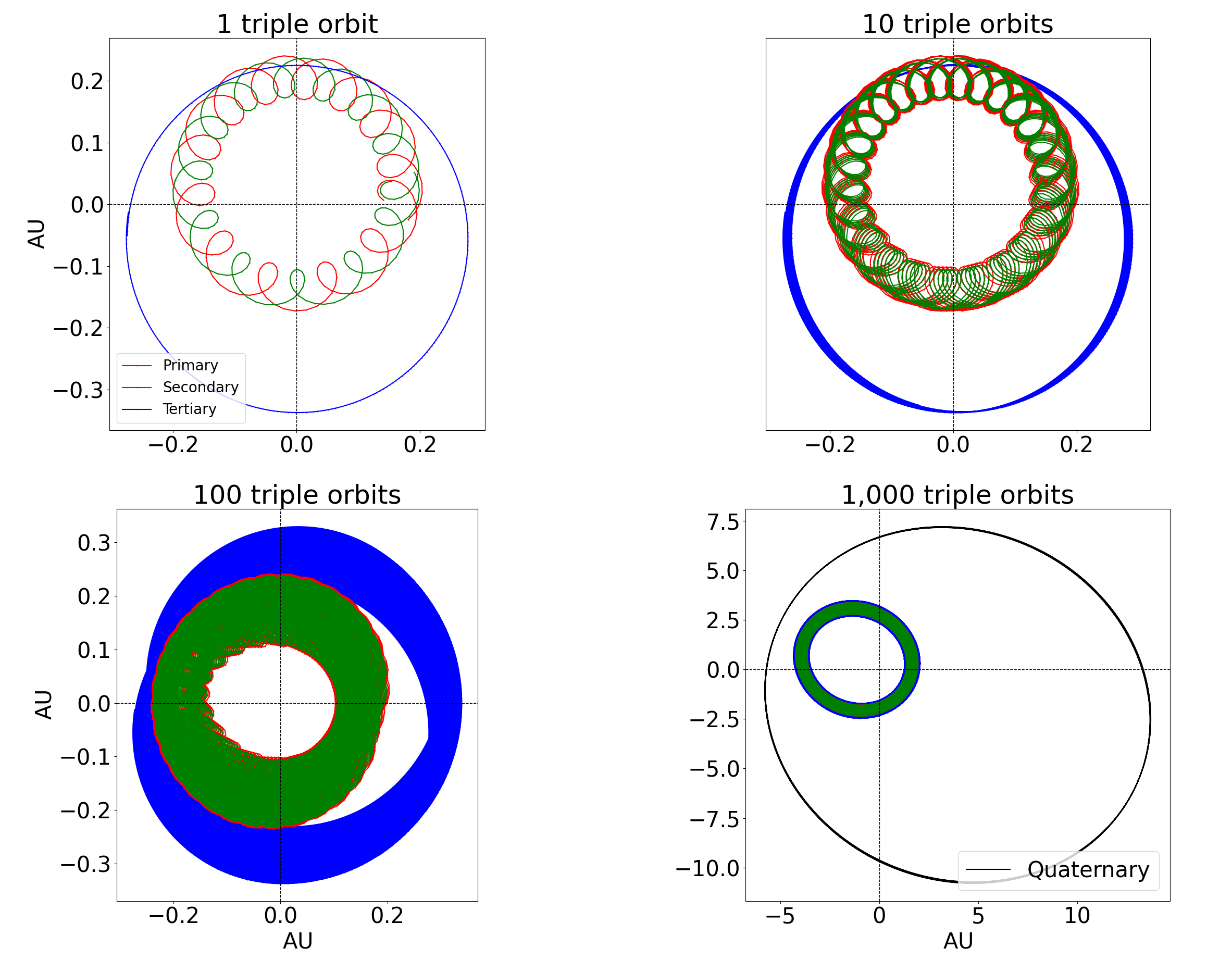}
    \includegraphics[width=0.95\textwidth]{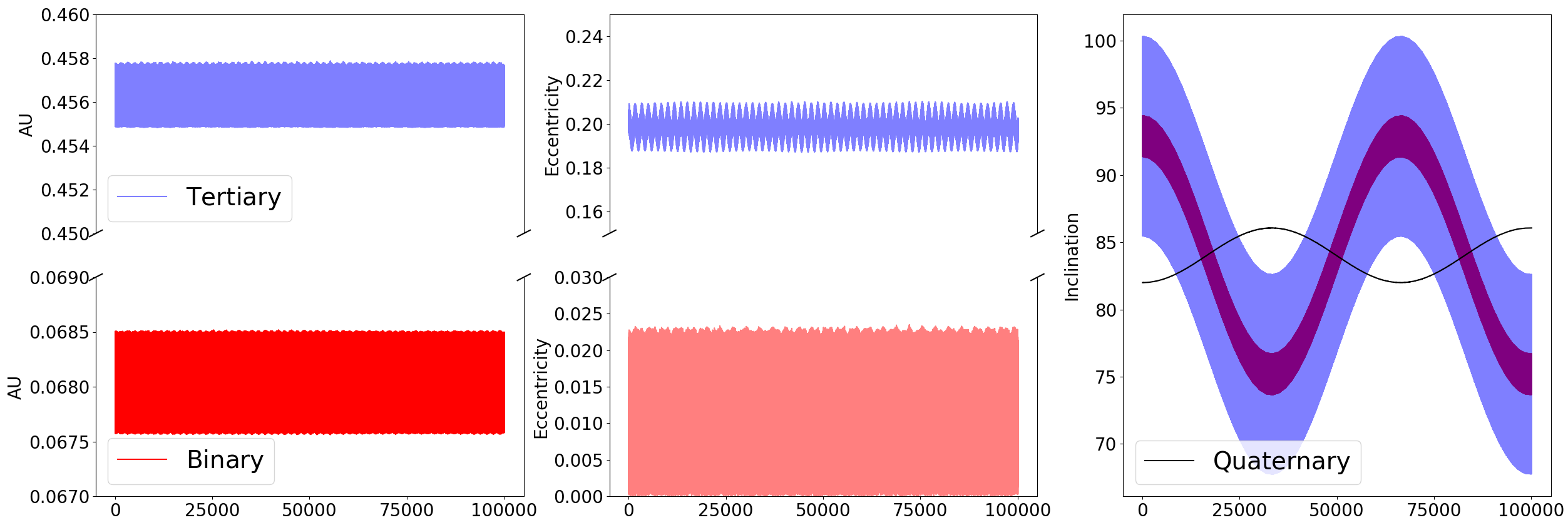}   
    \caption{{\it Upper and middle panels}: Orbital configuration of TIC 290061484 as seen from above the outer orbital plane over the course of 1, 10, 100, and 1000 triple orbits (upper left, upper right, middle left, middle right panels, respectively). The individual components are color-coded as indicated in the legend. The observer is in the x-y plane, looking along the $y$ direction. To highlight the rapid orbital precession of the triple subsystem, the upper left, upper right, and middle left panels are shown in the center of mass reference frame of that system. The middle right panel is shown in the center of mass reference frame of the entire quadruple system. {\it Lower panels}: Dynamical evolution of the system over the course of 100,000 triple orbits ($\sim 2.5$ million days) showing the semi-major axes (left), eccentricity (middle) and inclination (right) for the inner binary (red) and for the outer tertiary (blue). The parameters are plotted once a day for viewing purposes. There are no indications for chaotic motion and the system is dynamically-stable for the duration of the integrations.}
    \label{fig:top_view}
\end{figure*}

For completeness, we numerically integrated the orbits of TIC 290061484 over the next 1 million years ($\sim 15$ million outer orbits), utilizing the \textsc{REBOUND} N-body code \citep{Rein12}, \textsc{REBOUNDx} for the treatment of tidal effects \citep{Tamayo2020}, and using the best-fit parameters from the photodynamical solution as initial conditions at the reference time. The configuration of the system as seen from above is shown in the upper panels of Fig.~\ref{fig:top_view} for 1, 10, 100, and 1000 outer orbits ($\sim$25, $\sim$250, $\sim$2500, and $\sim$25,000 days, respectively), showcasing the relatively rapid precession of the inner and outer orbits. 

The dynamical evolution of the system's orbital parameters over the course of 100,000 outer orbits is shown in the lower panels of Fig.~\ref{fig:top_view}. The inner and outer semi-major axes oscillate by no more than $\sim1.5\%$ and $\sim0.7\%$, respectively. The eccentricity of the inner/outer orbits does not exceed $\sim0.03/0.21$, respectively. As seen from the figure, the most notable oscillations are in the inner and outer inclinations, where the former varies by up to $\sim32^\circ$ (i.e., from 68$^\circ$ to 100$^\circ$). Overall, the simulations demonstrate that, indeed, the orbital architecture of the system remains largely unchanged, without any indications for chaotic motion for the duration of the numerical integrations. 

\subsection{Future Evolution of the TIC 290061484 System}
\label{sec:evolution}

Currently, star B fills some 11\% of its Roche lobe in the outer orbit of the triple, while star Aa fills 54\% of its Roche lobe in the inner binary. Thus, it is something of a competition as to whether star B, which is 12\% more massive than Aa but fills far less of its Roche lobe, will start to transfer mass to the inner binary before Aa, which fills a larger fraction of its Roche lobe in the inner binary, starts mass transfer to Ab.  

\begin{figure}[h]
    \centering
    \includegraphics[width=1.01\columnwidth]{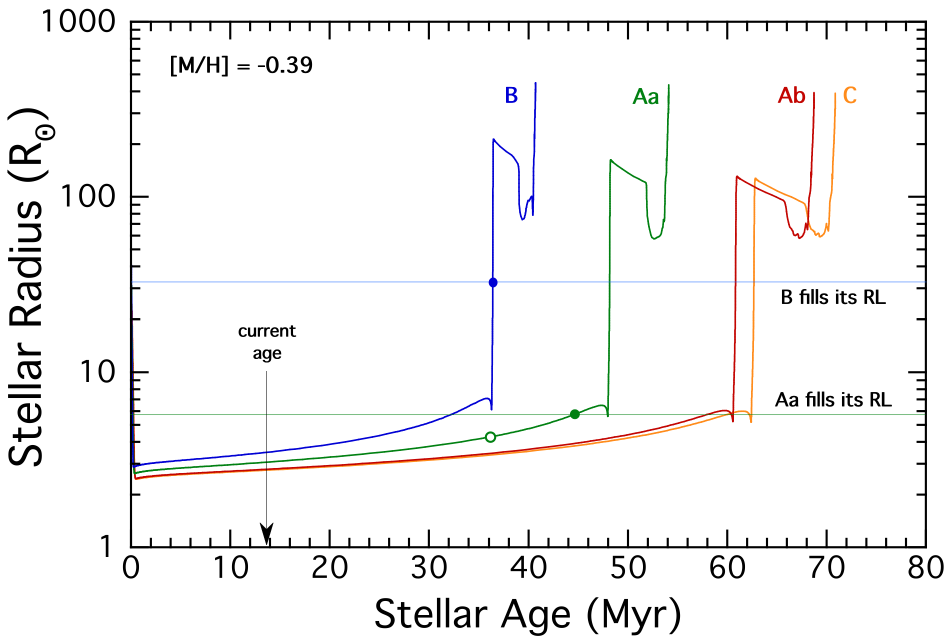}
    \caption{MIST radius evolution vs.~time for the four stars in the TIC 290061484 system. (The MIST tracks are from \citet{Dotter2016}, \citet{Choi2016}, \citet{Paxton2011}, \citet{Paxton2013}, \citet{Paxton2015}, and \citet{Paxton2018}).  The system is currently $\sim$ 14 Myr old (marked with an arrow). The most massive star, B, will ascend the giant branch at $\sim$36 Myr and begin to transfer mass to the inner binary (blue dot) shortly before star Aa fills its Roche lobe in the inner binary ($\sim$45 Myr; green dot). What may happen during both of the mass transfer events is described in Sect.~\ref{sec:evolution}.}
    \label{fig:evolution}  
\end{figure}

In Figure \ref{fig:evolution} we show the evolution of the stellar radii for all four stars as a function of time. Stars B and Aa ascend the giant branch at ages 36 Myr and 48 Myr, respectively. However, after 36 Myr, when star B will overflow its Roche lobe in the outer orbit, star B will have an evolved radius of close to 4.2 R$_\odot$ and will be filling 75\% of its Roche lobe while orbiting star Ab. It does appear that star B will be the first to overflow its Roche lobe and commence mass transfer onto the inner binary.   

For the case of the tertiary overflowing its Roche lobe first, we look to the work of \citet{deVries2014} for guidance, who modeled mass transfer in several much wider triples (including $\xi$ Tau with $P_{\rm out} = 145$ d).
The authors found that most of the mass transferred from the tertiary to the inner binary was ejected from the system without accreting onto the inner binary, and that the ejected matter left the system from the L2 and L3 points. If indeed the tertiary star in TIC 290061484 evolves to fill its Roche lobe first, and this is followed by the lose of its entire envelope, it will leave behind a $\sim$2 M$_\odot$ remnant core.   

The inner binaries in the \citet{deVries2014} calculations tended to shrink by some 5-10\% during the early part of the mass transfer process. In the context of TIC 290061484, this would drive the inner EB, with a current orbital period of 1.792 days, nearly into Roche-lobe contact. Either the inner binary would be driven into actual Roche lobe contact by the episode of the tertiary transferring mass or, within a few Myr thereafter, star Aa will naturally evolve to Roche lobe overflow via its nuclear evolution. Therefore, sooner or later, mass transfer within the inner binary will occur. 
Such mass transfer is likely to lead to the production of a merged star that is $\sim$12.9 M$_\odot$, and massive enough to undergo a type II supernova.


However, before doing that, the merged inner binary will have expanded more than sufficiently to overflow its Roche lobe in the outer binary and engulf the tertiary star.  The latter event would not unbind the envelope of the now merged 12.9 M$_\odot$ star whose envelope star B is entering, and the merged system would further evolve as a 15 M$_\odot$ star, leaving behind a neutron star. 

This is a long, complex, and interesting road to the formation of an isolated neutron star! 


\section{Acknowledgements}

This paper includes data collected by the {\em TESS} mission, which are publicly available from the Mikulski Archive for Space Telescopes (MAST). Funding for the {\em TESS} mission is provided by NASA's Science Mission directorate. 

This research has made use of the Exoplanet Follow-up Observation Program website, which is operated by the California Institute of Technology, under contract with the National Aeronautics and Space Administration under the Exoplanet Exploration Program. 

V.\,B.\,K., S.\,R., and B.\,P. acknowledge financial support of the NASA Citizen Science Seed Funding Program, grant number 22-CSSFP22-0004. V.\,B.\,K. is grateful for financial support from NASA grant 80NSSC21K0631 and from NSF grant AST-2206814.

Resources supporting this work were provided by the NASA High-End Computing (HEC) Program through the NASA Center for Climate Simulation (NCCS) at Goddard Space Flight Center. 

This work has made use of data from the European Space Agency (ESA) mission {\it Gaia} (\url{https://www.cosmos.esa.int/gaia}), processed by the {\it Gaia} Data Processing and Analysis Consortium (DPAC, \url{https://www.cosmos.esa.int/web/gaia/dpac/consortium}). Funding for the DPAC has been provided by national institutions, in particular the institutions participating in the {\it Gaia} Multilateral Agreement. 

This project has received funding from the HUN-REN Hungarian Research Network. T.\,B. acknowledges the financial support of the Hungarian National Research, Development and Innovation Office -- NKFIH Grant K-147131. A.\,P. acknowledges the financial support of the Hungarian National Research, Development and Innovation Office -- NKFIH Grant K-138962. 

Some of the observations in this paper made use of the High-Resolution Imaging instrument ‘Alopeke and were obtained under Gemini LLP Proposal Number: GN/S-2021A-LP-105. ‘Alopeke was funded by the NASA Exoplanet Exploration Program and built at the NASA Ames Research Center by Steve B. Howell, Nic Scott, Elliott P. Horch, and Emmett Quigley. Alopeke was mounted on the Gemini North telescope of the international Gemini Observatory, a program of NSF’s OIR Lab, which is managed by the Association of Universities for Research in Astronomy (AURA) under a cooperative agreement with the National Science Foundation. on behalf of the Gemini partnership: the National Science Foundation (United States), National Research Council (Canada), Agencia Nacional de Investigación y Desarrollo (Chile), Ministerio de Ciencia, Tecnología e Innovación (Argentina), Ministério da Ciência, Tecnologia, Inovações e Comunicações (Brazil), and Korea Astronomy and Space Science Institute (Republic of Korea).

A portion of the research in this work was carried out at the Jet Propulsion Laboratory, California Institute of Technology, under a contract with the National Aeronautics and Space Administration (80NM0018D0004).



\facilities{
\emph{Gaia},
MAST,
TESS,
ATLAS,
Gemini,
NCCS
}

\software{
{\tt Astropy} \citep{astropy2013,astropy2018}, 
{\tt Eleanor} \citep{eleanor},
{\tt IPython} \citep{ipython},
{\tt Keras} \citep{keras},
{\tt LcTools} \citep{Schmitt2019,2021arXiv210310285S},
{\tt lightcurvefactory} \citep{Borkovits2019,Borkovits2020},
{\tt Lightkurve} \citep{lightkurve},
{\tt Matplotlib} \citep{matplotlib},
{\tt Mpi4py} \citep{mpi4py2008},
{\tt NumPy} \citep{numpy}, 
{\tt Pandas} \citep{pandas},
{\tt Scikit-learn} \citep{scikit-learn},
{\tt SciPy} \citep{scipy},
{\tt Tensorflow} \citep{tensorflow},
{\tt Tess-point} \citep{tess-point}
}

\bibliography{refs}{}

\begin{thebibliography}{}
\expandafter\ifx\csname natexlab\endcsname\relax\def\natexlab#1{#1}\fi
\providecommand{\url}[1]{\href{#1}{#1}}
\providecommand{\dodoi}[1]{doi:~\href{http://doi.org/#1}{\nolinkurl{#1}}}
\providecommand{\doeprint}[1]{\href{http://ascl.net/#1}{\nolinkurl{http://ascl.net/#1}}}
\providecommand{\doarXiv}[1]{\href{https://arxiv.org/abs/#1}{\nolinkurl{https://arxiv.org/abs/#1}}}

\bibitem[{Abadi {et~al.}(2015)Abadi, Agarwal, Barham, Brevdo, Chen, Citro,
  Corrado, Davis, Dean, Devin, Ghemawat, Goodfellow, Harp, Irving, Isard, Jia,
  Jozefowicz, Kaiser, Kudlur, Levenberg, Man\'{e}, Monga, Moore, Murray, Olah,
  Schuster, Shlens, Steiner, Sutskever, Talwar, Tucker, Vanhoucke, Vasudevan,
  Vi\'{e}gas, Vinyals, Warden, Wattenberg, Wicke, Yu, \& Zheng}]{tensorflow}
Abadi, M., Agarwal, A., Barham, P., {et~al.} 2015, {TensorFlow}: Large-Scale
  Machine Learning on Heterogeneous Systems.
\newblock \url{https://www.tensorflow.org/}

\bibitem[{{Alonso} {et~al.}(2015){Alonso}, {Deeg}, {Hoyer}, {Lodieu}, {Palle},
  \& {Sanchis-Ojeda}}]{Alonso2015}
{Alonso}, R., {Deeg}, H.~J., {Hoyer}, S., {et~al.} 2015, \aap, 584, L8,
  \dodoi{10.1051/0004-6361/201527109}

\bibitem[{{Applegate}(1992)}]{Applegate1992}
{Applegate}, J.~H. 1992, \apj, 385, 621, \dodoi{10.1086/170967}

\bibitem[{{Astropy Collaboration} {et~al.}(2013){Astropy Collaboration},
  {Robitaille}, {Tollerud}, {Greenfield}, {Droettboom}, {Bray}, {Aldcroft},
  {Davis}, {Ginsburg}, {Price-Whelan}, {Kerzendorf}, {Conley}, {Crighton},
  {Barbary}, {Muna}, {Ferguson}, {Grollier}, {Parikh}, {Nair}, {Unther},
  {Deil}, {Woillez}, {Conseil}, {Kramer}, {Turner}, {Singer}, {Fox}, {Weaver},
  {Zabalza}, {Edwards}, {Azalee Bostroem}, {Burke}, {Casey}, {Crawford},
  {Dencheva}, {Ely}, {Jenness}, {Labrie}, {Lim}, {Pierfederici}, {Pontzen},
  {Ptak}, {Refsdal}, {Servillat}, \& {Streicher}}]{astropy2013}
{Astropy Collaboration}, {Robitaille}, T.~P., {Tollerud}, E.~J., {et~al.} 2013,
  \aap, 558, A33, \dodoi{10.1051/0004-6361/201322068}

\bibitem[{{Astropy Collaboration} {et~al.}(2018){Astropy Collaboration},
  {Price-Whelan}, {Sip{\H{o}}cz}, {G{\"u}nther}, {Lim}, {Crawford}, {Conseil},
  {Shupe}, {Craig}, {Dencheva}, {Ginsburg}, {Vand erPlas}, {Bradley},
  {P{\'e}rez-Su{\'a}rez}, {de Val-Borro}, {Aldcroft}, {Cruz}, {Robitaille},
  {Tollerud}, {Ardelean}, {Babej}, {Bach}, {Bachetti}, {Bakanov}, {Bamford},
  {Barentsen}, {Barmby}, {Baumbach}, {Berry}, {Biscani}, {Boquien}, {Bostroem},
  {Bouma}, {Brammer}, {Bray}, {Breytenbach}, {Buddelmeijer}, {Burke},
  {Calderone}, {Cano Rodr{\'\i}guez}, {Cara}, {Cardoso}, {Cheedella}, {Copin},
  {Corrales}, {Crichton}, {D'Avella}, {Deil}, {Depagne}, {Dietrich}, {Donath},
  {Droettboom}, {Earl}, {Erben}, {Fabbro}, {Ferreira}, {Finethy}, {Fox},
  {Garrison}, {Gibbons}, {Goldstein}, {Gommers}, {Greco}, {Greenfield},
  {Groener}, {Grollier}, {Hagen}, {Hirst}, {Homeier}, {Horton}, {Hosseinzadeh},
  {Hu}, {Hunkeler}, {Ivezi{\'c}}, {Jain}, {Jenness}, {Kanarek}, {Kendrew},
  {Kern}, {Kerzendorf}, {Khvalko}, {King}, {Kirkby}, {Kulkarni}, {Kumar},
  {Lee}, {Lenz}, {Littlefair}, {Ma}, {Macleod}, {Mastropietro}, {McCully},
  {Montagnac}, {Morris}, {Mueller}, {Mumford}, {Muna}, {Murphy}, {Nelson},
  {Nguyen}, {Ninan}, {N{\"o}the}, {Ogaz}, {Oh}, {Parejko}, {Parley}, {Pascual},
  {Patil}, {Patil}, {Plunkett}, {Prochaska}, {Rastogi}, {Reddy Janga},
  {Sabater}, {Sakurikar}, {Seifert}, {Sherbert}, {Sherwood-Taylor}, {Shih},
  {Sick}, {Silbiger}, {Singanamalla}, {Singer}, {Sladen}, {Sooley},
  {Sornarajah}, {Streicher}, {Teuben}, {Thomas}, {Tremblay}, {Turner},
  {Terr{\'o}n}, {van Kerkwijk}, {de la Vega}, {Watkins}, {Weaver}, {Whitmore},
  {Woillez}, {Zabalza}, \& {Astropy Contributors}}]{astropy2018}
{Astropy Collaboration}, {Price-Whelan}, A.~M., {Sip{\H{o}}cz}, B.~M., {et~al.}
  2018, \aj, 156, 123, \dodoi{10.3847/1538-3881/aabc4f}

\bibitem[{{Bailer-Jones} {et~al.}(2021){Bailer-Jones}, {Rybizki}, {Fouesneau},
  {Demleitner}, \& {Andrae}}]{bailer-jonesetal21}
{Bailer-Jones}, C.~A.~L., {Rybizki}, J., {Fouesneau}, M., {Demleitner}, M., \&
  {Andrae}, R. 2021, \aj, 161, 147, \dodoi{10.3847/1538-3881/abd806}

\bibitem[{{Bate}(1998)}]{Bate1998}
{Bate}, M.~R. 1998, \apjl, 508, L95, \dodoi{10.1086/311719}

\bibitem[{{Bate}(2012)}]{Bate2012}
---. 2012, \mnras, 419, 3115, \dodoi{10.1111/j.1365-2966.2011.19955.x}

\bibitem[{{Bate} {et~al.}(2002){Bate}, {Bonnell}, \& {Bromm}}]{Bate2002}
{Bate}, M.~R., {Bonnell}, I.~A., \& {Bromm}, V. 2002, \mnras, 336, 705,
  \dodoi{10.1046/j.1365-8711.2002.05775.x}

\bibitem[{{Bate} {et~al.}(2010){Bate}, {Lodato}, \& {Pringle}}]{Bate2010}
{Bate}, M.~R., {Lodato}, G., \& {Pringle}, J.~E. 2010, \mnras, 401, 1505,
  \dodoi{10.1111/j.1365-2966.2009.15773.x}

\bibitem[{{Borkovits}(2022)}]{Borkovits_2022a}
{Borkovits}, T. 2022, Galaxies, 10, 9, \dodoi{10.3390/galaxies10010009}

\bibitem[{{Borkovits} {et~al.}(2016){Borkovits}, {Hajdu}, {Sztakovics},
  {Rappaport}, {Levine}, {B{\'\i}r{\'o}}, \& {Klagyivik}}]{2016MNRAS.455.4136B}
{Borkovits}, T., {Hajdu}, T., {Sztakovics}, J., {et~al.} 2016, \mnras, 455,
  4136, \dodoi{10.1093/mnras/stv2530}

\bibitem[{{Borkovits} \& {Mitnyan}(2023)}]{BorkovitsMitnyan2023}
{Borkovits}, T., \& {Mitnyan}, T. 2023, Universe, 9, 485,
  \dodoi{10.3390/universe9110485}

\bibitem[{{Borkovits} {et~al.}(2015){Borkovits}, {Rappaport}, {Hajdu}, \&
  {Sztakovics}}]{borkovitsetal15}
{Borkovits}, T., {Rappaport}, S., {Hajdu}, T., \& {Sztakovics}, J. 2015,
  \mnras, 448, 946, \dodoi{10.1093/mnras/stv015}

\bibitem[{{Borkovits} {et~al.}(2020{\natexlab{a}}){Borkovits}, {Rappaport},
  {Hajdu}, {Maxted}, {P{\'a}l}, {Forg{\'a}cs-Dajka}, {Klagyivik}, \&
  {Mitnyan}}]{Borkovits2020}
{Borkovits}, T., {Rappaport}, S.~A., {Hajdu}, T., {et~al.} 2020{\natexlab{a}},
  \mnras, 493, 5005, \dodoi{10.1093/mnras/staa495}

\bibitem[{{Borkovits} {et~al.}(2020{\natexlab{b}}){Borkovits}, {Rappaport},
  {Hajdu}, {Maxted}, {P{\'a}l}, {Forg{\'a}cs-Dajka}, {Klagyivik}, \&
  {Mitnyan}}]{Borkovits2020b}
---. 2020{\natexlab{b}}, \mnras, 493, 5005, \dodoi{10.1093/mnras/staa495}

\bibitem[{{Borkovits} {et~al.}(2022{\natexlab{a}}){Borkovits}, {Rappaport},
  {Toonen}, {Moe}, {Mitnyan}, \& {Cs{\'a}nyi}}]{Borkovits2022}
{Borkovits}, T., {Rappaport}, S.~A., {Toonen}, S., {et~al.} 2022{\natexlab{a}},
  \mnras, 515, 3773, \dodoi{10.1093/mnras/stac1983}

\bibitem[{{Borkovits} {et~al.}(2019{\natexlab{a}}){Borkovits}, {Sperauskas},
  {Tokovinin}, {Latham}, {Cs{\'a}nyi}, {Hajdu}, \&
  {Moln{\'a}r}}]{2019MNRAS.487.4631B}
{Borkovits}, T., {Sperauskas}, J., {Tokovinin}, A., {et~al.}
  2019{\natexlab{a}}, \mnras, 487, 4631, \dodoi{10.1093/mnras/stz1510}

\bibitem[{{Borkovits} {et~al.}(2013){Borkovits}, {Derekas}, {Kiss},
  {Kir{\'a}ly}, {Forg{\'a}cs-Dajka}, {B{\'\i}r{\'o}}, {Bedding}, {Bryson},
  {Huber}, \& {Szab{\'o}}}]{borkovitsetal13}
{Borkovits}, T., {Derekas}, A., {Kiss}, L.~L., {et~al.} 2013, \mnras, 428,
  1656, \dodoi{10.1093/mnras/sts146}

\bibitem[{{Borkovits} {et~al.}(2018){Borkovits}, {Albrecht}, {Rappaport},
  {Nelson}, {Vanderburg}, {Gary}, {Tan}, {Justesen}, {Kristiansen}, {Jacobs},
  {LaCourse}, {Ngo}, {Wallack}, {Ruane}, {Mawet}, {Howell}, \&
  {Tronsgaard}}]{Borkovits2018a}
{Borkovits}, T., {Albrecht}, S., {Rappaport}, S., {et~al.} 2018, \mnras, 478,
  5135, \dodoi{10.1093/mnras/sty1386}

\bibitem[{{Borkovits} {et~al.}(2019{\natexlab{b}}){Borkovits}, {Rappaport},
  {Kaye}, {Isaacson}, {Vanderburg}, {Howard}, {Kristiansen}, {Omohundro},
  {Schwengeler}, {Terentev}, {Shporer}, {Relles}, {Villanueva}, {Tan},
  {Col{\'o}n}, {Blex}, {Haas}, {Cochran}, \& {Endl}}]{Borkovits2019}
{Borkovits}, T., {Rappaport}, S., {Kaye}, T., {et~al.} 2019{\natexlab{b}},
  \mnras, 483, 1934, \dodoi{10.1093/mnras/sty3157}

\bibitem[{{Borkovits} {et~al.}(2022{\natexlab{b}}){Borkovits}, {Mitnyan},
  {Rappaport}, {Pribulla}, {Powell}, {Kostov}, {B{\'\i}r{\'o}}, {Cs{\'a}nyi},
  {Garai}, {Gary}, {Kaye}, {Kom{\v{z}}{\'\i}k}, {Terentev}, {Omohundro},
  {Gagliano}, {Jacobs}, {Kristiansen}, {LaCourse}, {Schwengeler}, {Czavalinga},
  {Seli}, {Huang}, {P{\'a}l}, {Vanderburg}, {Rodriguez}, \&
  {Stevens}}]{Borkovits2022b}
{Borkovits}, T., {Mitnyan}, T., {Rappaport}, S.~A., {et~al.}
  2022{\natexlab{b}}, \mnras, 510, 1352, \dodoi{10.1093/mnras/stab3397}

\bibitem[{{Boss}(1986)}]{Boss1986}
{Boss}, A.~P. 1986, \apjs, 62, 519, \dodoi{10.1086/191150}

\bibitem[{{Burke} {et~al.}(2020){Burke}, {Levine}, {Fausnaugh}, {Vanderspek},
  {Barclay}, {Libby-Roberts}, {Morris}, {Sipocz}, {Owens}, {Feinstein}, \&
  {Camacho}}]{tess-point}
{Burke}, C.~J., {Levine}, A., {Fausnaugh}, M., {et~al.} 2020, {TESS-Point: High
  precision TESS pointing tool}, Astrophysics Source Code Library.
\newblock \doeprint{2003.001}

\bibitem[{{Carter} {et~al.}(2011){Carter}, {Fabrycky}, {Ragozzine}, {Holman},
  {Quinn}, {Latham}, {Buchhave}, {Van Cleve}, {Cochran}, {Cote}, {Endl},
  {Ford}, {Haas}, {Jenkins}, {Koch}, {Li}, {Lissauer}, {MacQueen}, {Middour},
  {Orosz}, {Rowe}, {Steffen}, \& {Welsh}}]{2011Sci...331..562C}
{Carter}, J.~A., {Fabrycky}, D.~C., {Ragozzine}, D., {et~al.} 2011, Science,
  331, 562, \dodoi{10.1126/science.1201274}

\bibitem[{{Castelli} \& {Kurucz}(2003)}]{Castelli2003}
{Castelli}, F., \& {Kurucz}, R.~L. 2003, in Modelling of Stellar Atmospheres,
  ed. N.~{Piskunov}, W.~W. {Weiss}, \& D.~F. {Gray}, Vol. 210, A20,
  \dodoi{10.48550/arXiv.astro-ph/0405087}

\bibitem[{{Chambers} {et~al.}(2016){Chambers}, {Magnier}, {Metcalfe},
  {Flewelling}, {Huber}, {Waters}, {Denneau}, {Draper}, {Farrow}, {Finkbeiner},
  {Holmberg}, {Koppenhoefer}, {Price}, {Rest}, {Saglia}, {Schlafly}, {Smartt},
  {Sweeney}, {Wainscoat}, {Burgett}, {Chastel}, {Grav}, {Heasley}, {Hodapp},
  {Jedicke}, {Kaiser}, {Kudritzki}, {Luppino}, {Lupton}, {Monet}, {Morgan},
  {Onaka}, {Shiao}, {Stubbs}, {Tonry}, {White}, {Ba{\~n}ados}, {Bell},
  {Bender}, {Bernard}, {Boegner}, {Boffi}, {Botticella}, {Calamida},
  {Casertano}, {Chen}, {Chen}, {Cole}, {Deacon}, {Frenk}, {Fitzsimmons},
  {Gezari}, {Gibbs}, {Goessl}, {Goggia}, {Gourgue}, {Goldman}, {Grant},
  {Grebel}, {Hambly}, {Hasinger}, {Heavens}, {Heckman}, {Henderson}, {Henning},
  {Holman}, {Hopp}, {Ip}, {Isani}, {Jackson}, {Keyes}, {Koekemoer}, {Kotak},
  {Le}, {Liska}, {Long}, {Lucey}, {Liu}, {Martin}, {Masci}, {McLean}, {Mindel},
  {Misra}, {Morganson}, {Murphy}, {Obaika}, {Narayan}, {Nieto-Santisteban},
  {Norberg}, {Peacock}, {Pier}, {Postman}, {Primak}, {Rae}, {Rai}, {Riess},
  {Riffeser}, {Rix}, {R{\"o}ser}, {Russel}, {Rutz}, {Schilbach}, {Schultz},
  {Scolnic}, {Strolger}, {Szalay}, {Seitz}, {Small}, {Smith}, {Soderblom},
  {Taylor}, {Thomson}, {Taylor}, {Thakar}, {Thiel}, {Thilker}, {Unger},
  {Urata}, {Valenti}, {Wagner}, {Walder}, {Walter}, {Watters}, {Werner},
  {Wood-Vasey}, \& {Wyse}}]{panstarrs2016}
{Chambers}, K.~C., {Magnier}, E.~A., {Metcalfe}, N., {et~al.} 2016, arXiv
  e-prints, arXiv:1612.05560, \dodoi{10.48550/arXiv.1612.05560}

\bibitem[{{Choi} {et~al.}(2016){Choi}, {Dotter}, {Conroy}, {Cantiello},
  {Paxton}, \& {Johnson}}]{Choi2016}
{Choi}, J., {Dotter}, A., {Conroy}, C., {et~al.} 2016, \apj, 823, 102,
  \dodoi{10.3847/0004-637X/823/2/102}

\bibitem[{Chollet {et~al.}(2015)}]{keras}
Chollet, F., {et~al.} 2015, Keras, \url{https://keras.io}

\bibitem[{{Claret}(2023)}]{2023A&A...674A..67C}
{Claret}, A. 2023, \aap, 674, A67, \dodoi{10.1051/0004-6361/202346250}

\bibitem[{{Cutri} {et~al.}(2012){Cutri}, {Wright}, {Conrow}, {Bauer},
  {Benford}, {Brandenburg}, {Dailey}, {Eisenhardt}, {Evans}, {Fajardo-Acosta},
  {Fowler}, {Gelino}, {Grillmair}, {Harbut}, {Hoffman}, {Jarrett},
  {Kirkpatrick}, {Leisawitz}, {Liu}, {Mainzer}, {Marsh}, {Masci}, {McCallon},
  {Padgett}, {Ressler}, {Royer}, {Skrutskie}, {Stanford}, {Wyatt}, {Tholen},
  {Tsai}, {Wachter}, {Wheelock}, {Yan}, {Alles}, {Beck}, {Grav}, {Masiero},
  {McCollum}, {McGehee}, {Papin}, \& {Wittman}}]{WISE}
{Cutri}, R.~M., {Wright}, E.~L., {Conrow}, T., {et~al.} 2012, {Explanatory
  Supplement to the WISE All-Sky Data Release Products}, Explanatory Supplement
  to the WISE All-Sky Data Release Products

\bibitem[{Dalcin {et~al.}(2008)Dalcin, Paz, Storti, \& D'Elia}]{mpi4py2008}
Dalcin, L., Paz, R., Storti, M., \& D'Elia, J. 2008, Journal of Parallel and
  Distributed Computing, 68, 655,
  \dodoi{http://dx.doi.org/10.1016/j.jpdc.2007.09.005}

\bibitem[{{de Vries} {et~al.}(2014){de Vries}, {Portegies Zwart}, \&
  {Figueira}}]{deVries2014}
{de Vries}, N., {Portegies Zwart}, S., \& {Figueira}, J. 2014, \mnras, 438,
  1909, \dodoi{10.1093/mnras/stt1688}

\bibitem[{{Docobo} {et~al.}(2021){Docobo}, {Piccotti}, {Abad}, \&
  {Campo}}]{Docobo2021}
{Docobo}, J.~A., {Piccotti}, L., {Abad}, A., \& {Campo}, P.~P. 2021, \aj, 161,
  43, \dodoi{10.3847/1538-3881/abc94e}

\bibitem[{{Dotter}(2016)}]{Dotter2016}
{Dotter}, A. 2016, \apjs, 222, 8, \dodoi{10.3847/0067-0049/222/1/8}

\bibitem[{{Ebbighausen} \& {Struve}(1956)}]{1956ApJ...124..507E}
{Ebbighausen}, E.~G., \& {Struve}, O. 1956, \apj, 124, 507,
  \dodoi{10.1086/146254}

\bibitem[{{Eisner} {et~al.}(2022){Eisner}, {Johnston}, {Toonen}, {Frost},
  {Janssens}, {Lintott}, {Aigrain}, {Sana}, {Abdul-Masih},
  {Arellano-C{\'o}rdova}, {Beck}, {Bordier}, {Cannon}, {Escorza}, {Fabry},
  {Hermansson}, {Howell}, {Miller}, {Sheyte}, {Alhassan}, {Baeten}, {Barnet},
  {Bean}, {Bernau}, {Bundy}, {Di Fraia}, {Emralino}, {Goodwin}, {Hermes},
  {Hoffman}, {Huten}, {Jan{\'\i}{\v{c}}ek}, {Lee}, {Mazzucato}, {Rogers},
  {Rout}, {Sejpka}, {Tanner}, {Terentev}, \& {Urvoy}}]{Eisner2022}
{Eisner}, N.~L., {Johnston}, C., {Toonen}, S., {et~al.} 2022, \mnras, 511,
  4710, \dodoi{10.1093/mnras/stab3619}

\bibitem[{{El-Badry} {et~al.}(2019){El-Badry}, {Rix}, {Tian}, {Duch{\^e}ne}, \&
  {Moe}}]{ElBadry2019}
{El-Badry}, K., {Rix}, H.-W., {Tian}, H., {Duch{\^e}ne}, G., \& {Moe}, M. 2019,
  \mnras, 489, 5822, \dodoi{10.1093/mnras/stz2480}

\bibitem[{{Fabrycky} \& {Tremaine}(2007)}]{Fabrycky2007}
{Fabrycky}, D., \& {Tremaine}, S. 2007, \apj, 669, 1298, \dodoi{10.1086/521702}

\bibitem[{{Farris} {et~al.}(2014){Farris}, {Duffell}, {MacFadyen}, \&
  {Haiman}}]{Farris2014}
{Farris}, B.~D., {Duffell}, P., {MacFadyen}, A.~I., \& {Haiman}, Z. 2014, \apj,
  783, 134, \dodoi{10.1088/0004-637X/783/2/134}

\bibitem[{{Feinstein} {et~al.}(2019){Feinstein}, {Montet}, {Foreman-Mackey},
  {Bedell}, {Saunders}, {Bean}, {Christiansen}, {Hedges}, {Luger}, {Scolnic},
  \& {Cardoso}}]{eleanor}
{Feinstein}, A.~D., {Montet}, B.~T., {Foreman-Mackey}, D., {et~al.} 2019,
  \pasp, 131, 094502, \dodoi{10.1088/1538-3873/ab291c}

\bibitem[{{Ford}(2005)}]{ford05}
{Ford}, E.~B. 2005, \aj, 129, 1706, \dodoi{10.1086/427962}

\bibitem[{{Gaia Collaboration} {et~al.}(2021){Gaia Collaboration}, {Brown},
  {Vallenari}, {Prusti}, {de Bruijne}, {Babusiaux}, {Biermann}, {Creevey},
  {Evans}, {Eyer}, {Hutton}, {Jansen}, {Jordi}, {Klioner}, {Lammers},
  {Lindegren}, {Luri}, {Mignard}, {Panem}, {Pourbaix}, {Randich}, {Sartoretti},
  {Soubiran}, {Walton}, {Arenou}, {Bailer-Jones}, {Bastian}, {Cropper},
  {Drimmel}, {Katz}, {Lattanzi}, {van Leeuwen}, {Bakker}, {Cacciari},
  {Casta{\~n}eda}, {De Angeli}, {Ducourant}, {Fabricius}, {Fouesneau},
  {Fr{\'e}mat}, {Guerra}, {Guerrier}, {Guiraud}, {Jean-Antoine Piccolo},
  {Masana}, {Messineo}, {Mowlavi}, {Nicolas}, {Nienartowicz}, {Pailler},
  {Panuzzo}, {Riclet}, {Roux}, {Seabroke}, {Sordo}, {Tanga}, {Th{\'e}venin},
  {Gracia-Abril}, {Portell}, {Teyssier}, {Altmann}, {Andrae}, {Bellas-Velidis},
  {Benson}, {Berthier}, {Blomme}, {Brugaletta}, {Burgess}, {Busso}, {Carry},
  {Cellino}, {Cheek}, {Clementini}, {Damerdji}, {Davidson}, {Delchambre},
  {Dell'Oro}, {Fern{\'a}ndez-Hern{\'a}ndez}, {Galluccio}, {Garc{\'\i}a-Lario},
  {Garcia-Reinaldos}, {Gonz{\'a}lez-N{\'u}{\~n}ez}, {Gosset}, {Haigron},
  {Halbwachs}, {Hambly}, {Harrison}, {Hatzidimitriou}, {Heiter},
  {Hern{\'a}ndez}, {Hestroffer}, {Hodgkin}, {Holl}, {Jan{\ss}en}, {Jevardat de
  Fombelle}, {Jordan}, {Krone-Martins}, {Lanzafame}, {L{\"o}ffler}, {Lorca},
  {Manteiga}, {Marchal}, {Marrese}, {Moitinho}, {Mora}, {Muinonen}, {Osborne},
  {Pancino}, {Pauwels}, {Petit}, {Recio-Blanco}, {Richards}, {Riello},
  {Rimoldini}, {Robin}, {Roegiers}, {Rybizki}, {Sarro}, {Siopis}, {Smith},
  {Sozzetti}, {Ulla}, {Utrilla}, {van Leeuwen}, {van Reeven}, {Abbas}, {Abreu
  Aramburu}, {Accart}, {Aerts}, {Aguado}, {Ajaj}, {Altavilla}, {{\'A}lvarez},
  {{\'A}lvarez Cid-Fuentes}, {Alves}, {Anderson}, {Anglada Varela}, {Antoja},
  {Audard}, {Baines}, {Baker}, {Balaguer-N{\'u}{\~n}ez}, {Balbinot}, {Balog},
  {Barache}, {Barbato}, {Barros}, {Barstow}, {Bartolom{\'e}}, {Bassilana},
  {Bauchet}, {Baudesson-Stella}, {Becciani}, {Bellazzini}, {Bernet}, {Bertone},
  {Bianchi}, {Blanco-Cuaresma}, {Boch}, {Bombrun}, {Bossini}, {Bouquillon},
  {Bragaglia}, {Bramante}, {Breedt}, {Bressan}, {Brouillet}, {Bucciarelli},
  {Burlacu}, {Busonero}, {Butkevich}, {Buzzi}, {Caffau}, {Cancelliere},
  {C{\'a}novas}, {Cantat-Gaudin}, {Carballo}, {Carlucci}, {Carnerero},
  {Carrasco}, {Casamiquela}, {Castellani}, {Castro-Ginard}, {Castro Sampol},
  {Chaoul}, {Charlot}, {Chemin}, {Chiavassa}, {Cioni}, {Comoretto}, {Cooper},
  {Cornez}, {Cowell}, {Crifo}, {Crosta}, {Crowley}, {Dafonte}, {Dapergolas},
  {David}, {David}, {de Laverny}, {De Luise}, {De March}, {De Ridder}, {de
  Souza}, {de Teodoro}, {de Torres}, {del Peloso}, {del Pozo}, {Delbo},
  {Delgado}, {Delgado}, {Delisle}, {Di Matteo}, {Diakite}, {Diener},
  {Distefano}, {Dolding}, {Eappachen}, {Edvardsson}, {Enke}, {Esquej}, {Fabre},
  {Fabrizio}, {Faigler}, {Fedorets}, {Fernique}, {Fienga}, {Figueras},
  {Fouron}, {Fragkoudi}, {Fraile}, {Franke}, {Gai}, {Garabato},
  {Garcia-Gutierrez}, {Garc{\'\i}a-Torres}, {Garofalo}, {Gavras}, {Gerlach},
  {Geyer}, {Giacobbe}, {Gilmore}, {Girona}, {Giuffrida}, {Gomel}, {Gomez},
  {Gonzalez-Santamaria}, {Gonz{\'a}lez-Vidal}, {Granvik},
  {Guti{\'e}rrez-S{\'a}nchez}, {Guy}, {Hauser}, {Haywood}, {Helmi}, {Hidalgo},
  {Hilger}, {H{\l}adczuk}, {Hobbs}, {Holland}, {Huckle}, {Jasniewicz},
  {Jonker}, {Juaristi Campillo}, {Julbe}, {Karbevska}, {Kervella}, {Khanna},
  {Kochoska}, {Kontizas}, {Kordopatis}, {Korn}, {Kostrzewa-Rutkowska},
  {Kruszy{\'n}ska}, {Lambert}, {Lanza}, {Lasne}, {Le Campion}, {Le Fustec},
  {Lebreton}, {Lebzelter}, {Leccia}, {Leclerc}, {Lecoeur-Taibi}, {Liao},
  {Licata}, {Lindstr{\o}m}, {Lister}, {Livanou}, {Lobel}, {Madrero Pardo},
  {Managau}, {Mann}, {Marchant}, {Marconi}, {Marcos Santos}, {Marinoni},
  {Marocco}, {Marshall}, {Martin Polo}, {Mart{\'\i}n-Fleitas}, {Masip},
  {Massari}, {Mastrobuono-Battisti}, {Mazeh}, {McMillan}, {Messina},
  {Michalik}, {Millar}, {Mints}, {Molina}, {Molinaro}, {Moln{\'a}r},
  {Montegriffo}, {Mor}, {Morbidelli}, {Morel}, {Morris}, {Mulone}, {Munoz},
  {Muraveva}, {Murphy}, {Musella}, {Noval}, {Ord{\'e}novic}, {Orr{\`u}},
  {Osinde}, {Pagani}, {Pagano}, {Palaversa}, {Palicio}, {Panahi}, {Pawlak},
  {Pe{\~n}alosa Esteller}, {Penttil{\"a}}, {Piersimoni}, {Pineau}, {Plachy},
  {Plum}, {Poggio}, {Poretti}, {Poujoulet}, {Pr{\v{s}}a}, {Pulone}, {Racero},
  {Ragaini}, {Rainer}, {Raiteri}, {Rambaux}, {Ramos}, {Ramos-Lerate}, {Re
  Fiorentin}, {Regibo}, {Reyl{\'e}}, {Ripepi}, {Riva}, {Rixon}, {Robichon},
  {Robin}, {Roelens}, {Rohrbasser}, {Romero-G{\'o}mez}, {Rowell}, {Royer},
  {Rybicki}, {Sadowski}, {Sagrist{\`a} Sell{\'e}s}, {Sahlmann}, {Salgado},
  {Salguero}, {Samaras}, {Sanchez Gimenez}, {Sanna}, {Santove{\~n}a},
  {Sarasso}, {Schultheis}, {Sciacca}, {Segol}, {Segovia}, {S{\'e}gransan},
  {Semeux}, {Shahaf}, {Siddiqui}, {Siebert}, {Siltala}, {Slezak}, {Smart},
  {Solano}, {Solitro}, {Souami}, {Souchay}, {Spagna}, {Spoto}, {Steele},
  {Steidelm{\"u}ller}, {Stephenson}, {S{\"u}veges}, {Szabados}, {Szegedi-Elek},
  {Taris}, {Tauran}, {Taylor}, {Teixeira}, {Thuillot}, {Tonello}, {Torra},
  {Torra}, {Turon}, {Unger}, {Vaillant}, {van Dillen}, {Vanel}, {Vecchiato},
  {Viala}, {Vicente}, {Voutsinas}, {Weiler}, {Wevers}, {Wyrzykowski}, {Yoldas},
  {Yvard}, {Zhao}, {Zorec}, {Zucker}, {Zurbach}, \& {Zwitter}}]{Gaia2021}
{Gaia Collaboration}, {Brown}, A.~G.~A., {Vallenari}, A., {et~al.} 2021, \aap,
  649, A1, \dodoi{10.1051/0004-6361/202039657}

\bibitem[{{Gaulme} {et~al.}(2022){Gaulme}, {Borkovits}, {Appourchaux},
  {Pavlovski}, {Spada}, {Gehan}, {Ong}, {Miglio}, {Tkachenko}, {Mosser},
  {Vrard}, {Benbakoura}, {Drew Chojnowski}, {Perkins}, {Hedlund}, \&
  {Jackiewicz}}]{Gaulme2022}
{Gaulme}, P., {Borkovits}, T., {Appourchaux}, T., {et~al.} 2022, \aap, 668,
  A173, \dodoi{10.1051/0004-6361/202244373}

\bibitem[{{Graham} {et~al.}(2019){Graham}, {Kulkarni}, {Bellm}, {Adams},
  {Barbarino}, {Blagorodnova}, {Bodewits}, {Bolin}, {Brady}, {Cenko}, {Chang},
  {Coughlin}, {De}, {Eadie}, {Farnham}, {Feindt}, {Franckowiak}, {Fremling},
  {Gezari}, {Ghosh}, {Goldstein}, {Golkhou}, {Goobar}, {Ho}, {Huppenkothen},
  {Ivezi{\'c}}, {Jones}, {Juric}, {Kaplan}, {Kasliwal}, {Kelley}, {Kupfer},
  {Lee}, {Lin}, {Lunnan}, {Mahabal}, {Miller}, {Ngeow}, {Nugent}, {Ofek},
  {Prince}, {Rauch}, {van Roestel}, {Schulze}, {Singer}, {Sollerman}, {Taddia},
  {Yan}, {Ye}, {Yu}, {Barlow}, {Bauer}, {Beck}, {Belicki}, {Biswas}, {Brinnel},
  {Brooke}, {Bue}, {Bulla}, {Burruss}, {Connolly}, {Cromer}, {Cunningham},
  {Dekany}, {Delacroix}, {Desai}, {Duev}, {Feeney}, {Flynn}, {Frederick},
  {Gal-Yam}, {Giomi}, {Groom}, {Hacopians}, {Hale}, {Helou}, {Henning},
  {Hover}, {Hillenbrand}, {Howell}, {Hung}, {Imel}, {Ip}, {Jackson}, {Kaspi},
  {Kaye}, {Kowalski}, {Kramer}, {Kuhn}, {Landry}, {Laher}, {Mao}, {Masci},
  {Monkewitz}, {Murphy}, {Nordin}, {Patterson}, {Penprase}, {Porter},
  {Rebbapragada}, {Reiley}, {Riddle}, {Rigault}, {Rodriguez}, {Rusholme}, {van
  Santen}, {Shupe}, {Smith}, {Soumagnac}, {Stein}, {Surace}, {Szkody}, {Terek},
  {Van Sistine}, {van Velzen}, {Vestrand}, {Walters}, {Ward}, {Zhang}, \&
  {Zolkower}}]{Graham2019}
{Graham}, M.~J., {Kulkarni}, S.~R., {Bellm}, E.~C., {et~al.} 2019, \pasp, 131,
  078001, \dodoi{10.1088/1538-3873/ab006c}

\bibitem[{{Grether} \& {Lineweaver}(2006)}]{Grether2006}
{Grether}, D., \& {Lineweaver}, C.~H. 2006, \apj, 640, 1051,
  \dodoi{10.1086/500161}

\bibitem[{{G{\"u}ver} \& {{\"O}zel}(2009)}]{Guver2009}
{G{\"u}ver}, T., \& {{\"O}zel}, F. 2009, \mnras, 400, 2050,
  \dodoi{10.1111/j.1365-2966.2009.15598.x}

\bibitem[{{Handler} {et~al.}(2020){Handler}, {Kurtz}, {Rappaport}, {Saio},
  {Fuller}, {Jones}, {Guo}, {Chowdhury}, {Sowicka}, {Kahraman
  Ali{\c{c}}avu{\c{s}}}, {Streamer}, {Murphy}, {Gagliano}, {Jacobs}, \&
  {Vanderburg}}]{Handler2020}
{Handler}, G., {Kurtz}, D.~W., {Rappaport}, S.~A., {et~al.} 2020, Nature
  Astronomy, 4, 684, \dodoi{10.1038/s41550-020-1035-1}

\bibitem[{Harris {et~al.}(2020)Harris, Millman, van~der Walt, Gommers,
  Virtanen, Cournapeau, Wieser, Taylor, Berg, Smith, Kern, Picus, Hoyer, van
  Kerkwijk, Brett, Haldane, del R{\'{i}}o, Wiebe, Peterson,
  G{\'{e}}rard-Marchant, Sheppard, Reddy, Weckesser, Abbasi, Gohlke, \&
  Oliphant}]{numpy}
Harris, C.~R., Millman, K.~J., van~der Walt, S.~J., {et~al.} 2020, Nature, 585,
  357, \dodoi{10.1038/s41586-020-2649-2}

\bibitem[{{Heinze} {et~al.}(2018){Heinze}, {Tonry}, {Denneau}, {Flewelling},
  {Stalder}, {Rest}, {Smith}, {Smartt}, \& {Weiland}}]{2018AJ....156..241H}
{Heinze}, A.~N., {Tonry}, J.~L., {Denneau}, L., {et~al.} 2018, \aj, 156, 241,
  \dodoi{10.3847/1538-3881/aae47f}

\bibitem[{{Howell} {et~al.}(2011){Howell}, {Everett}, {Sherry}, {Horch}, \&
  {Ciardi}}]{howell2011}
{Howell}, S.~B., {Everett}, M.~E., {Sherry}, W., {Horch}, E., \& {Ciardi},
  D.~R. 2011, \aj, 142, 19, \dodoi{10.1088/0004-6256/142/1/19}

\bibitem[{Hunter(2007)}]{matplotlib}
Hunter, J.~D. 2007, Computing in science \& engineering, 9, 90

\bibitem[{{Innes}(1917)}]{Innes_1917}
{Innes}, R.~T.~A. 1917, Circular of the Union Observatory Johannesburg, 40, 331

\bibitem[{{Kervella} {et~al.}(2017){Kervella}, {Th{\'e}venin}, \&
  {Lovis}}]{Kervella_2017}
{Kervella}, P., {Th{\'e}venin}, F., \& {Lovis}, C. 2017, \aap, 598, L7,
  \dodoi{10.1051/0004-6361/201629930}

\bibitem[{{Kiseleva} {et~al.}(1998){Kiseleva}, {Eggleton}, \&
  {Mikkola}}]{Kiseleva1998}
{Kiseleva}, L.~G., {Eggleton}, P.~P., \& {Mikkola}, S. 1998, \mnras, 300, 292,
  \dodoi{10.1046/j.1365-8711.1998.01903.x}

\bibitem[{{Kochanek} {et~al.}(2017){Kochanek}, {Shappee}, {Stanek}, {Holoien},
  {Thompson}, {Prieto}, {Dong}, {Shields}, {Will}, {Britt}, {Perzanowski}, \&
  {Pojma{\'n}ski}}]{2017PASP..129j4502K}
{Kochanek}, C.~S., {Shappee}, B.~J., {Stanek}, K.~Z., {et~al.} 2017, \pasp,
  129, 104502, \dodoi{10.1088/1538-3873/aa80d9}

\bibitem[{{Kostov} {et~al.}(2022){Kostov}, {Powell}, {Rappaport}, {Borkovits},
  {Gagliano}, {Jacobs}, {Kristiansen}, {LaCourse}, {Omohundro}, {Orosz},
  {Schmitt}, {Schwengeler}, {Terentev}, {Torres}, {Barclay}, {Friedman},
  {Kruse}, {Olmschenk}, {Vanderburg}, \& {Welsh}}]{2022ApJS..259...66K}
{Kostov}, V.~B., {Powell}, B.~P., {Rappaport}, S.~A., {et~al.} 2022, \apjs,
  259, 66, \dodoi{10.3847/1538-4365/ac5458}

\bibitem[{{Kostov} {et~al.}(2024){Kostov}, {Powell}, {Rappaport}, {Borkovits},
  {Gagliano}, {Jacobsy}, {Jayaraman}, {Kristiansen}, {LaCourse}, {Mitnyan},
  {Omohundro}, {Orosz}, {P{\'a}l}, {Schmitt}, {Schwengeler}, {Terentev},
  {Torres}, {Barclay}, {Vanderburg}, \& {Welsh}}]{Kostov2024}
---. 2024, \mnras, 527, 3995, \dodoi{10.1093/mnras/stad2947}

\bibitem[{{Kov{\'a}cs} {et~al.}(2002){Kov{\'a}cs}, {Zucker}, \&
  {Mazeh}}]{kovacs02}
{Kov{\'a}cs}, G., {Zucker}, S., \& {Mazeh}, T. 2002, \aap, 391, 369,
  \dodoi{10.1051/0004-6361:20020802}

\bibitem[{{Kratter} \& {Lodato}(2016)}]{Kratter2016}
{Kratter}, K., \& {Lodato}, G. 2016, \araa, 54, 271,
  \dodoi{10.1146/annurev-astro-081915-023307}

\bibitem[{{Kristiansen} {et~al.}(2022){Kristiansen}, {Rappaport}, {Vanderburg},
  {Jacobs}, {Martin Schwengeler}, {Gagliano}, {Terentev}, {LaCourse},
  {Omohundro}, {Schmitt}, {Powell}, \& {Kostov}}]{Kristiansen2022}
{Kristiansen}, M. H.~K., {Rappaport}, S.~A., {Vanderburg}, A.~M., {et~al.}
  2022, \pasp, 134, 074401, \dodoi{10.1088/1538-3873/ac6e06}

\bibitem[{{Lightkurve Collaboration} {et~al.}(2018){Lightkurve Collaboration},
  {Cardoso}, {Hedges}, {Gully-Santiago}, {Saunders}, {Cody}, {Barclay}, {Hall},
  {Sagear}, {Turtelboom}, {Zhang}, {Tzanidakis}, {Mighell}, {Coughlin}, {Bell},
  {Berta-Thompson}, {Williams}, {Dotson}, \& {Barentsen}}]{lightkurve}
{Lightkurve Collaboration}, {Cardoso}, J.~V.~d.~M., {Hedges}, C., {et~al.}
  2018, {Lightkurve: Kepler and TESS time series analysis in Python},
  Astrophysics Source Code Library.
\newblock \doeprint{1812.013}

\bibitem[{{Mardling} \& {Aarseth}(2001)}]{Mardling2001}
{Mardling}, R.~A., \& {Aarseth}, S.~J. 2001, \mnras, 321, 398,
  \dodoi{10.1046/j.1365-8711.2001.03974.x}

\bibitem[{McKinney(2010)}]{pandas}
McKinney, W. 2010, in Proceedings of the 9th Python in Science Conference, ed.
  S.~van~der Walt \& J.~Millman, 51 -- 56

\bibitem[{{Mikkola}(2008)}]{Mikkola2008}
{Mikkola}, S. 2008, in The Cambridge N-Body Lectures, ed. S.~J. {Aarseth},
  C.~A. {Tout}, \& R.~A. {Mardling}, Vol. 760, 31,
  \dodoi{10.1007/978-1-4020-8431-7_2}

\bibitem[{{Miller} {et~al.}(2020){Miller}, {Maxted}, \& {Smalley}}]{Miller2020}
{Miller}, N.~J., {Maxted}, P.~F.~L., \& {Smalley}, B. 2020, \mnras, 497, 2899,
  \dodoi{10.1093/mnras/staa2167}

\bibitem[{{Mitnyan} {et~al.}(2020){Mitnyan}, {Borkovits}, {Rappaport},
  {P{\'a}l}, \& {Maxted}}]{Mitnyan2020}
{Mitnyan}, T., {Borkovits}, T., {Rappaport}, S.~A., {P{\'a}l}, A., \& {Maxted},
  P.~F.~L. 2020, \mnras, 498, 6034, \dodoi{10.1093/mnras/staa2762}

\bibitem[{{Moe} \& {Di Stefano}(2017)}]{Moe2017}
{Moe}, M., \& {Di Stefano}, R. 2017, \apjs, 230, 15,
  \dodoi{10.3847/1538-4365/aa6fb6}

\bibitem[{{Moe} \& {Kratter}(2018)}]{Moe2018}
{Moe}, M., \& {Kratter}, K.~M. 2018, \apj, 854, 44,
  \dodoi{10.3847/1538-4357/aaa6d2}

\bibitem[{{Moharana} {et~al.}(2024){Moharana}, {He{\l}miniak}, {Marcadon},
  {Pawar}, {Pawar}, {Konacki}, {Jord{\'a}n}, {Brahm}, \&
  {Espinoza}}]{Moharana2024}
{Moharana}, A., {He{\l}miniak}, K.~G., {Marcadon}, F., {et~al.} 2024, arXiv
  e-prints, arXiv:2405.12136, \dodoi{10.48550/arXiv.2405.12136}

\bibitem[{{Nanouris} {et~al.}(2011){Nanouris}, {Kalimeris}, {Antonopoulou}, \&
  {Rovithis-Livaniou}}]{2011A&A...535A.126N}
{Nanouris}, N., {Kalimeris}, A., {Antonopoulou}, E., \& {Rovithis-Livaniou}, H.
  2011, \aap, 535, A126, \dodoi{10.1051/0004-6361/201116707}

\bibitem[{{Nanouris} {et~al.}(2015){Nanouris}, {Kalimeris}, {Antonopoulou}, \&
  {Rovithis-Livaniou}}]{2015A&A...575A..64N}
---. 2015, \aap, 575, A64, \dodoi{10.1051/0004-6361/201323136}

\bibitem[{{Naoz}(2016)}]{Naoz2016}
{Naoz}, S. 2016, \araa, 54, 441, \dodoi{10.1146/annurev-astro-081915-023315}

\bibitem[{{Nguyen} {et~al.}(2022){Nguyen}, {Costa}, {Girardi}, {Volpato},
  {Bressan}, {Chen}, {Marigo}, {Fu}, \& {Goudfrooij}}]{PARSEC}
{Nguyen}, C.~T., {Costa}, G., {Girardi}, L., {et~al.} 2022, \aap, 665, A126,
  \dodoi{10.1051/0004-6361/202244166}

\bibitem[{{Offner} {et~al.}(2023){Offner}, {Moe}, {Kratter}, {Sadavoy},
  {Jensen}, \& {Tobin}}]{Offner2023}
{Offner}, S.~S.~R., {Moe}, M., {Kratter}, K.~M., {et~al.} 2023, in Astronomical
  Society of the Pacific Conference Series, Vol. 534, Protostars and Planets
  VII, ed. S.~{Inutsuka}, Y.~{Aikawa}, T.~{Muto}, K.~{Tomida}, \& M.~{Tamura},
  275, \dodoi{10.48550/arXiv.2203.10066}

\bibitem[{{Orosz}(2023)}]{Orosz2023}
{Orosz}, J.~A. 2023, Universe, 9, 505, \dodoi{10.3390/universe9120505}

\bibitem[{{P{\'a}l}(2012)}]{Pal2012}
{P{\'a}l}, A. 2012, \mnras, 421, 1825, \dodoi{10.1111/j.1365-2966.2011.19813.x}

\bibitem[{{Paxton} {et~al.}(2011){Paxton}, {Bildsten}, {Dotter}, {Herwig},
  {Lesaffre}, \& {Timmes}}]{Paxton2011}
{Paxton}, B., {Bildsten}, L., {Dotter}, A., {et~al.} 2011, \apjs, 192, 3,
  \dodoi{10.1088/0067-0049/192/1/3}

\bibitem[{{Paxton} {et~al.}(2013){Paxton}, {Cantiello}, {Arras}, {Bildsten},
  {Brown}, {Dotter}, {Mankovich}, {Montgomery}, {Stello}, {Timmes}, \&
  {Townsend}}]{Paxton2013}
{Paxton}, B., {Cantiello}, M., {Arras}, P., {et~al.} 2013, \apjs, 208, 4,
  \dodoi{10.1088/0067-0049/208/1/4}

\bibitem[{{Paxton} {et~al.}(2015){Paxton}, {Marchant}, {Schwab}, {Bauer},
  {Bildsten}, {Cantiello}, {Dessart}, {Farmer}, {Hu}, {Langer}, {Townsend},
  {Townsley}, \& {Timmes}}]{Paxton2015}
{Paxton}, B., {Marchant}, P., {Schwab}, J., {et~al.} 2015, \apjs, 220, 15,
  \dodoi{10.1088/0067-0049/220/1/15}

\bibitem[{{Paxton} {et~al.}(2018){Paxton}, {Schwab}, {Bauer}, {Bildsten},
  {Blinnikov}, {Duffell}, {Farmer}, {Goldberg}, {Marchant}, {Sorokina},
  {Thoul}, {Townsend}, \& {Timmes}}]{Paxton2018}
{Paxton}, B., {Schwab}, J., {Bauer}, E.~B., {et~al.} 2018, \apjs, 234, 34,
  \dodoi{10.3847/1538-4365/aaa5a8}

\bibitem[{Pedregosa {et~al.}(2011)Pedregosa, Varoquaux, Gramfort, Michel,
  Thirion, Grisel, Blondel, Prettenhofer, Weiss, Dubourg, Vanderplas, Passos,
  Cournapeau, Brucher, Perrot, \& Duchesnay}]{scikit-learn}
Pedregosa, F., Varoquaux, G., Gramfort, A., {et~al.} 2011, Journal of Machine
  Learning Research, 12, 2825

\bibitem[{P\'erez \& Granger(2007)}]{ipython}
P\'erez, F., \& Granger, B.~E. 2007, Computing in Science and Engineering, 9,
  21, \dodoi{10.1109/MCSE.2007.53}

\bibitem[{{Powell} {et~al.}(2021{\natexlab{a}}){Powell}, {Kostov}, {Rappaport},
  {Borkovits}, {Zasche}, {Tokovinin}, {Kruse}, {Latham}, {Montet}, {Jensen},
  {Jayaraman}, {Collins}, {Ma{\v{s}}ek}, {Hellier}, {Evans}, {Tan},
  {Schlieder}, {Torres}, {Smale}, {Friedman}, {Barclay}, {Gagliano},
  {Quintana}, {Jacobs}, {Gilbert}, {Kristiansen}, {Col{\'o}n}, {LaCourse},
  {Olmschenk}, {Omohundro}, {Schnittman}, {Schwengeler}, {Barry}, {Terentev},
  {Boyd}, {Schmitt}, {Quinn}, {Vanderburg}, {Palle}, {Armstrong}, {Ricker},
  {Vanderspek}, {Seager}, {Winn}, {Jenkins}, {Caldwell}, {Wohler}, {Shiao},
  {Burke}, {Daylan}, \& {Villase{\~n}or}}]{2021AJ....161..162P}
{Powell}, B.~P., {Kostov}, V.~B., {Rappaport}, S.~A., {et~al.}
  2021{\natexlab{a}}, \aj, 161, 162, \dodoi{10.3847/1538-3881/abddb5}

\bibitem[{{Powell} {et~al.}(2021{\natexlab{b}}){Powell}, {Kostov}, {Rappaport},
  {Tokovinin}, {Shporer}, {Collins}, {Corbett}, {Borkovits}, {Gary}, {Chiang},
  {Rodriguez}, {Law}, {Barclay}, {Gagliano}, {Vanderburg}, {Olmschenk},
  {Kruse}, {Schlieder}, {Soto}, {Goeke}, {Jacobs}, {Kristiansen}, {LaCourse},
  {Omohundro}, {Schwengeler}, {Terentev}, \& {Schmitt}}]{2021AJ....162..299P}
---. 2021{\natexlab{b}}, \aj, 162, 299, \dodoi{10.3847/1538-3881/ac2c81}

\bibitem[{{Powell} {et~al.}(2022{\natexlab{a}}){Powell}, {Kruse}, {Montet},
  {Feinstein}, {Lewis}, {Foreman-Mackey}, {Barclay}, {Quintana}, {Col{\'o}n},
  {Kostov}, {Boyd}, {Smale}, {Mullally}, {Schlieder}, {Schnittman}, {Carroll},
  {Carriere}, {Salmon}, {Strong}, {Acks}, {Pfaff}, {Gerner}, \&
  {Burch}}]{Powell2022_eleanor_lite}
{Powell}, B.~P., {Kruse}, E., {Montet}, B.~T., {et~al.} 2022{\natexlab{a}},
  Research Notes of the American Astronomical Society, 6, 111,
  \dodoi{10.3847/2515-5172/ac74c4}

\bibitem[{{Powell} {et~al.}(2022{\natexlab{b}}){Powell}, {Rappaport},
  {Borkovits}, {Kostov}, {Torres}, {Jayaraman}, {Latham},
  {Ku{\v{c}}{\'a}kov{\'a}}, {Garai}, {Pribulla}, {Vanderburg}, {Kruse},
  {Barclay}, {Olmschenk}, {Kristiansen}, {Gagliano}, {Jacobs}, {LaCourse},
  {Omohundro}, {Schwengeler}, {Terentev}, \& {Schmitt}}]{Powell2022}
{Powell}, B.~P., {Rappaport}, S.~A., {Borkovits}, T., {et~al.}
  2022{\natexlab{b}}, \apj, 938, 133, \dodoi{10.3847/1538-4357/ac8934}

\bibitem[{{Powell, Brian}(2022)}]{eleanor_lite}
{Powell, Brian}. 2022, TESS FFI-Based Light Curves from the GSFC Team
  ("GSFC-ELEANOR-LITE"),  STScI/MAST, \dodoi{10.17909/J2YT-T417}

\bibitem[{{Pribulla} {et~al.}(2023){Pribulla}, {Borkovits}, {Jayaraman},
  {Rappaport}, {Mitnyan}, {Zasche}, {Kom{\v{z}}{\'\i}k}, {P{\'a}l},
  {Uhla{\v{r}}}, {Ma{\v{s}}ek}, {Henzl}, {B{\'\i}r{\'o}}, {Cs{\'a}nyi},
  {Stuik}, {Kristiansen}, {Schwengeler}, {Gagliano}, {Jacobs}, {Omohundro},
  {Kostov}, {Powell}, {Terentev}, {Vanderburg}, {LaCourse}, {Rodriguez},
  {Bakos}, {Csubry}, \& {Hartman}}]{2023MNRAS.524.4220P}
{Pribulla}, T., {Borkovits}, T., {Jayaraman}, R., {et~al.} 2023, \mnras, 524,
  4220, \dodoi{10.1093/mnras/stad2015}

\bibitem[{{Raghavan} {et~al.}(2010){Raghavan}, {McAlister}, {Henry}, {Latham},
  {Marcy}, {Mason}, {Gies}, {White}, \& {ten Brummelaar}}]{Raghavan2010}
{Raghavan}, D., {McAlister}, H.~A., {Henry}, T.~J., {et~al.} 2010, \apjs, 190,
  1, \dodoi{10.1088/0067-0049/190/1/1}

\bibitem[{{Rappaport} {et~al.}(2013){Rappaport}, {Deck}, {Levine}, {Borkovits},
  {Carter}, {El Mellah}, {Sanchis-Ojeda}, \& {Kalomeni}}]{rappaport2013}
{Rappaport}, S., {Deck}, K., {Levine}, A., {et~al.} 2013, \apj, 768, 33,
  \dodoi{10.1088/0004-637X/768/1/33}

\bibitem[{{Rappaport} {et~al.}(2022){Rappaport}, {Borkovits}, {Gagliano},
  {Jacobs}, {Kostov}, {Powell}, {Terentev}, {Omohundro}, {Torres},
  {Vanderburg}, {Mitnyan}, {Kristiansen}, {LaCourse}, {Schwengeler}, {Kaye},
  {P{\'a}l}, {Pribulla}, {B{\'\i}r{\'o}}, {Cs{\'a}nyi}, {Garai}, {Zasche},
  {Maxted}, {Rodriguez}, \& {Stevens}}]{rappaport22a}
{Rappaport}, S.~A., {Borkovits}, T., {Gagliano}, R., {et~al.} 2022, \mnras,
  513, 4341, \dodoi{10.1093/mnras/stac957}

\bibitem[{{Rappaport} {et~al.}(2023){Rappaport}, {Borkovits}, {Gagliano},
  {Jacobs}, {Tokovinin}, {Mitnyan}, {Kom{\v{z}}{\'\i}k}, {Kostov}, {Powell},
  {Torres}, {Terentev}, {Omohundro}, {Pribulla}, {Vanderburg}, {Kristiansen},
  {Latham}, {Schwengeler}, {LaCourse}, {B{\'\i}r{\'o}}, {Cs{\'a}nyi},
  {Czavalinga}, {Garai}, {P{\'a}l}, {Rodriguez}, \& {Stevens}}]{Rappaport2023a}
---. 2023, \mnras, 521, 558, \dodoi{10.1093/mnras/stad367}

\bibitem[{{Rappaport} {et~al.}(2024){Rappaport}, {Borkovits}, {Mitnyan},
  {Gagliano}, {Eisner}, {Jacobs}, {Tokovinin}, {Powell}, {Kostov}, {Omohundro},
  {Kristiansen}, {Jayaraman}, {Terentev}, {Schwengeler}, {LaCourse}, {Garai},
  {Pribulla}, {Maxted}, {B{\'\i}r{\'o}}, {Cs{\'a}nyi}, {P{\'a}l}, \&
  {Vanderburg}}]{Rappaport2024}
{Rappaport}, S.~A., {Borkovits}, T., {Mitnyan}, T., {et~al.} 2024, \aap, 686,
  A27, \dodoi{10.1051/0004-6361/202449273}

\bibitem[{Rein \& Liu(2012)}]{Rein12}
Rein, H., \& Liu, S.-F. 2012, Astronomy \& Astrophysics, 537, A128

\bibitem[{{Schmitt} \& {Vanderburg}(2021)}]{2021arXiv210310285S}
{Schmitt}, A., \& {Vanderburg}, A. 2021, arXiv e-prints, arXiv:2103.10285.
\newblock \doarXiv{2103.10285}

\bibitem[{{Schmitt} {et~al.}(2019){Schmitt}, {Hartman}, \&
  {Kipping}}]{Schmitt2019}
{Schmitt}, A.~R., {Hartman}, J.~D., \& {Kipping}, D.~M. 2019, arXiv e-prints,
  arXiv:1910.08034.
\newblock \doarXiv{1910.08034}

\bibitem[{{Scott} {et~al.}(2021){Scott}, {Howell}, {Gnilka}, {Stephens},
  {Salinas}, {Matson}, {Furlan}, {Horch}, {Everett}, {Ciardi}, {Mills}, \&
  {Quigley}}]{scott2021}
{Scott}, N.~J., {Howell}, S.~B., {Gnilka}, C.~L., {et~al.} 2021, Frontiers in
  Astronomy and Space Sciences, 8, 138, \dodoi{10.3389/fspas.2021.716560}

\bibitem[{{Skrutskie} {et~al.}(2006){Skrutskie}, {Cutri}, {Stiening},
  {Weinberg}, {Schneider}, {Carpenter}, {Beichman}, {Capps}, {Chester},
  {Elias}, {Huchra}, {Liebert}, {Lonsdale}, {Monet}, {Price}, {Seitzer},
  {Jarrett}, {Kirkpatrick}, {Gizis}, {Howard}, {Evans}, {Fowler}, {Fullmer},
  {Hurt}, {Light}, {Kopan}, {Marsh}, {McCallon}, {Tam}, {Van Dyk}, \&
  {Wheelock}}]{2MASS}
{Skrutskie}, M.~F., {Cutri}, R.~M., {Stiening}, R., {et~al.} 2006, \aj, 131,
  1163, \dodoi{10.1086/498708}

\bibitem[{{Stassun} \& {Torres}(2016)}]{stassun16}
{Stassun}, K.~G., \& {Torres}, G. 2016, \aj, 152, 180,
  \dodoi{10.3847/0004-6256/152/6/180}

\bibitem[{{Stassun} {et~al.}(2018){Stassun}, {Oelkers}, {Pepper}, {Paegert},
  {De Lee}, {Torres}, {Latham}, {Charpinet}, {Dressing}, {Huber}, {Kane},
  {L{\'e}pine}, {Mann}, {Muirhead}, {Rojas-Ayala}, {Silvotti}, {Fleming},
  {Levine}, \& {Plavchan}}]{TIC}
{Stassun}, K.~G., {Oelkers}, R.~J., {Pepper}, J., {et~al.} 2018, \aj, 156, 102,
  \dodoi{10.3847/1538-3881/aad050}

\bibitem[{{Tamayo} {et~al.}(2020){Tamayo}, {Rein}, {Shi}, \&
  {Hernandez}}]{Tamayo2020}
{Tamayo}, D., {Rein}, H., {Shi}, P., \& {Hernandez}, D.~M. 2020, \mnras, 491,
  2885, \dodoi{10.1093/mnras/stz2870}

\bibitem[{{Tokovinin}(2014)}]{Tokovinin2014}
{Tokovinin}, A. 2014, \aj, 147, 87, \dodoi{10.1088/0004-6256/147/4/87}

\bibitem[{{Tokovinin}(2017{\natexlab{a}})}]{Tokovinin2017b}
---. 2017{\natexlab{a}}, \apj, 844, 103, \dodoi{10.3847/1538-4357/aa7746}

\bibitem[{{Tokovinin}(2017{\natexlab{b}})}]{Tokovinin2017}
---. 2017{\natexlab{b}}, \mnras, 468, 3461, \dodoi{10.1093/mnras/stx707}

\bibitem[{{Tokovinin}(2021)}]{Tokovinin2021}
---. 2021, Universe, 7, 352, \dodoi{10.3390/universe7090352}

\bibitem[{{Tokovinin} \& {Moe}(2020)}]{Tokovinin2020}
{Tokovinin}, A., \& {Moe}, M. 2020, \mnras, 491, 5158,
  \dodoi{10.1093/mnras/stz3299}

\bibitem[{{Tokovinin} {et~al.}(2006){Tokovinin}, {Thomas}, {Sterzik}, \&
  {Udry}}]{Tokovinin2006}
{Tokovinin}, A., {Thomas}, S., {Sterzik}, M., \& {Udry}, S. 2006, \aap, 450,
  681, \dodoi{10.1051/0004-6361:20054427}

\bibitem[{{Tokovinin}(2000)}]{Tokovinin2000}
{Tokovinin}, A.~A. 2000, \aap, 360, 997

\bibitem[{{Valtonen} \& {Karttunen}(2006)}]{Valtonen2006}
{Valtonen}, M., \& {Karttunen}, H. 2006, {The Three-Body Problem}

\bibitem[{{Virtanen} {et~al.}(2020){Virtanen}, {Gommers}, {Oliphant},
  {Haberland}, {Reddy}, {Cournapeau}, {Burovski}, {Peterson}, {Weckesser},
  {Bright}, {van der Walt}, {Brett}, {Wilson}, {Jarrod Millman}, {Mayorov},
  {Nelson}, {Jones}, {Kern}, {Larson}, {Carey}, {Polat}, {Feng}, {Moore},
  {VanderPlas}, {Laxalde}, {Perktold}, {Cimrman}, {Henriksen}, {Quintero},
  {Harris}, {Archibald}, {Ribeiro}, {Pedregosa}, {van Mulbregt}, \&
  {Contributors}}]{scipy}
{Virtanen}, P., {Gommers}, R., {Oliphant}, T.~E., {et~al.} 2020, Nature
  Methods, \dodoi{https://doi.org/10.1038/s41592-019-0686-2}

\bibitem[{{Young} \& {Clarke}(2015)}]{Young2015}
{Young}, M.~D., \& {Clarke}, C.~J. 2015, \mnras, 452, 3085,
  \dodoi{10.1093/mnras/stv1512}

\end{thebibliography}
\bibliographystyle{aasjournal}

\appendix

\section{Eclipse Times for the Inner 1.792 Day Binary in TIC 290061484}
\label{app:eclipse_times}

Table \ref{Tab:TIC_290061484_ToM} tabulates the mid-times of the regular binary eclipses of TIC 290061484 during the six sectors of TESS observations. Primary eclipses are designated by integer cycle numbers, while secondary eclipses have half integer values.

\begin{table*}
 \centering
\caption{Eclipse Times for the Inner Binary in TIC 290061484, extracted from the \textsc{FITSH} light curve}
 \label{Tab:TIC_290061484_ToM}
\begin{tabular}{@{}lrllrllrllrl}
\hline
BJD & Cycle  & std. dev. & BJD & Cycle  & std. dev. & BJD & Cycle  & std. dev. & BJD & Cycle  & std. dev. \\ 
$-2\,400\,000$ & no. &   \multicolumn{1}{c}{$(d)$} & $-2\,400\,000$ & no. &   \multicolumn{1}{c}{$(d)$} & $-2\,400\,000$ & no. &   \multicolumn{1}{c}{$(d)$} & $-2\,400\,000$ & no. &   \multicolumn{1}{c}{$(d)$} \\ 
\hline
58712.12602 &    -0.5 & 0.00030 & 58754.25374 &    23.0 & 0.00026 & 59828.10505 &   622.0 & 0.00021 & 60356.96993 &   917.0 & 0.00033  \\ 
58713.02277 &     0.0 & 0.00033 & 58755.15171 &    23.5 & 0.00030 & 59829.00452 &   622.5 & 0.00025 & 60357.86671 &   917.5 & 0.00040  \\ 
58713.91819 &     0.5 & 0.00034 & 58756.05013 &    24.0 & 0.00036 & 59829.89724 &   623.0 & 0.00025 & 60358.76081 &   918.0 & 0.00037  \\ 
58714.81379 &     1.0 & 0.00038 & 58756.94418 &    24.5 & 0.00035 & 59830.79770 &   623.5 & 0.00027 & 60359.66057 &   918.5 & 0.00030  \\ 
58715.70989 &     1.5 & 0.00038 & 58757.84118 &    25.0 & 0.00032 & 59831.69366 &   624.0 & 0.00037 & 60360.55246 &   919.0 & 0.00033  \\ 
58716.60597 &     2.0 & 0.00040 & 58758.73779 &    25.5 & 0.00039 & 59832.59484 &   624.5 & 0.00026 & 60361.44858 &   919.5 & 0.00032  \\ 
58717.49993 &     2.5 & 0.00032 & 58759.63555 &    26.0 & 0.00030 & 59833.49147 &   625.0 & 0.00024 & 60362.34170 &   920.0 & 0.00029  \\ 
58718.39713 &     3.0 & 0.00033 & 58760.53045 &    26.5 & 0.00040 & 59834.39058 &   625.5 & 0.00026 & 60363.23957 &   920.5 & 0.00033  \\ 
58719.29179 &     3.5 & 0.00040 & 58761.42510 &    27.0 & 0.00037 & 59835.28514 &   626.0 & 0.00022 & 60364.13210 &   921.0 & 0.00030  \\ 
58720.18866 &     4.0 & 0.00036 & 58762.32104 &    27.5 & 0.00054 & 59836.18435 &   626.5 & 0.00027 & 60365.02953 &   921.5 & 0.00031  \\ 
58721.08454 &     4.5 & 0.00046 & 58763.21913 &    28.0 & 0.00060 & 59837.07734 &   627.0 & 0.00023 & 60365.92250 &   922.0 & 0.00027  \\ 
58721.97972 &     5.0 & 0.00054 & 59797.63700 &   605.0 & 0.00026 & 59837.97882 &   627.5 & 0.00024 & 60366.82050 &   922.5 & 0.00031  \\ 
58722.87403 &     5.5 & 0.00056 & 59798.53506 &   605.5 & 0.00032 & 59838.87225 &   628.0 & 0.00021 & 60367.71511 &   923.0 & 0.00074  \\ 
58723.77121 &     6.0 & 0.00040 & 59799.42739 &   606.0 & 0.00030 & 59839.77318 &   628.5 & 0.00021 & 60368.61254 &   923.5 & 0.00032  \\ 
58725.56272 &     7.0 & 0.00026 & 59800.32441 &   606.5 & 0.00031 & 59840.66742 &   629.0 & 0.00024 & 60369.50614 &   924.0 & 0.00033  \\ 
58726.45751 &     7.5 & 0.00025 & 59801.21684 &   607.0 & 0.00030 & 59841.56638 &   629.5 & 0.00025 & 60370.40529 &   924.5 & 0.00036  \\ 
58727.35651 &     8.0 & 0.00032 & 59802.11742 &   607.5 & 0.00034 & 59842.45776 &   630.0 & 0.00023 & 60371.30248 &   925.0 & 0.00035  \\ 
58728.25469 &     8.5 & 0.00038 & 59803.00789 &   608.0 & 0.00033 & 59843.35765 &   630.5 & 0.00024 & 60372.20108 &   925.5 & 0.00037  \\ 
58729.15509 &     9.0 & 0.00030 & 59803.90699 &   608.5 & 0.00036 & 59844.24903 &   631.0 & 0.00019 & 60373.09690 &   926.0 & 0.00034  \\ 
58730.05272 &     9.5 & 0.00040 & 59804.80041 &   609.0 & 0.00036 & 59845.14801 &   631.5 & 0.00023 & 60373.99589 &   926.5 & 0.00036  \\ 
58730.95139 &    10.0 & 0.00034 & 59805.69847 &   609.5 & 0.00036 & 59846.93883 &   632.5 & 0.00027 & 60374.89047 &   927.0 & 0.00029  \\ 
58731.84611 &    10.5 & 0.00037 & 59806.59422 &   610.0 & 0.00029 & 59847.83099 &   633.0 & 0.00024 & 60375.78906 &   927.5 & 0.00031  \\ 
58732.74441 &    11.0 & 0.00035 & 59807.49445 &   610.5 & 0.00036 & 59848.73015 &   633.5 & 0.00028 & 60376.68503 &   928.0 & 0.00027  \\ 
58733.63745 &    11.5 & 0.00045 & 59808.39066 &   611.0 & 0.00024 & 59849.62115 &   634.0 & 0.00025 & 60377.58547 &   928.5 & 0.00029  \\ 
58734.53561 &    12.0 & 0.00044 & 59809.29369 &   611.5 & 0.00026 & 59850.51989 &   634.5 & 0.00026 & 60379.38123 &   929.5 & 0.00028  \\ 
58735.43090 &    12.5 & 0.00051 & 59811.08573 &   612.5 & 0.00023 & 59851.41039 &   635.0 & 0.00024 & 60380.27567 &   930.0 & 0.00027  \\ 
58736.32880 &    13.0 & 0.00053 & 59811.98073 &   613.0 & 0.00021 & 59852.31166 &   635.5 & 0.00029 & 60381.17286 &   930.5 & 0.00029  \\ 
58737.22428 &    13.5 & 0.00046 & 59812.87843 &   613.5 & 0.00021 & 60339.93283 &   907.5 & 0.00028 & 60382.06628 &   931.0 & 0.00025  \\ 
58739.01538 &    14.5 & 0.00033 & 59813.77544 &   614.0 & 0.00026 & 60340.82678 &   908.0 & 0.00034 & 60382.96473 &   931.5 & 0.00028  \\ 
58739.91179 &    15.0 & 0.00031 & 59814.67400 &   614.5 & 0.00029 & 60341.72206 &   908.5 & 0.00036 & 60383.85827 &   932.0 & 0.00034  \\ 
58740.80702 &    15.5 & 0.00031 & 59815.56759 &   615.0 & 0.00025 & 60342.61705 &   909.0 & 0.00039 & 60384.75522 &   932.5 & 0.00035  \\ 
58741.70402 &    16.0 & 0.00032 & 59816.46730 &   615.5 & 0.00029 & 60343.51497 &   909.5 & 0.00037 & 60385.64880 &   933.0 & 0.00033  \\ 
58742.59803 &    16.5 & 0.00032 & 59817.36029 &   616.0 & 0.00024 & 60344.40906 &   910.0 & 0.00037 & 60386.54476 &   933.5 & 0.00033  \\ 
58743.49538 &    17.0 & 0.00035 & 59818.26084 &   616.5 & 0.00029 & 60347.10142 &   911.5 & 0.00039 & 60387.43746 &   934.0 & 0.00032  \\ 
58744.39082 &    17.5 & 0.00035 & 59819.15238 &   617.0 & 0.00030 & 60347.99960 &   912.0 & 0.00036 & 60388.33693 &   934.5 & 0.00071  \\ 
58745.28569 &    18.0 & 0.00037 & 59820.05140 &   617.5 & 0.00031 & 60348.89766 &   912.5 & 0.00034 & 60389.22890 &   935.0 & 0.00032  \\ 
58746.18120 &    18.5 & 0.00039 & 59820.94295 &   618.0 & 0.00024 & 60349.79219 &   913.0 & 0.00030 & 60390.12632 &   935.5 & 0.00030  \\ 
58747.07682 &    19.0 & 0.00040 & 59821.84083 &   618.5 & 0.00034 & 60350.69039 &   913.5 & 0.00032 & 60391.01885 &   936.0 & 0.00033  \\ 
58747.97316 &    19.5 & 0.00051 & 59822.73260 &   619.0 & 0.00029 & 60351.58626 &   914.0 & 0.00029 & 60391.91816 &   936.5 & 0.00034  \\ 
58748.86842 &    20.0 & 0.00063 & 59823.63252 &   619.5 & 0.00028 & 60352.48570 &   914.5 & 0.00030 & 60392.81115 &   937.0 & 0.00027  \\ 
58749.76556 &    20.5 & 0.00068 & 59825.42357 &   620.5 & 0.00029 & 60353.38337 &   915.0 & 0.00028 & 60393.70943 &   937.5 & 0.00027  \\ 
58752.45519 &    22.0 & 0.00031 & 59826.31564 &   621.0 & 0.00022 & 60355.17814 &   916.0 & 0.00030 & 60394.60643 &   938.0 & 0.00024  \\ 
58753.35304 &    22.5 & 0.00041 & 59827.21296 &   621.5 & 0.00025 & 60356.07552 &   916.5 & 0.00035  \\ 
\hline
\end{tabular}
\end{table*}

\section{Population Synthesis of Compact Multiples}
\label{sec:popsyn}

The toy model of \citet{Tokovinin2020} not only simulates close binaries, but also synthesizes compact triples and 2+1+1 quadruples while accounting for pre-MS mergers and disruptions as a result of dynamical instability. In their model, a brown dwarf companion fragments from the cool outer disk near $a$ = 200 au, the expected seed mass and location of disk fragmentation \citep{Kratter2016}. The binary subsequently accretes from a circumbinary disk, driving the companion inward and increasing the mass ratio. Mass accretion and inward migration are treated in a stochastic manner, thereby producing a variety of multiple systems but with clear correlations, e.g., very close companions that underwent significant circumbinary accretion are more likely to be twins. In their model, the protostellar disk can fragment at random times and more than once during the main accretion phase, producing a rich diversity of binaries, triples, and quadruples. 

To gain insight into the rarity of TIC~290061484, we run the population synthesis code of \citet{Tokovinin2020} for 10$^7$ systems. Instead of assuming circular orbits, we draw outer components from a uniform eccentricity distribution across $e_{\rm out}$ = 0.0\,-\,0.8, consistent with the observed eccentricity distribution of companions beyond $P$ $>$ 20 days that have not been tidally circularized \citep{Tokovinin2014,Moe2017}. We then impose the eccentricity-dependent stability criterion of Eqn.~\ref{eqn:stableP1}, and assume that outer components that migrate within this limit are ejected. We adopt the baseline solar-type model parameters from \citet{Tokovinin2020} except that we set the input parameter that regulates the probability of disk fragmentation to $f_{\rm bin}$ = 0.12, which reproduces several observed properties of solar-type multiples. Specifically, this model yields a very close binary fraction within $P$ $<$ 10 days of 2\%, a binary fraction within $P$ $<$ 1,000 days of 11\%, and a triple fraction within $P_{\rm out}$ $<$ 10,000 days of 0.3\%.

\begin{figure}
    \centering
    \includegraphics[width=0.5\textwidth]{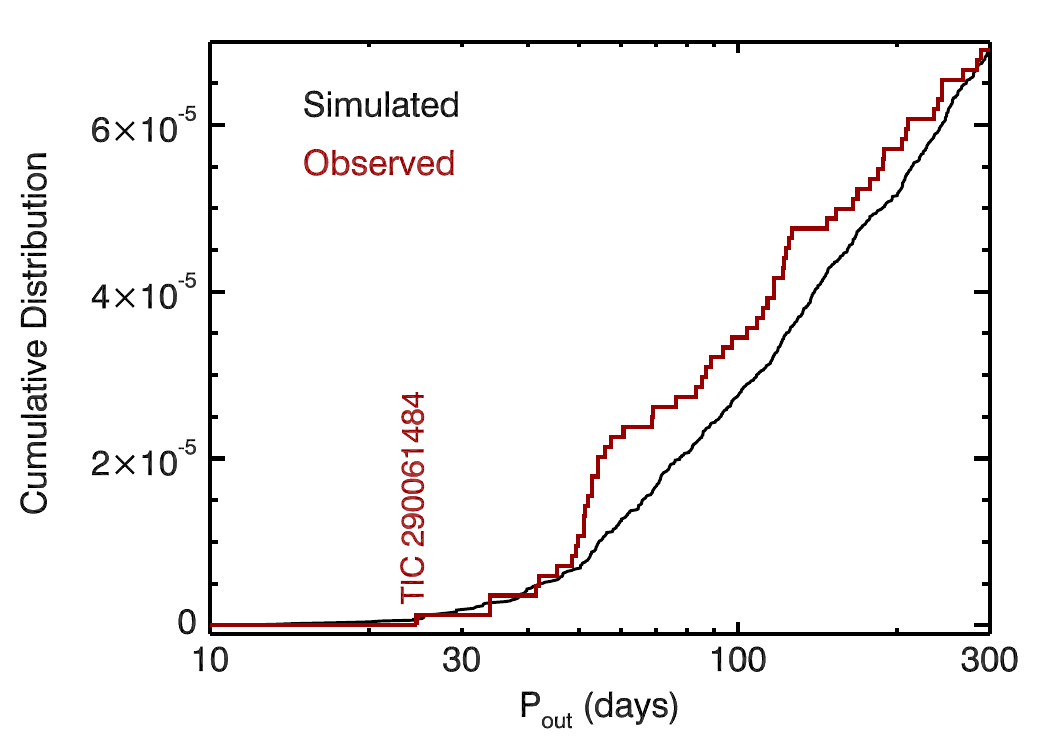}
   \caption{Cumulative distributions of tertiary periods $P_{\rm out}$ for the simulated population (black) and 59 observed compact triples with $P_{\rm out}$ $<$ 300 days (red).}
    \label{fig:Poutdist}
\end{figure}

In our simulation, the fractions of final systems with tertiaries below $P_{\rm out}$ $<$ 300, 50, and 25 days are 7$\times$10$^{-5}$, 6$\times$10$^{-6}$, and 8$\times$10$^{-7}$ respectively (see Fig.~\ref{fig:Poutdist}). These results roughly match the observed ratios 59:9:1 in our current sample\footnote{ We have added 15 additional triples to the set of 44 systems shown in Fig.~\ref{fig:massratios} with $P_{\rm out}$ $<$ 300 days but without individually measured component masses.  This brings the total  number of compact triples in our sample to 59 objects.}, demonstrating that the \citet{Tokovinin2020} toy model can approximately reproduce the tertiary period distribution of compact triples. About 0.3\% of our simulated systems resulted in pre-MS mergers and another 0.8\% were triples that became dynamically disrupted. Indeed, the formation of an ultra-compact triple is a delicate process. For every triple like TIC~290061484 that survived the hurdle of maintaining dynamical stability while migrating to $P_{\rm out}$ $<$ 25 days, we expect $\sim$10$^4$ triples to become dynamically unstable and disrupted during their formation process.  More compact configurations are possible but substantially rarer. The tightest triple in our simulation of 10$^7$ systems has $P_{\rm out}$ = 10~days, consistent with our discussion in Section~\ref{sec:stability}.

Our simulation predicts there is one ultra-compact triple like TIC~290061484 for every 1.3 million star systems. Accounting for the 10\% probability that the tertiary is oriented across $i$ $\approx$ 84$^{\circ}$\,-\,96$^{\circ}$ to be detected as a triply eclipsing triple, then TIC~290061484 represents a unique object out of $\sim$13 million systems. Given the finite sensitivity and selection biases of TESS, it is thus not surprising it required searching through $\sim$100 million stars and $\sim$1 million eclipsing binaries to find a rare ultra-compact triple such as TIC~290061484.

Our simulation also yielded 47 2+1+1 quadruples ($F_{\rm quad}$ $\approx$ 5$\times$10$^{-6}$). The observed quadruple fraction of solar-type stars is considerably larger at 4\% \citep{Tokovinin2014}. Wider companions more likely derive from core fragmentation \citep{Tokovinin2017}, which is not encapsulated in the \citet{Tokovinin2020} toy model of disk fragmentation, migration, and accretion. Indeed, the formation of the relatively common 2+2 quadruple architecture requires that the outer pair first formed via core fragmentation and then both of the resulting protostellar disks could subsequently fragment. 

Of the six 2+1+1 quadruples in our simulation with $P_{\rm quad}$ $<$ 30,000 days, all have extremely compact tertiaries with $P_{\rm tert}$ $<$ 300 days. As discussed above, formation of very close binaries requires substantially migration in a massive disk/envelope, which is also more prone to fragment twice into a triple system. Similarly, the formation and migration of extremely compact triples with $P_{\rm tert}$ $<$ 50 days requires an even more massive disk/envelope, which is more likely to fragment three times into a 2+1+1 quadruple. It is therefore not surprising that TIC~290061484, the most compact triple yet discovered, also contains an additional component at $P$ $\approx$ 3,200 days. 



\end{document}